%% file: main.tex
\documentclass[aps,pre,superscriptaddress,twocolumn,showpacs]{revtex4}
\usepackage{graphicx}
\usepackage{times}
\usepackage{amsmath}

\newcommand\order{O} 
\newcommand\ie{i.e.,~} 

\newcommand\erfc{\mbox{erfc}} 

\newcommand\sign{\,\mbox{sign}}
\newcommand{\lss}[2]{\ensuremath{#1_{\mbox{\scriptsize #2}}}}

\begin{document}

\newcommand{\affOne}{\affiliation{Laboratoire de Physique Th\'eorique et
  Mod\`eles Statistiques, Universit\'e Paris-Sud, B\^atiment 100, 91405
  Orsay Cedex, France}}

\newcommand{\affTwo}{\affiliation{Universit\'e Pierre et Marie Curie - Paris
  6, Institut Henri Poincar\'e, 11 rue Pierre et Marie Curie, Paris,
  F-75005, France}}

%%%%%%%%%%%%%%%%%%%%
%%% FRONTMATTER %%%%
%%%%%%%%%%%%%%%%%%%%

\title{Statistical Properties of Functionals of the Paths of a Particle\\
Diffusing in a One-Dimensional Random Potential}

\author{Sanjib Sabhapandit} \affOne \affTwo

\author{Satya N. Majumdar} \affOne

\author{Alain Comtet} \affOne \affTwo

\date{\today}
 
\begin{abstract}
We present a formalism for obtaining the statistical properties of
functionals and inverse functionals of the paths of a particle diffusing in
a one-dimensional quenched random potential. We demonstrate the
implementation of the formalism in two specific examples: (1) where the
functional corresponds to the local time spent by the particle around the
origin and (2) where the functional corresponds to the occupation time spent
by the particle on the positive side of the origin, within an observation
time window of size $t$.  We compute the disorder average distributions of
the local time, the inverse local time, the occupation time and the inverse
occupation time, and show that in many cases disorder modifies the behavior
drastically.
\end{abstract}

\pacs{05.40.-a, 02.50.-r, 46.65.+g}

\maketitle
%\tableofcontents

%%%%%%%%%%%%%%%%%%%%%%%%%%%
\section{Introduction}
\label{introduction}
%%%%%%%%%%%%%%%%%%%%%%%%%%%%

\begin{figure}
\centering
\includegraphics[width=.9\hsize]{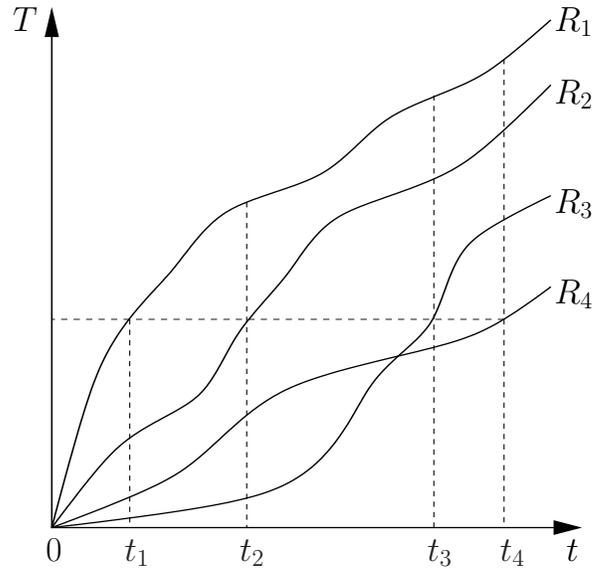}
\caption{\label{T-t} Schematic plots of $T$ defined by Eq.~(\ref{definition
of T}) as a function of $t$, corresponding to four different realizations of
the paths $[\{x(\tau)\}, ~\mbox{for}~ 0\le\tau\le t]$ denoted by $R_1, R_2,
R_3$ and $R_4$ respectively. For fixed $t$ ($t_1$ or $t_2$ or $t_3$ or
$t_4$, shown by vertical dashed lines), $T$ takes different value for
different realizations. On the other hand for a fixed $T$ (horizontal dashed
line) corresponding $t$ is different for different realizations: $t_1$ for
$R_1$, $t_2$ for $R_2$, $t_3$ for $R_3$ and $t_4$ for $R_4$.}
\end{figure}

The statistical properties of functionals of a one dimensional Brownian
motion have been extensively studied and have found numerous applications in
diverse fields ranging from probability
theory~\cite{MR0027960,MR0045333,yor:1992(book)},
finance~\cite{yor:2000(book),MR1129194,geman:1993}, mesocopic
physics~\cite{0305-4470-38-37-R01}, computer
science~\cite{majumdar:0510064}, and in understanding weather
records~\cite{majumdar:051112}. The position $x(\tau)$ of a one dimensional
Brownian motion evolves with time $\tau$ via the Langevin equation
\begin{equation}
\frac{dx}{d\tau}=\eta(\tau),
\label{eq.1}
\end{equation}
starting from $x(0)=x_0$, where $\eta(\tau)$ is a thermal Gaussian white
noise with mean $\langle \eta(\tau)\rangle=0$ and a correlator $\langle
\eta(\tau)\eta(\tau')\rangle = \delta(\tau-\tau')$.  A functional $T$ is
simply the integral up to time $t$
\begin{equation}
T = \int_0^t V\biglb(x(\tau)\bigrb)\, d\tau,
\label{definition of T}
\end{equation}
where $V(x)$ is a prescribed non-negative function whose choice depends on
the specific observable of interest.  For a fixed initial position $x_0$ of
the Brownian motion and a fixed observation time $t$, the value of $T$
varies from one history or realization of the Brownian path $\{x(\tau)\}$ to
another (see Fig.~\ref{T-t}) and a natural question is: what is the
probability density function (pdf) $P(T|t,x_0)$?

%%%%%%%%%%%%%%%%%%%%%%%%%%%%%%%%%%%%%
%%%%%%%%%%%%%%%%%%%%%%%%%%%%%%%%%%%%%%%

Following the path integral methods devised by Feynman~\cite{Feynman-Hibbs},
Kac showed~\cite{MR0027960,MR0045333} that the calculation of the pdf
$P(T|t,x_0)$ can essentially be reduced to a quantum mechanics problem,
namely solving a single particle Shr\"odinger equation in an external
potential $V(x)$. This formalism is known in the literature as the
celebrated Feynman-Kac formula. Subsequently, this method has been widely
used to calculate the pdf of $T$ with different choices of $V(x)$ as
demanded by specific applications.  This has been reviewed recently in
Ref.~\cite{majumdar:0510064}. In particular, two most popular applications
correspond respectively to the choices, $V(x)= \delta(x-a)$ and
$V(x)=\theta(x)$, where $\delta(x)$ is the Dirac's delta function and
$\theta(x)$ is the Heaviside step function. In the former case, the
corresponding functional $T(a)=\int_0^t \delta(x(\tau)-a)\, d\tau$ has the
following physical meaning: $T(a)\,da$ is just the time spent by the
particle in the vicinity of the point $a$ in space, \ie in the region
$[a,a+da]$, out of the total observation time $t$. Note that, by definition,
$\int T(a)\,da=t$. The functional $T(a)$ is known as the `local time'
(density) in the literature.  In the second case $V(x)=\theta(x)$, the
functional $T=\int_0^t \theta(x(\tau))\,d\tau$ measures the time spent by
the particle on the positive side of the origin out of the total time $t$
and is known as the `occupation' time.  The probability distribution of the
occupation time was originally computed by L\'evy~\cite{levy:1939}, $
\int_0^T P(T'|t,0)\,dT' = \frac{2}{\pi} \arcsin \left(\sqrt{T/t}\right) $,
and is known as the arcsine law of L\'evy. Since then, the local and the
occupation times for pure diffusion, have been studied extensively by
mathematicians~\cite{darling:1957, lamperti:1958, ito:1974, watanabe:1995,
yor:2003, borodin:2002}.  Recently, the study of the occupation time has
seen a revival in physics literature and has been used in understanding the
dynamics out of equilibrium in coarsening systems~\cite{0305-4470-31-24-004,
newman:1998}, ergodicity properties in anomalously diffusive
processes~\cite{dhar:6413, 0305-4470-34-7-303}, in renewal
processes~\cite{godreche:jsp:2001}, in models related to spin
glasses~\cite{majumdar:041102}, in simple models of blinking quantum
dots~\cite{margolin:080601}, and also in the context of
persistence~\cite{ehrhardt:016106, majumdar:persistence}.  Local and
Occupation times have been also studied in the context of stochastic
ergodicity breaking~\cite{bel:240602}, first passage
time~\cite{barkai:cond-mat/0601143}, diffusion controlled reactions
activated by catalytic sites~\cite{0305-4470-38-33-001} and diffusion on
graphs~\cite{0305-4470-35-46-101, 0305-4470-35-47-102}.  In polymer science,
a long flexible polymer of length $t$ is often modeled by a Brownian path up
to time $t$.  In this context, the local time at a position $\vec{r}$ is
proportional to the concentration of monomers at $\vec{r}$ in a polymer of
length $t$.

A natural and important question is how to generalize the Feynman-Kac
formalism to calculate the statistical properties of the functionals of the
type in Eq.~(\ref{definition of T}) when $x(\tau)$ is not just a pure
diffusion process, but it represents the position of a particle in an
external random medium. While various properties of diffusion in random
media have been widely studied in the past~\cite{1990PhR...195..127B,
bouchaud:ap:1990, 1987PhR...150..263H, havlin:advp:1987}, the study of
statistical properties of functionals in random media is yet to receive its
much deserved attention. In this paper we undertake this task. More
precisely, we are interested in calculating the pdf $P(T|t,x_0)$ of a
functional $T$ as in Eq.~(\ref{definition of T}) where $x(\tau)$ now evolves
via the Langevin equation
\begin{equation}
\frac{dx}{d\tau} = F\biglb(x(\tau)\bigrb) + \eta(\tau)
\label{Langevin}
\end{equation}
where $\eta(\tau)$ represents the thermal noise as in Eq.~(\ref{eq.1}) and
$F(x)=-dU/dx$ represents the external force, the derivative of the potential
$U(x)$, felt by the particle.  Most generally, the external potential
consists of two parts, $U(x)=U_d(x)+U_r(x)$, a deterministic part $U_d(x)$
and a random part $U_r(x)$. The random part of the potential $U_r(x)$ is
`quenched' in the sense that it does not change during the time evolution of
the particle, but fluctuates from one sample to another according to some
prescribed probability distribution.  Consequently, the pdf $P(T|t,x_0)$
will also fluctuate from one sample of the random potential to another and
the goal is to compute the disorder averaged pdf $\overline {P(T|t,x_0)}$
where the $\overline {\ldots}$ denote the average over the distribution of
the random potential.  A popular model for the random potential is the
celebrated Sinai model~\cite{sinai:tpa:1982}, where various disorder
averaged physical quantities can be computed
analytically~\cite{1990PhR...195..127B, bouchaud:ap:1990, comtet:jpa:1998,
majumdar:061105, fisher:1998, 2005PhR...412..277I, 1990PhyA..164...52A,
shi:pan.2001}, and yet the results exhibit rich and nontrivial behaviors and
also capture many of the qualitative behaviors of more complex realistic
disordered systems.  The Sinai model assumes that $U_r(x)=\sqrt{\sigma}
B(x)$ where $B(x)$ represents a Brownian motion in space, \ie
\begin{equation}
\frac{dB}{dx}= \xi(x)
\label{eq.4}
\end{equation}
where $\xi(x)$ is a Gaussian white noise with mean $\langle \xi(x)\rangle=0$
and a correlator $\langle \xi(x)\xi(x')\rangle =\delta(x-x')$. The constant
$\sigma$ represents the strength of the random potential.

\begin{figure*}
\centering

\includegraphics[width=0.3\hsize]{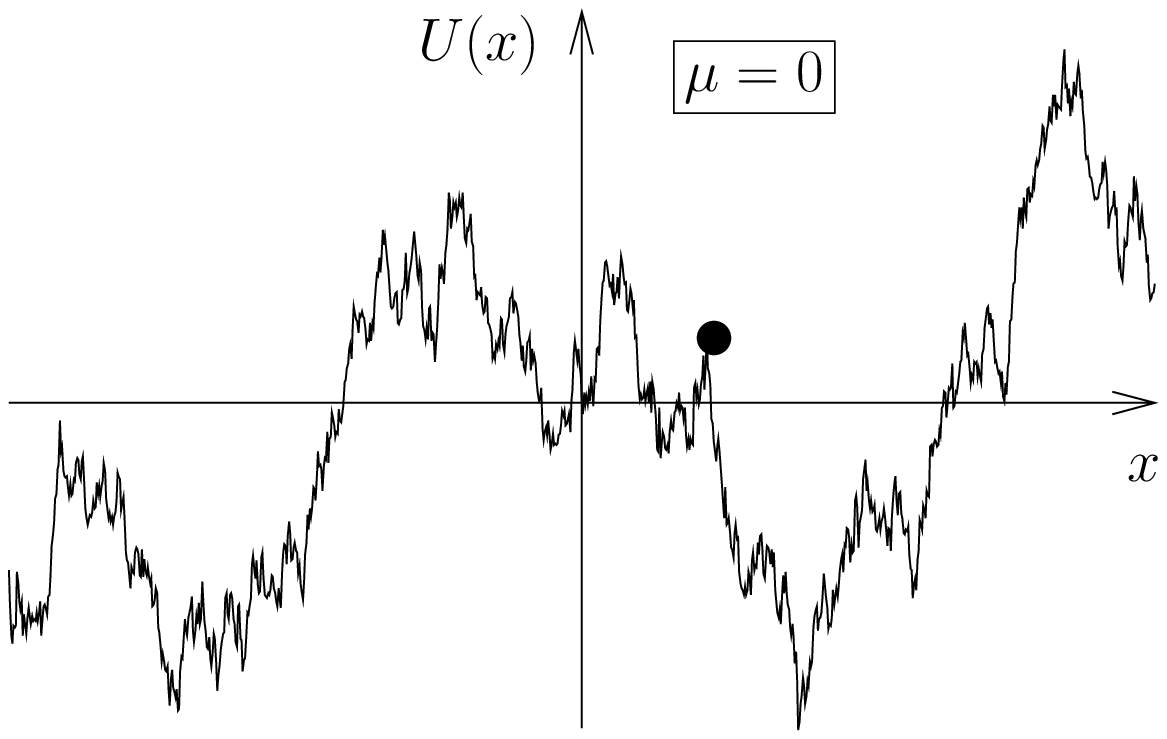}\hfill
\includegraphics[width=0.3\hsize]{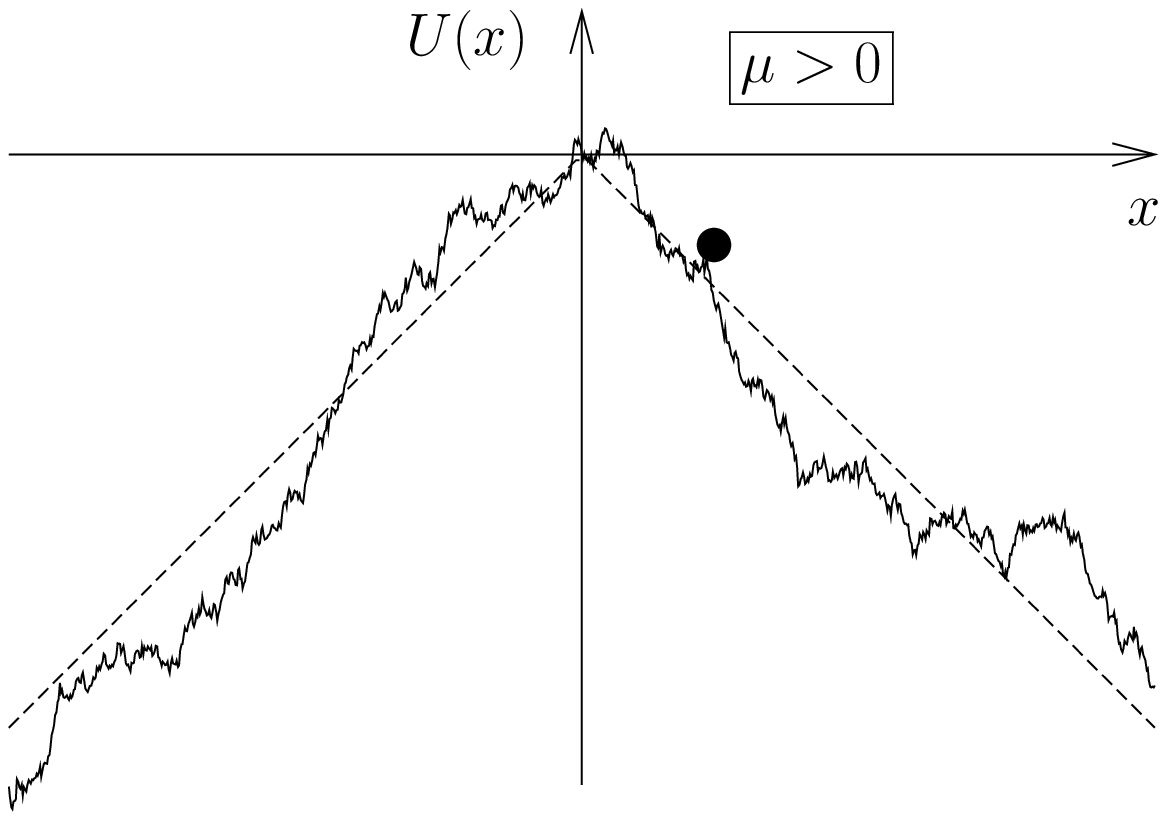}\hfill
\includegraphics[width=0.3\hsize]{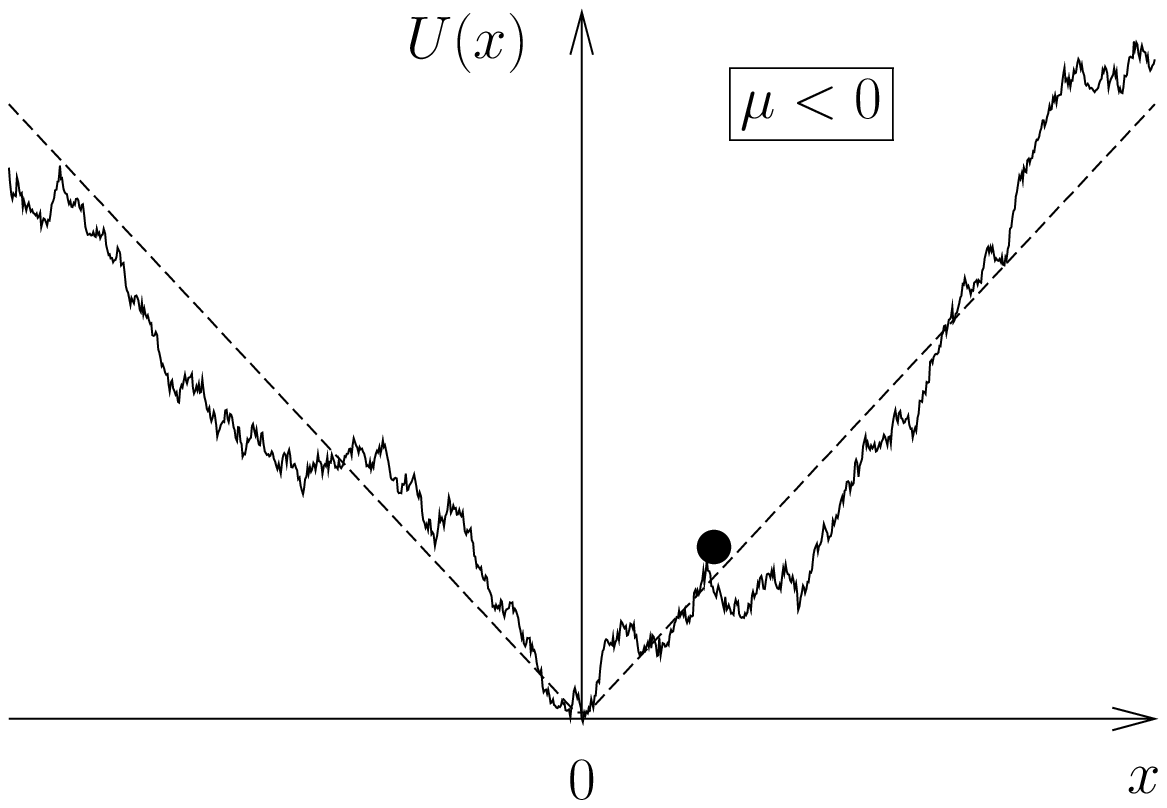}
\caption{\label{potential} A classical particle (represented by $\bullet$)
diffusing in a typical realization of the potential $U(x)=-\mu |x| +
\sqrt{\sigma} B(x)$, where $B(x)$ represents the trajectory of a Brownian
motion in space with $B(0)=0$. The three figures are for $\mu=0$, $\mu>0$
and $\mu<0$ respectively.  The dash lines show the potential for
$\sigma=0$.}
\end{figure*}

In this paper, we first present a generalization of the Feynman-Kac
formalism to calculate the pdf $P(T|t,x_0)$ in presence of an arbitrary
external potential $U(x)$. To obtain explicit results using this formalism,
we next assume that the random part of the potential is as in the Sinai
model, \ie our external potential is of the form $U(x)=U_d(x) +
\sqrt{\sigma} B(x)$, where $B(x)$ is a Brownian motion in space and $U_d(x)$
is the non-random deterministic part of the potential. It turns out that the
asymptotic behavior of the disorder averaged pdf $\overline {P(T|t,x_0)}$,
quite generically, has three different qualitative behavior depending on the
curvature of the deterministic potential $U_d(x)$, \ie whether $U_d(x)$ has
a convex (concave upward) shape representing a stable potential (\ie
attractive force towards the origin), a concave (concave downward) shape
representing  unstable potential (a repulsive force away from the origin)
or just flat indicating the absence of any external potential. To facilitate
explicit calculation, we model the deterministic potential simply by,
$U_d(x) = -\mu |x|$, so that $\mu<0$ represents a stable potential, $\mu>0$
represents an unstable potential and $\mu=0$ represents a flat potential.
This specific choice facilitates explicit calculation, but the results are
qualitatively similar if one chooses another form of this potential.  Thus,
in our model, we will consider the external potential as
\begin{equation}
U(x) = -\mu |x| + \sqrt{\sigma} B(x)
\label{U(x)}
\end{equation}
where $B(x)= \int_0^{x} \xi(x')\,dx'$ is the trajectory of a Brownian motion
in space (see Fig.~\ref{potential}).  Note that the case $\mu=0$ corresponds
to the pure Sinai model.  Figure~(\ref{potential}) shows typical realization
of potentials for $\mu=0$, $\mu>0$ and $\mu<0$.  The corresponding force in
Eq.~(\ref{Langevin}) is simply given by
\begin{equation}
F(x)=\mu\sign(x)+\sqrt{\sigma}\xi(x).
\label{F(x)}
\end{equation}

%%%%%%%%%%%%%%%%%%%%%%%%%%%%%%%%%%%%%%%%%
%%%%%%%%%%%%%%%%%%%%%%%%%%%%%%%%%%%%%%%%

We will demonstrate how to calculate explicitly, using our generalized
Feynman-Kac formalism, the disorder averaged pdf $\overline {P(T|t,x_0)}$
when the external potential is of the form given by Eq.~(\ref{U(x)}).
Despite the simplicity of the choice of the external potential, a variety of
rich and interesting behavior can be obtained by tuning the parameter
$\mu/\sigma$, as shown in this paper.  We will present detailed results for
the two functionals, namely for the local time and the occupation time
corresponding to the choices $V(x)=\delta(x)$ and $V(x)=\theta(x)$
respectively in Eq.~(\ref{definition of T}). Also, to keep the discussion
simple, we will present our final results for $x_0=0$ corresponding to the
particle starting at the origin. However, our method is not limited only to
this specific choice. Some of these results were briefly announced in a
previous letter~\cite{majumdar:060601}.

In addition, in this paper we also introduce the notion of `inverse
functional', which is defined as follows. If $V(x)$ in Eq.~(\ref{definition
of T}) is non-negative, then for each path $\{x(\tau)\}$, $T$ is a
non-decreasing function of $t$, which we formally denote by
$T=g\biglb(t|\{x(\tau)\},x_0\bigrb)$. Therefore for a given realization of
path $\{x(\tau)\}$ and given $T$ there is a unique value of $t$ (see
Fig.~\ref{T-t}), which we formally write as the inverse of the functional
$g$~\footnote{Strictly speaking this inverse does not exist always over a
dense set of points. The inverse functional is properly defined in
Ref.~\cite{bertoin} p113, in the context of local time}
\begin{equation}
t=g^{-1}\biglb(T|\{x(\tau)\}, x_0\bigrb).
\label{inverse time}
\end{equation}
This inverse time $t$ physically means the observation time that is required
for any given path $\{x(\tau)\}$ in order to produce a prescribed value of
$T$.  Of course, for the same value $T$, for a different path $\{x(\tau)\}$,
the value of $t$ will be different. Thus, $t$ is a random variable for a
fixed $T$, which takes different values for different realizations of paths
and we would like to compute its pdf, which we denote by $I(t|T,x_0)$ and by
definition $\int_0^\infty I(t|T,0)\, dt = 1$. Clearly, this pdf will also
differ from sample to sample of the external potential in Eq.~(\ref{U(x)})
and our goal is to obtain the disorder averaged distribution
$\overline{I(t|T,x_0)}$. In this paper, we present detailed results for
$\overline{I(t|T,0)}$ again for the two choices of $V(x)=\delta(x)$ and
$V(x)=\theta(x)$ corresponding to the local time and the occupation time
respectively. The inverse local and occupation times are useful for
experimentalists as they provide an estimate on the required measurement
time. For example, in the context of polymers, the inverse local time is the
typical length of a polymer required to obtain a certain monomer
concentration.

The rest of the paper is organized as follows.  In Sec.~\ref{general
approach}, we present our general approach for computing the pdf $P(T|t,x)$
of the functional $T$ defined by Eq.~(\ref{definition of T}) for a given
$t$, and the pdf $I(t|T,x)$ of the inverse functional defined by
Eq.~(\ref{inverse time}) for a given $T$, for a given sample of the random
potential, for arbitrary starting position of the particle $x(0)=x$ and for
arbitrary but non-negative $V(x)$. After this section we consider the
specific examples of local time and occupation time by setting
$V(x)=\delta(x)$ and $V(x)=\theta(x)$ respectively.  We will use different
notations for the pdfs in the two examples to avoid any misunderstanding. In
the first example, where $T$ is the local time, we denote the pdf of the
local time $P(T|t,0)$ for a given $t$ by $\lss{P}{loc}(T|t)$, and the pdf of
the inverse local time $I(t|T,0)$ for a given $T$ by $\lss{I}{loc}(t|T)$. In
the second example, where $T$ is the occupation time we denote the pdf of
the occupation time $P(T|t,0)$ for a given $t$ by $\lss{P}{occ}(T|t)$ and
the pdf of the inverse occupation time $I(t|T,0)$ for a given $T$ by
$\lss{I}{occ}(t|T)$.  While our final goal is to obtain the disorder
averaged distributions $\overline{\lss{P}{loc}(T|t)}$,
$\overline{\lss{I}{loc}(t|T)}$, $\overline{\lss{P}{occ} (T|t)}$ and
$\overline{\lss{I}{occ}(t|T)}$, it is however, instructive to study the pure
case first, before tackling the problem with disorder which is obviously
harder.  In the same spirit, we have presented the detailed discussions on
the local time, inverse local time, occupation time and inverse occupation
time for the pure case ($\sigma=0$) in Secs.~\ref{local time}, \ref{inverse
local time}, \ref{occupation time} and \ref{inverse occupation time}
respectively, before computing their disorder average in Secs.~\ref{local
time disorder}, \ref{inverse local time disorder}, \ref{occupation time
disorder} and \ref{inverse occupation time disorder} respectively.
Sec.~\ref{summary} contains some concluding remarks. Some of the details are
relegated to the appendixes. The results are summarized in
Tables~\ref{table-flat}, \ref{table-unstable}, and \ref{table-stable}.

%%%%%%%%%%%%%%%%%%%%%%%%%%%%%%%%%%%%%%%%%%%%%%%%%%%%%%%
\section{General Approach}
\label{general approach}
%%%%%%%%%%%%%%%%%%%%%%%%%%%%%%%%%%%%%%%%%%%%%%%%%%%%%%%

In this section we will show how to compute the pdfs $P(T|t,x)$ and
$I(t|T,x)$ for arbitrary non-negative $V(x)$ and arbitrary starting position
$x(0)=x$, for each realization of random force $F(x)$, by using a backward
Fokker-Planck equation approach.  In the following discussion we will denote
the functional defined in Eq.~(\ref{definition of T}) by
$g\biglb(t|\{x(\tau)\},x_0\bigrb)$, and use $T$ as the value of the
functional for a given path $[\{x(\tau)\}, ~\mbox{for}~ 0\le\tau\le t]$.

Since $V(x)$ is considered to be non-negative, $T$ defined by
Eq.~(\ref{definition of T}) has only positive support. Therefore, a natural
step is to introduce the Laplace transform of the pdf $P(T|t,x)$ with
respect to $T$
\begin{align} 
Q_p(x,t)&=\int_0^\infty P(T|t,x) e^{-pT}\,dT \nonumber\\*
&=\left\langle
e^{-p g\biglb(t|\{x(\tau)\},x\bigrb)}\right\rangle_{x(0)=x} \nonumber\\*
&=\left\langle \exp\left\{-p\int_0^t
V[x(t^\prime)]\,dt^\prime\right\}\right\rangle_{x(0)=x},
\label{def-Q}
\end{align}
where $\langle~\rangle_{x(0)=x}$ denotes the average over all paths that
start at the position $x(0)=x$ and propagate up to time $t$.  Our aim is to
derive a backward Fokker-Planck equation for $Q_p(x,t)$ with respect to the
initial position $x(0)=x$.

We consider a particle starting from the initial position $x$, evolves via
Eq.~(\ref{Langevin}) up to time $t+\Delta t$.  Now we split the time
interval $[0,t+\Delta t]$ into two parts: an infinitesimal interval
$[0,\Delta t]$, over which the particle experiences an infinitesimal
displacement $\Delta x$ from its initial position $x$ and the remaining
interval $[\Delta t, t+\Delta t]$ in which the particle evolves from a
starting position $x+\Delta x$.  Since $x(0)=x$, one gets $\int_0^{\Delta t}
V[x(t^\prime)]\,dt^\prime = V(x) \Delta t + \order[(\Delta t)^2]$.
Therefore using Eq.~(\ref{def-Q}), and splitting the integral over $t'$ into
the above two time intervals we obtain
\begin{align}
 Q_p(x,t+\Delta t)&= \left\langle \exp\left\{-p\int_0^{t+\Delta t}
    V[x(t^\prime)]\,dt^\prime\right\}\right\rangle_{x(0)=x}\nonumber \\* &=
    e^{-pV(x)\Delta t} \Biglb\langle Q_p(x+\Delta x,t)\Bigrb\rangle_{\Delta
    x},
\label{Q_dt}
\end{align}
where $\langle~\rangle_{\Delta x}$ denotes the average with respect to all
possible displacements $\Delta x$.

Now, in the limit $\Delta t\rightarrow 0$, integrating Eq.~(\ref{Langevin})
one gets,
\begin{equation}
\Delta x = F(x) \Delta t + \int_0^{\Delta t}\eta(\tau)\,d\tau+\order[(\Delta
t)^2].
\end{equation}
Hence, using the zero mean and the uncorrelated properties of the noise we
get
\begin{equation}
\lim_{\Delta t\rightarrow 0}\frac{\langle \Delta x\rangle}{\Delta t} = F(x)
\quad \mbox{and}\quad \lim_{\Delta t\rightarrow 0}\frac{\langle (\Delta x)^2
\rangle}{\Delta t} = 1,
\end{equation}
Therefore, by Taylor expanding Eq.~(\ref{Q_dt}) for small $\Delta
x$, and taking the limit $\Delta t\rightarrow 0$, one arrives at the
`backward' Fokker-Planck equation,
\begin{equation}
{\partial Q_p\over\partial t}={1\over 2}{\partial^2Q_p\over\partial x^2}
+F(x){\partial Q_p\over\partial x}-pV(x)Q_p,
\label{partial-Q}
\end{equation}
with the initial condition $Q_p(x,0)=1$, which is easily checked by
Eq.~(\ref{def-Q}). The advantage of the above equation over the usual
Feynman-Kac formalism~\cite{MR0027960,MR0045333} is that, in the later case
one has a `forward' Fokker-Planck equation (spatial derivative with respect
to the final position) where after obtaining the solution of the
differential equation, one has to again perform an additional step of
integration over the final position. In contrast, Eq.~(\ref{partial-Q})
involves the spatial derivatives with respect to the initial position of the
particle, and hence no additional step of integration over the final
position is required.

The standard practice of attacking the partial differential equations of
above type is by using the method of Laplace transform.  We define the
Laplace transform of $Q_p(x,t)$ with respect to $t$
\begin{align}
 u(x)&=\int_0^\infty Q_p(x,t) e^{-\alpha t}\,dt \nonumber\\*
&=\int_0^\infty dt\, e^{-\alpha t} \int_0^\infty dT \, e^{-pT}
P(T|t,x), 
\label{LL-P}
\end{align}
where for notational convenience, we have suppressed the $\alpha$ and $p$
dependence of $u(x)$.  Now by taking Laplace transform of
Eq.~(\ref{partial-Q}) with respect to $t$ we obtain the ordinary
differential equation
\begin{equation}
{1\over 2} u^{\prime\prime}(x) + F(x) u^\prime(x) - [\alpha +p V(x)]u(x)=-1,
\label{ordinary-u}
\end{equation}
where $u^\prime(x)=du/dx$. The appropriate boundary conditions
$u(x\rightarrow\pm\infty)$ are to be derived from the observation that if
the particle starts at $x\rightarrow\pm\infty$ it will never cross the
origin in finite time.  Note that Eq.~(\ref{ordinary-u}) is valid for each
sample of the quenched random force $F(x)$. Thus in principle, from the
solution $u(x)$ one obtains $P(T|t,x)$ by inverting the double Laplace
transform in Eq.~(\ref{LL-P}) for each sample of quenched random potential,
and then takes the average over the disorder.

Our next goal in this section, is to show how to compute the pdf $I(t|T,x)$
for a given sample of the quenched random force $F(x)$. It turns out that
$I(t|T,x)$ is related to the pdf $P(T|t,x)$ in their Laplace space as shown
below.  By definition we have,
\begin{equation}
I(t|T,x)=\left\langle \delta\left(t-g^{-1}\biglb(T|\{x(\tau)\},
x\bigrb)\right)\right\rangle_x.
\end{equation}
However, it is elementary that for each realization of path $\{x(\tau)\}$
\begin{equation}
\delta\Biglb(t-g^{-1}\biglb(T|\{x(\tau)\}, x\bigrb)\Bigrb) =
\delta\Biglb(T-g\biglb(t|\{x(\tau)\},x\bigrb)\Bigrb)\left|\frac{dT}{dt}
\right|,
\end{equation}
where $|dT/dt|$ is the usual Jacobian of the transformation, which is simply
$dT/dt$ as both $T$ and $t$ have only positive support. It immediately
follows from the above two equations that
\begin{equation}
I(t|T,x)=\left\langle
\delta\Biglb(T-g\biglb(t|\{x(\tau)\},x\bigrb)\Bigrb)\frac{dT}{dt}
\right\rangle_x .
\end{equation}
Therefore, Laplace transform of $I(t|T,x)$ with respect to $T$ reads
\begin{align}
\int_0^\infty dT\, e^{-pT} I(t|T,x) &=\left\langle
e^{-pg\biglb(t|\{x(\tau)\},x\bigrb)}\,\frac{dg}{dt}\right\rangle_x
\nonumber\\* &=-{1\over p}{\partial\over\partial t} Q_p(x,t),
\label{P-I relation (1)}
\end{align}
where $Q_p(x,t)$ is given by Eq.~(\ref{def-Q}).  Now taking a further
Laplace transform in Eq.~(\ref{P-I relation (1)}) with respect to $t$, it is
straightforward to obtain
\begin{equation}
\int_0^\infty dt\,e^{-\alpha t}\int_0^\infty dT\, e^{-pT} I(t|T,x) =
{1-\alpha u(x)\over p}.
\label{LL-I}
\end{equation}
Thus, we have established via Eqs.~(\ref{P-I relation (1)}) and
(\ref{LL-I}), the relationships between the Laplace transforms of the pdf of
the functional $T$ defined by Eq.~(\ref{definition of T}) and the pdf of the
inverse functional defined by Eq.~(\ref{inverse time}).  Hence, again in
principle, from the solution $u(x)$ of the ordinary differential
equation~(\ref{ordinary-u}), one obtains $I(t|T,x)$ by inverting the double
Laplace transform in Eq.~(\ref{LL-I}) for each sample of quenched random
potential, and then takes the average over the disorder.  Note that putting
$\alpha=0$ in Eq.~(\ref{LL-I}) and inverting the Laplace transform with
respect to $p$ immediately verifies the normalization condition
$\int_0^\infty I(t|T,x)\,dt=1$.

In the rest of the paper, we will demonstrate how to implement this
formalism for the particular examples of the local time corresponding to the
choice $V(x)=\delta(x)$ and the occupation time corresponding to the choice
$V(x)=\theta(x)$. Since in these examples we consider the starting position
of the particle to be the origin, we need to only find the solution $u(0)$
of the differential equation~(\ref{ordinary-u}).  In each example, we will
consider the pure cases ($\sigma=0$) first, which help us anticipate the
general features of the results in the disordered case ($\sigma>0$) studied
later.

%%%%%%%%%%%%%%%%%%%%%%%%%%%%%%%%%%%%%%%%%%%%%%%%%%
\section{Local Time without disorder ($\sigma=0$)}
\label{local time}
%%%%%%%%%%%%%%%%%%%%%%%%%%%%%%%%%%%%%%%%%%%%%%%%%%%

In this case $V(x)=\delta(x)$, corresponds to $T$ in Eq.~(\ref{definition of
T}) being the local time in the vicinity of the origin and $P(T|t,0)$ in
Eq.~(\ref{LL-P}) being $\lss{P}{loc}(T|t)$ --- the pdf of the local time $T$
for a given observation time window of size $t$ and the starting position of
the particle $x(0)=0$. For our purpose we only need the solution $u(0)$ of
the differential equation~(\ref{ordinary-u}), which corresponds to the
starting position of the particle being the origin. However, to obtain
$u(0)$ we have to solve Eq.~(\ref{ordinary-u}) in the entire region of $x$
with the boundary conditions $u(x\rightarrow\pm\infty)$ which are derived
from the following observation. If the initial position $x\rightarrow \pm
\infty$, the particle can not reach the origin in finite time, which means
that the local time $T=0$. Therefore, by substituting $P(T|t,x\rightarrow\pm
\infty)\rightarrow \delta(T)$ in Eq.~(\ref{LL-P}) one obtains the boundary
conditions
\begin{equation}
u(x\rightarrow\pm\infty) = \frac{1}{\alpha}.
\end{equation}
We have to obtain the solutions $u(x)=u_+(x)$ for $x>0$ and $u(x)=u_-(x)$
for $x<0$ by solving Eq.~(\ref{ordinary-u}) separately in the respective two
regions,
\begin{equation}
{1\over 2} u^{\prime\prime}_\pm(x) + F(x) u^\prime_\pm(x) - \alpha
u_\pm(x)=-1,
\label{ordinary-u local}
\end{equation}
with the boundary conditions $u_+(x\rightarrow\infty)=1/\alpha$ and
$u_-(x\rightarrow-\infty)=1/\alpha$, and then matching the two solutions
$u_+(x)$ and $u_-(x)$ at $x=0$.  The matching conditions are
\begin{equation}
u_+(0)=u_-(0)=u(0), \quad u_+^\prime(0)-u_-^\prime(0)=2pu(0).
\label{match-local}
\end{equation}
The first condition follows from the fact that the solution must be
continuous at $x=0$ and the second one is derived by integrating
Eq.~(\ref{ordinary-u}) across $x=0$.

By making a constant shift $u_\pm(x)=1/\alpha + A_\pm y_\pm(x)$, from
Eq.~(\ref{ordinary-u local}) one finds that $y_\pm(x)$ satisfy the
$p$-independent homogeneous equation
\begin{equation}
{1\over 2} y_\pm^{\prime\prime}(x) + F(x) y_\pm^\prime(x) -\alpha y_\pm(x)=0
\label{ordinary-y}
\end{equation}
with the boundary conditions $y_+(x\rightarrow \infty)\rightarrow 0$ and
$y_-(x\rightarrow -\infty) \rightarrow 0$. The constants $A_\pm$ are
determined by the matching conditions given in Eq.~(\ref{match-local}),
which can be rewritten as
\begin{subequations}
\begin{align}
& A_+y_+(0)=A_-y_-(0)=u(0)-{1\over\alpha},\\* & A_+y_+^\prime(0)
-A_-y_-^\prime(0) = 2pu(0).
\end{align}
\label{match-local.1}
\end{subequations}
Eliminating the constants $A_\pm$ from Eq.~(\ref{match-local.1}), we obtain
the Laplace transform $u(0)$, defined by Eq.~(\ref{LL-P}) with
$P(T|t,0)\equiv \lss{P}{loc}(T|t)$, as
\begin{align}
 u(0) &= \int_0^\infty dt\,e^{-\alpha t}\int_0^\infty dT\, e^{-pT}
\lss{P}{loc}(T|t) \nonumber\\* &=
{\lambda(\alpha)\over\alpha[p+\lambda(\alpha)]} .
\label{u(0)-local}
\end{align}
where $\lambda(\alpha)$ is simply given by
\begin{equation} 
\lambda(\alpha)={z_-(0)-z_+(0)\over 2} ~~~\mbox{with}~~~
z_\pm(x)={y_\pm^\prime(x)\over y_\pm(x)}.
\label{lambda-alpha}
\end{equation}
Note that putting $p=0$ in Eq.~(\ref{u(0)-local}) and then inverting the
Laplace transform with respect to $\alpha$ readily verifies the
normalization condition
\begin{equation} 
\int_0^\infty \lss{P}{loc}(T|t)\,dT=1.  
\end{equation}
Since $\lambda(\alpha)$ is independent of $p$, inverting the Laplace
transform in Eq.~(\ref{u(0)-local}) with respect to $p$ yields
\begin{align}
 G(\alpha)&= \int_0^\infty\,dt\,e^{-\alpha t}\lss{P}{loc}(T|t) \nonumber \\*
&= {\lambda(\alpha)\over\alpha} \exp\left[-\lambda(\alpha)T\right],
\label{laplace-a}
\end{align}
which is valid for any arbitrary force $F(x)$. In the following subsections
we will consider qualitatively different types of deterministic potentials
to derive more explicit results.

%%%%%%%%%%%%%%%%%%%%%%%%%%%%%%%%%%%%%%%
\subsection{Flat potential}
\label{local time mu=0}
%%%%%%%%%%%%%%%%%%%%%%%%%%%%%%%%%%%%%%%

We first consider the simple Brownian motion without any external force,
$F(x)=0$. In this case the solutions of Eq.~(\ref{ordinary-y}) are obtained
as
\begin{equation} 
y_\pm(x)=y_\pm(0)\exp\left[\mp x\sqrt{2\alpha}\right]. 
\end{equation}
Using the solutions in Eq.~(\ref{lambda-alpha}) one gets
$\lambda(\alpha)=\sqrt{2\alpha}$ and hence the Laplace transform $G(\alpha)$
in Eq.~(\ref{laplace-a}) becomes
\begin{equation} 
G(\alpha)=\int_0^\infty\,dt\,e^{-\alpha t} \lss{P}{loc}(T|t)=
{\sqrt{2}\over\sqrt{\alpha}}e^{-\sqrt{2\alpha}T}.
\end{equation}
Now inverting the Laplace transform with respect to $\alpha$, one finds that
the distribution of the local time is Gaussian for all $T$ and $t$,
\begin{equation} 
\lss{P}{loc}(T|t)={\sqrt{2}\over\sqrt{\pi t}} \exp\left[-{T^2\over
2t}\right].
\label{local-1}
\end{equation}

%%%%%%%%%%%%%%%%%%%%%%%%%%%%%%%%%%%%%%%%%%%
\subsection{Unstable potential}
\label{local time mu>0}
%%%%%%%%%%%%%%%%%%%%%%%%%%%%%%%%%%%%%%%%%%%

Now we consider the case of a Brownian particle moving in an unstable
potential $U(x)$ such that $U(x\rightarrow\pm\infty) \rightarrow
-\infty$. The corresponding repulsive force $F(x)$ drives the particle
eventually either to $+\infty$ or to $-\infty$. The pdf of the local time
$\lss{P}{loc}(T|t)$ in the case of an unstable potential tends to a steady
distribution $\lss{P}{loc}(T)$ as $t\rightarrow\infty$, which can be
computed explicitly.  To see this consider the function $G(\alpha)$ in
Eq.~(\ref{laplace-a}). By making a change of variable $\tau=\alpha t$, it
follows from Eq.~(\ref{laplace-a}),
\begin{equation} 
G(\alpha)={1\over\alpha}\int_0^\infty d\tau\,
\lss{P}{loc}\left(T\left|\frac{\tau}{\alpha}\right.\right).  
\end{equation}
Assuming $\lss{P}{loc}(T|t\rightarrow\infty)=\lss{P}{loc}(T)$, we find form
the above equation that $G(\alpha)\rightarrow \lss{P}{loc}(T)/\alpha$ as
$\alpha\rightarrow 0$. Comparing this behavior with Eq.~(\ref{laplace-a})
gives
\begin{equation} 
\lss{P}{loc}(T)=\lambda(0) \exp\left[-\lambda(0) T\right],
\label{local mu>0}
\end{equation}
provided $\lambda(0)$ is a finite positive number.  Thus generically for all
repulsive force $F(x)$, the local time distribution has a universal Poisson
distribution in the limit $t\rightarrow\infty$. The only dependence on the
precise form of the force $F(x)$ is through the rate constant $\lambda(0)$.

The rate constant $\lambda(0)$ can be expressed in terms of the force $F(x)$
in a more explicit manner. Putting $\alpha=0$ in Eq.~(\ref{ordinary-y}) and
solving the resulting equation with the boundary conditions
$y_+(x\rightarrow\infty)\rightarrow 0$ and $y_-(x\rightarrow-\infty)
\rightarrow 0$ we get
\begin{alignat}{2}
 y_+(x) &= y_+(0){\int_x^\infty \psi^2(y)\,dy \over \int_0^\infty
\psi^2(y)\, dy}, &\quad& x>0, \\ y_-(x) &= y_-(0){\int_{-\infty}^x
\psi^2(y)\,dy \over \int_{-\infty}^0 \psi^2(y)\, dy}, && x<0, 
\end{alignat}
where $\psi(y)=\exp[-\int_0^y F(x)\,dx]$. Substituting these results in
Eq.~(\ref{lambda-alpha}) gives the rate constant as
\begin{equation} 
\lambda(0)={1\over 2} \left[{1\over\int_{-\infty}^0 \psi^2(y)\, dy} +
{1\over \int_0^\infty \psi^2(y)\, dy} \right].
\label{lambda(0)}
\end{equation}

Let us now consider a simple example where the potential $U(x)=-\mu|x|$ with
$\mu>0$, corresponding to the repulsive force $F(x)=\mu\sign(x)$ from the
origin. In this case $\psi(y)=\exp[-\mu|y|]$ and hence from
Eq.~(\ref{lambda(0)}) we get $\lambda(0)=2\mu$.

%%%%%%%%%%%%%%%%%%%%%%%%%%%%%%%%%%%%%%%%%%%
\subsection{Stable potential}
\label{local time mu<0}
%%%%%%%%%%%%%%%%%%%%%%%%%%%%%%%%%%%%%%%%%%%

We now turn our attention to the complementary situation when the potential
$U(x)$ is stable, \ie $U(x\rightarrow\pm\infty)\rightarrow\infty$. In this
case the force $F(x)$ is attractive towards the origin so that the system
eventually reaches a well defined stationary state. The stationary
probability distribution $p(x)$ for the position of the particle is given by
the Gibbs measure
\begin{equation}
p(x)={e^{-2U(x)}\over Z},
\label{local Gibbs}
\end{equation}
where $U(x)=-\int_0^xF(x')\,dx'$ and $Z$ is the partition function,
\begin{equation}
Z=\int_{-\infty}^\infty e^{-2U(x)}\, dx.
\label{local partition}
\end{equation}
In this case the Laplace transform $G(\alpha)$ of the pdf of the local time
$\lss{P}{loc}(T|t)$ is still given by Eq.~(\ref{laplace-a}). However, unlike
the unstable potential in the previous section, the distribution
$\lss{P}{loc}(T|t)$ does not approach a steady state as
$t\rightarrow\infty$. Instead it has a rather different asymptotic behavior.

To deduce this asymptotic behavior, let us first consider the average local
time $\langle T \rangle =\int_0^t \langle\delta[x(t')]\rangle\, dt'$. For
large $t'$, the average $\langle\delta [x(t')]\rangle$ approaches its
stationary value $\langle\delta[x(t')]\rangle \rightarrow p(0)$, where
$p(0)=1/Z$ from Eq.~(\ref{local Gibbs}). Hence as $t\rightarrow\infty$ the
ratio $T/t$ approaches the limit
\begin{equation}
{\langle T \rangle \over t} \rightarrow {1\over Z},
\end{equation}
where $Z$ is given by Eq.~(\ref{local partition}). Thus for large $t$, the
average local time scales linearly with time $t$, which indicates that the
natural scaling limit in this case is when $t\rightarrow\infty,
T\rightarrow\infty$ but keeping the ratio $r=T/t$ fixed. We will see that in
this scaling limit the local time distribution $\lss{P}{loc}(T|t)$ tends to
the following asymptotic form
\begin{equation}
\lss{P}{loc}(T|t) \sim \exp\left[-t \Phi\left({T\over t}\right) \right],
\label{asymptotic P_0 local stable}
\end{equation}
where $\Phi(r)$ is a large deviation function.

To compute the large deviation function we first substitute this presumed
asymptotic form of $\lss{P}{loc}(T|t)$ given by Eq.~(\ref{asymptotic P_0
local stable}) in the Laplace transform $G(\alpha)=\int_0^\infty e^{-\alpha
t} \lss{P}{loc}(T|t)\, dt$ and then make a change of variable $r=T/t$ in the
integration. The resulting integral can be evaluated in the large $T$ limit
by the method of steepest descent, which gives $G(\alpha) \sim
\exp[-TW(\alpha)]$ where $W(\alpha)=
\min_r[\{\alpha+\Phi(r)\}/r]$. Comparing this result with
Eq.~(\ref{laplace-a}) gives
\begin{equation}
{\min}_r \left[{\alpha+\Phi(r)\over r} \right] = \lambda(\alpha),
\end{equation}
where $\lambda(\alpha)$ is given by Eq.~(\ref{lambda-alpha}). Thus
$\lambda(\alpha)$ is just the Legendre transform of $\Phi(r)$. Inversion of
this transform gives the exact large deviation function
\begin{equation} 
\Phi(r)={\max}_\alpha [-\alpha +r\lambda(\alpha)],
\label{large deviation function local stable} 
\end{equation}
with $\lambda(\alpha)$ given by Eq.~(\ref{lambda-alpha}). This is a general
result valid for any confining potential $U(x)$.

We will now explicitly compute the large deviation $\Phi(r)$ for the
particular potential given by Eq.~(\ref{U(x)}) with $\mu<0$ and $\sigma=0$.
Substituting the corresponding force $F(x)=-|\mu| \sign(x)$ in
Eq.~(\ref{ordinary-y}) and solving the resulting differential equations with
the boundary conditions $y_+(x\rightarrow\infty)\rightarrow 0$ and
$y_-(x\rightarrow -\infty)\rightarrow 0$ we get
\begin{equation} 
y_\pm(x)=y_\pm(0) \exp \left[ \mp \left( -|\mu|+ \sqrt{\mu^2+2\alpha}
\right)x \right].
\end{equation}
Substituting these results in Eq.~(\ref{lambda-alpha}) we get
$\lambda(\alpha)=-|\mu|+\sqrt{\mu^2+2\alpha}$. From Eq.~(\ref{large deviation
function local stable}) one then gets the large deviation function
\begin{equation}
\Phi(r)={1\over 2} \left(r-|\mu|\right)^2.
\end{equation}

It turns out that for this particular form of the force
$F(x)=-|\mu|\sign(x)$, the Laplace transform in Eq.~(\ref{laplace-a}) can be
inverted to get the local time distribution $\lss{P}{loc}(T|t)$ exactly for
all $T$ and $t$. The calculations are presented in appendix~\ref{appendix
local time stable}.  We find that in the large $t$ limit, the distribution
reduces to the asymptotic form
\begin{equation} 
\lss{P}{loc}(T|t)\approx{1\over\sqrt{2\pi t}} \exp\left[-{t\over 2}
\left({T\over t}-|\mu|\right)^2\right],
\end{equation}
near the mean $\langle T \rangle = {|\mu|} t$, which verifies the result
obtained above by the large deviation function calculation.

In fact, the limiting Gaussian form of the distribution of the local time
near its mean value is quite generic for any stable potentials (where the
system eventually becomes ergodic) and is just the manifestation of the
central limit theorem. From the definition, $T-\langle T\rangle = \int_0^t
\{\delta[x(t')] -\langle \delta[x(t')] \rangle\}\, dt'$, it follows that
when $T\rightarrow \langle T\rangle $, the random variables $ \delta[x(t')]
-\langle \delta[x(t')] \rangle$ at different times $t'$ become only weakly
correlated. Then in the limit when $t$ is much larger than the correlation
time between these variables, one expects the central limit theorem to hold
which predicts a Gaussian form for $T$ near its mean value $\langle T
\rangle$.

%%%%%%%%%%%%%%%%%%%%%%%%%%%%%%%%%%%%%%%%%%%%%%%%%%%%%%%%%%%%%%%%%
\section{Local Time with disorder ($\sigma>0$)}
\label{local time disorder}
%%%%%%%%%%%%%%%%%%%%%%%%%%%%%%%%%%%%%%%%%%%%%%%%%%%%%%%%%%%%%%%%%

So far we have considered the case where the random part of the potential
was not present. In this section we will study the effect of the randomness
by adding a random part to the potential.  In particular, we will consider
the diffusive motion of the particle when the force $F(x)$ is given by
Eq.~(\ref{F(x)}) with $\sigma>0$.

Equation~(\ref{laplace-a}) still remains valid for each realization of the
force $F(x)$, \ie for each realization of $\{\xi(x)\}$. Our aim is to
compute the average of the pdf of the local time
$\overline{\lss{P}{loc}(T|t)}$ over the noise history $\{\xi(x)\}$.  From
Eq.~(\ref{laplace-a}), one needs to know the distribution of
$\lambda(\alpha)=[z_-(0)-z_+(0)]/2$, which is now a random variable since
$F(x)$ is random.  The variables $-z_+(0)$ and $z_-(0)$ are independent of
each other and therefore their joint probability distribution factorizes to
the individual distributions. The calculations of these distributions are
presented in appendix~\ref{z+- distributions}.  Using the distributions of
$z_\pm(0)$ from Eqs.~(\ref{P+ equilibrium}) and~(\ref{P- equilibrium})
respectively with $a_\pm=\alpha$, one gets
\begin{equation}
\overline{\exp[-\lambda(\alpha) T]}=\left[{q(T)\over q(0)}\right]^2,
\label{lambda - disorder}
\end{equation}
with
\begin{align}
 q(T)&=\int_0^\infty w^{\mu/\sigma -1} \exp\left[-{1\over 2\sigma}\left\{
w(1+\sigma T)+{2\alpha \over w}\right\} \right] \, dw
\label{q(T)-1} \\*
&=2(2\alpha)^{\mu/2\sigma} (1+\sigma
T)^{-\mu/2\sigma}
K_{\mu/\sigma}
\left({\sqrt{2\alpha(1+\sigma T)}\over\sigma} \right),
\label{q(T)-2}
\end{align}
where $K_\nu(x)$ is the modified Bessel function of order
$\nu$~\cite{gradshteyn} and $K_{-\nu}(x)=K_\nu(x)$. Averaging
Eq.~(\ref{laplace-a}) over disorder we finally get the exact formula
\begin{equation} 
\int_0^\infty dt\,e^{-\alpha t} \overline{\lss{P}{loc}(T|t)} =-{1\over
  \alpha q^2(0)}{d\over dT}\left[ q^2(T) \right].
\label{laplace-a disorder}
\end{equation}
However, it is not an easy task to invert the Laplace transform to get the
exact distribution $\overline{\lss{P}{loc}(T|t)}$ for all $T$ and $t$. In
the following subsections we will extract the asymptotic behaviors of
$\overline{\lss{P}{loc}(T|t)}$, for the three cases, when the deterministic
part of the potential is:  (i)~flat corresponding to $\mu=0$, (ii)~unstable
corresponding to $\mu>0$ and (iii)~stable corresponding to $\mu<0$.

%%%%%%%%%%%%%%%%%%%%%%%%%%%%%%%%%%%%%%%%%%%%%%%%%%%%%
\subsection{Flat potential $(\mu=0)$ -- Sinai model}
\label{local time disorder mu=0}
%%%%%%%%%%%%%%%%%%%%%%%%%%%%%%%%%%%%%%%%%%%%%%%%%%%%%

We first consider a particle diffusing in the continuous Sinai
potential, \ie $\mu=0$ in Eq.~(\ref{U(x)}). Our aim is to find out how
this random potential modifies the behavior of the local time.  In
this case substituting $q(T)$ and $q(0)$ from Eq.~(\ref{q(T)-2}) with
$\mu=0$ in Eq.~(\ref{laplace-a disorder}) we get the Laplace transform
of the disorder averaged local time distribution as
\begin{multline}
 \int_0^\infty dt\,e^{-\alpha t} \overline{\lss{P}{loc}(T|t)} = \\* -{1\over
\alpha K_0^2 \left(\sqrt{2\alpha}/\sigma\right)} {\partial\over\partial T}
K_0^2 \left({\sqrt{2\alpha(1+\sigma T)}\over\sigma} \right).
\label{laplace-a disorder mu=0}
\end{multline}
We will now consider the interesting limit where both $t$ and $T$ are large,
but the ratio $y=T/t$ is kept fixed. This corresponds to taking the limit
$\alpha\rightarrow 0$ with $\alpha T=s$ keeping fixed. In this limit
\begin{align}
&K_0 \left({\sqrt{2\alpha}\over\sigma} \right) \rightarrow - {1\over
2} \log\alpha \\*
\intertext{and}
&K_0\left({\sqrt{2\alpha(1+\sigma T)}\over\sigma}  \right) \rightarrow
K_0 \left({\sqrt{2 s}\over\sqrt{\sigma}} \right).
\end{align}
Therefore substituting $t=s/\alpha y$ and $T=s/\alpha$ in
Eq.~(\ref{laplace-a disorder mu=0}), in the limit $\alpha\rightarrow 0$ we
get
\begin{multline}
\int_0^\infty dy\,e^{-s/y} \left[{s\over\alpha
y^2}\overline{\lss{P}{loc}\left({s\over\alpha}\left|{s\over\alpha
y}\right.\right)}\right] = \\* -{4\over\log^2\alpha} {\partial\over\partial
s} K_0^2 \left({\sqrt{2 s}\over\sqrt{\sigma}} \right).
\label{laplace-a disorder mu=0 (1)}
\end{multline}
The above equation suggests that, in the limit $t\rightarrow\infty$ and
$T\rightarrow\infty$, while their ratio $T/t$ is kept fixed,
$\overline{\lss{P}{loc}(T|t)}$ should have the scaling form
\begin{equation} 
\overline{\lss{P}{loc}(T|t)} = {1\over t \log^2t} f_1(T/t).
\label{P_0 disorder mu=0 scaling}
\end{equation}

\begin{figure}
\centering
\includegraphics[width=0.9\hsize]{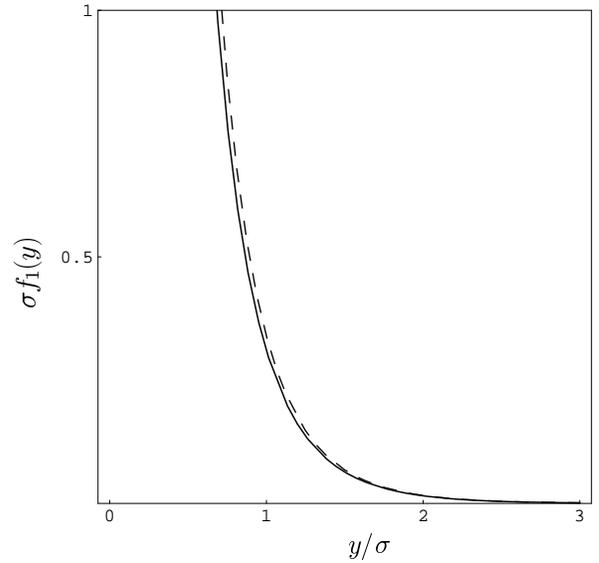}
\caption{\label{f1 plot}The scaling function $f_1(y)$ in Eq.~(\ref{P_0
disorder mu=0 scaling}). The solid line is plotted by using Eq.~(\ref{f
local time disorder mu=0}), and the dash line is plotted by using the
limiting form $f_1(y)\sim \sqrt{2\pi\sigma}\, y^{-3/2}
e^{-2y/\sigma}$ as $y\rightarrow\infty$.}
\end{figure}

Now substituting the above form in Eq.~(\ref{laplace-a disorder mu=0 (1)})
and making the change of variable $\tilde{y}=1/y$, we obtain after
straightforward simplification
\begin{equation} \int_0^\infty d\tilde{y}\,e^{-s\tilde{y}}
\left[{f_1(1/\tilde{y})\over\tilde{y}^2}\right] = 4 K_0^2
\left({\sqrt{2 s}\over\sqrt{\sigma}} \right).  
\label{laplace-y (1)}
\end{equation}
Note that the right hand side of the above equation is simply the Laplace
transform of the function $f_1(1/\tilde{y})/\tilde{y}^2$.  Therefore by
using the identity
\begin{equation}
\int_0^\infty \frac{e^{-a^2/4\omega}}{\omega}\, e^{-s\omega}\, d\omega =2
K_0\left(a\sqrt{s}\right)
\end{equation}
% inverse Laplace transform
% \begin{equation}
% \mathcal{L}^{-1}\left\{K_0\left(a\sqrt{s}\right) \right\}_s (\omega)
% ={1\over 2\omega} \exp\left(-a^2/4\omega\right)
% \end{equation}
%
and the convolution property of Laplace transform, we can invert the Laplace
transform in Eq.~(\ref{laplace-y (1)}) with respect to $s$.  Inverting the
Laplace transform and after simplification we finally get
\begin{equation}
f_1(1/\tilde{y})=2\tilde{y}\int_0^{1/2} {dx\over x(1-x)}
\exp\left[-{1\over 2\sigma\tilde{y} x(1-x)} \right].
\end{equation}
Therefore the scaling function $f_1(y)$ is simply given by 
\begin{equation}
f_1(y)={2\over y}\int_0^{1/2} {dx\over x(1-x)}
\exp\left[-{y\over 2\sigma x(1-x)} \right].
\end{equation}
By substituting $x(1-x)=1/z$ gives
\begin{equation} 
f_1(y)={2\over y}\int_4^\infty {dz\over \sqrt{z(z-4)}}\exp\left( -{y\over 2
\sigma}z\right),
\end{equation}
where the integral can be evaluated exactly~\cite{gradshteyn}, which finally
gives the scaling function in Eq.~(\ref{P_0 disorder mu=0 scaling}) as
\begin{equation}
f_1(y)={2\over y}\, e^{-y/\sigma} K_0\left(y/\sigma\right).
\label{f local time disorder mu=0}
\end{equation}
However the scaling given by Eq.~(\ref{P_0 disorder mu=0 scaling}) breaks
down for very small $y$ (very small $T$) when $y\ll\sigma$.  The scaling
function is displayed in Fig.~\ref{f1 plot}.  In the large $y$ limit, using
the asymptotic the behavior $K_\nu(x)\sim \sqrt{\pi/2x}\,e^{-x}$ from
Eq.~(\ref{f local time disorder mu=0}) we find that $f_1(y)\sim
\sqrt{2\pi\sigma}\, y^{-3/2} e^{-2y/\sigma}$.

%%%%%%%%%%%%%%%%%%%%%%%%%%%%%%%%%%%%%%%%%
\subsection{Unstable potential $(\mu>0)$}
\label{local time disorder mu>0}
%%%%%%%%%%%%%%%%%%%%%%%%%%%%%%%%%%%%%%%%%

In this case the behavior in the limit $t\rightarrow \infty$ can be obtained
by either setting $\alpha=0$ in the integral in Eq.~(\ref{q(T)-1}) or taking
the $\alpha\rightarrow 0$ limit in $K_\nu(.)$ in Eq.~(\ref{q(T)-2}), which
gives
\begin{equation} 
q(T)\rightarrow\Gamma(\mu/\sigma) (2\sigma)^{\mu/\sigma} (1+\sigma
T)^{-\mu/\sigma},
\end{equation}
where $\Gamma(x)$ is the Gamma function~\cite{abramowitz}. Substituting
$q(T)$ and $q(0)$ in Eq.~(\ref{laplace-a disorder}) and inverting the
Laplace transform with respect to $\alpha$ gives
\begin{equation}
\overline{\lss{P}{loc}(T|t)} = 2\mu(1+\sigma T)^{-2\mu/\sigma -1},
\end{equation}
\ie in the limit $t\rightarrow \infty$, the distribution
$\overline{\lss{P}{loc}(T|t)}$ tends to a steady state distribution
$\overline{\lss{P}{loc}(T)}$ for all $T\ge 0$.  The disorder averaged local
time distribution has a broad power law distribution even though for each
sample the local time has a narrow exponential distribution (see
Eq.(\ref{local mu>0}) in Sec.~\ref{local time mu>0}). This indicates wide
sample to sample fluctuations and lack of self-averaging.

%%%%%%%%%%%%%%%%%%%%%%%%%%%%%%%%%%%%%%%%
\subsection{Stable potential $(\mu<0)$}
\label{local time disorder mu<0}
%%%%%%%%%%%%%%%%%%%%%%%%%%%%%%%%%%%%%%%%

In this case substituting $q(T)$ and $q(0)$ from Eq.~(\ref{q(T)-2}) in
Eq.~(\ref{laplace-a disorder}) and denoting $\nu=|\mu|/\sigma$ we get
\begin{multline}
\int_0^\infty dt\,e^{-\alpha t} \overline{\lss{P}{loc}(T|t)} =
-\frac{1}{\alpha K_\nu^2 \left(\sqrt{2\alpha}/\sigma\right)}\\* \times
{\partial\over \partial T} \left[ (1+\sigma T)^{\nu/2} K_\nu
\left({\sqrt{2\alpha(1+\sigma T)}\over\sigma} \right)\right]^2.
\label{laplace-a disorder mu<0}
\end{multline}

We consider the scaling limit where both $t$ and $T$ are large, but their
ratio $y=T/t$ is kept fixed. This corresponds to taking the limit
$\alpha\rightarrow 0$ with keeping $\alpha T=s$ fixed, which gives the
following limiting forms
\begin{align}
&(1+\sigma T) \rightarrow {\sigma s\over\alpha},\\ &K_\nu
\left({\sqrt{2\alpha}\over\sigma} \right) \rightarrow {\Gamma(\nu)\over 2}
\left( \sigma\sqrt{2}\over\sqrt{\alpha}\right)^\nu,\\
&K_\nu\left({\sqrt{2\alpha(1+\sigma T)}\over\sigma} \right) \rightarrow
K_\nu \left({\sqrt{2 s}\over\sqrt{\sigma}} \right).
\end{align}
Substituting the above limits on the right hand side of Eq.~(\ref{laplace-a
disorder mu<0}) and making change of variables $t=s/\alpha y$ and
$T=s/\alpha$ on the left hand side, it is straightforward to get
\begin{multline} 
\int_0^\infty dy\,e^{-s/y} \left[{s\over\alpha
y^2}\overline{\lss{P}{loc}\left({s\over\alpha}\left|{s\over\alpha
y}\right.\right)}\right] = \\* -{4\over(2\sigma)^\nu \Gamma^2(\nu)}
{\partial\over\partial s} \left[s^{\nu/2} K_\nu \left({\sqrt{2
s}\over\sqrt{\sigma}} \right)\right]^2,
\label{laplace-a disorder mu<0 (1)}
\end{multline}
in the limit $\alpha\rightarrow 0$. This suggests the limiting form for
$\overline{\lss{P}{loc}(T|t)}$
\begin{equation} 
\overline{\lss{P}{loc}(T|t)} \rightarrow {1\over t} f_2(T/t),
\label{scaling local time disorder mu<0}
\end{equation}
in the scaling limit $t\rightarrow\infty$ and $T\rightarrow\infty$ with a
fixed ratio $y=T/t$.  To compute the scaling function we substitute the
above scaling form in Eq.~(\ref{laplace-a disorder mu<0 (1)}), and make the
change of variable $\tilde{y}=1/y$. Then Eq.~(\ref{laplace-a disorder mu<0
(1)}) simplifies to the Laplace transform
\begin{multline}
\int_0^\infty d\tilde{y}\,e^{-s\tilde{y}}\left[f_2(1/\tilde{y})\over
  \tilde{y}^2 \right]= \\* {4\over(2\sigma)^\nu \Gamma^2(\nu)}
  \left[s^{\nu/2} K_\nu \left({\sqrt{2 s}\over\sqrt{\sigma}}
  \right)\right]^2,
\label{laplace-y disorder mu<0}
\end{multline}
which can be inverted with respect to $s$, by using the identity
\begin{equation}
\int_0^\infty \left(\frac{a}{2}\right)^\nu
\frac{e^{-a^2/4\omega}}{\omega^{\nu+1}}\, e^{-s\omega}\, d\omega
= 2 s^{\nu/2}K_\nu\left(a\sqrt{s}\right)
\end{equation}
and the convolution property of the Laplace transform. After simplification,
the inverse Laplace transform gives
\begin{multline}
f_2(y)={2y^{2\nu-1}\over (2\sigma)^{2\nu}\Gamma^2(\nu)}\int_0^{1/2} {dx\over
  x^{\nu+1}(1-x)^{\nu+1}}\\* \times \exp\left[-{y\over 2\sigma x(1-x)}
  \right].
\end{multline}
By making a change of variable $x(1-x)=1/z$ in the above integral, it can be
presented in the form
\begin{multline}
f_2(y)={2y^{2\nu-1}\over (2\sigma)^{2\nu}\Gamma^2(\nu)} \int_4^\infty
  z^{\nu-1/2}(z-4)^{-1/2} \\* \times \exp\left( -{y\over 2
  \sigma}z\right)\,dz,
\end{multline}
which now can be expressed in more elegant forms~\cite{abramowitz} as
\begin{equation}
f_2(y) =\left[\frac{2\sqrt{\pi}}{\sigma \Gamma^2(\nu)}\right]
\left(\frac{y}{\sigma}\right)^{2\nu-1} e^{-2y/\sigma} U(1/2,1+\nu,2y/\sigma),
\label{f local time disorder mu<0}
\end{equation}
where $U(a,b,x)$ is the Confluent Hypergeometric Function of the Second Kind
(also known as Kummer's function of the second kind)~\cite{abramowitz},
which has the following limiting behaviors:
\begin{subequations}
\begin{alignat}{2}
&U(1/2,1+\nu,x)\approx \frac{\Gamma(\nu)}{\sqrt{\pi}} x^{-\nu} &\quad&
\mbox{for small} ~x,\\
&U(1/2,1+\nu,x)\sim \frac{1}{\sqrt{x}}
&& \mbox{for large} ~x.
\end{alignat}
\label{limiting CHF}
\end{subequations}

\begin{figure}
\centering \includegraphics[width=0.9\hsize]{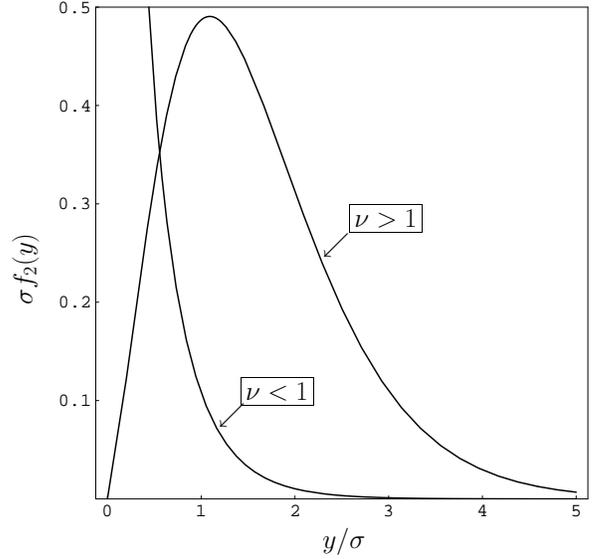}
\caption{\label{f2 plot}The scaling function $f_2(y)$ in Eq.~(\ref{scaling
local time disorder mu<0}), plotted by using Eq.~(\ref{f local time disorder
mu<0}).  $\nu=|\mu|/\sigma$.}
\end{figure}

The scaling function $f_2(y)$ is displayed in Fig.~\ref{f2 plot}.  Using the
limiting behaviors from Eq.~(\ref{limiting CHF}), one finds that the scaling
function decays as $f_2(y)\sim y^{(4\nu-3)/2} e^{-2y/\sigma}$ for large $y$.
For small $y$, the scaling function behaves as $f_2(y)\sim y^{\nu-1}$, which
increases with $y$ for $\nu>1$, however, diverges when $y\rightarrow 0$ for
$\nu<1$, a behavior qualitatively similar to the Sinai case (see
Fig.~\ref{f1 plot}). For $\nu<1$, the disorder wins over the strength of the
stable potential. In that situation when the particle gets trapped in the
wells of the random potential, the weak external deterministic force often
can not lift it out of the well and send towards the origin. Therefore, the
scaling function $f_2(y)$ carries very large weight near $y=0$ (which
corresponds to very small local time $T$ for a given observation time $t$).

Note that, for the particular value $\nu=1/2$, the scaling function has a
simple form $f_2(y)=\sqrt{2/\pi\sigma y}\, \exp({-2 y/\sigma})$.

%%%%%%%%%%%%%%%%%%%%%%%%%%%%%%%%%%%%%%%%%%%%%%%%%%%%%%%%%%
\section{Inverse Local time without disorder ($\sigma=0$)}
\label{inverse local time}
%%%%%%%%%%%%%%%%%%%%%%%%%%%%%%%%%%%%%%%%%%%%%%%%%%%%%%%%%%

\begin{figure}
\includegraphics[width=.9\hsize]{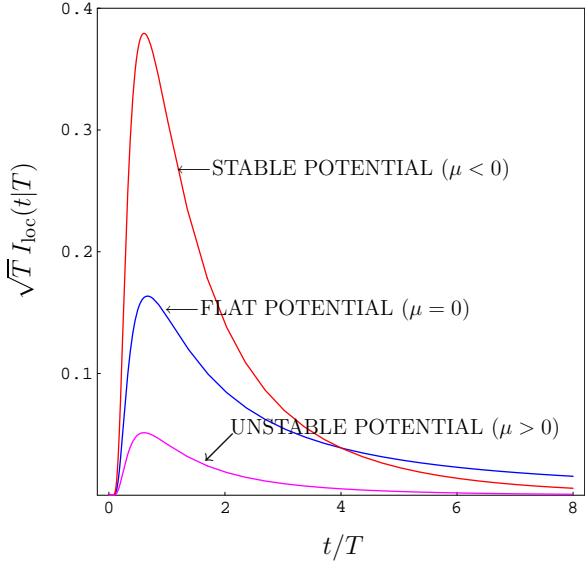}
\caption{\label{I_loc plot sigma=0}The pdfs of the inverse local time for
stable ($\mu=-1/2$), flat ($\mu=0$) and unstable ($\mu=1/2$) potentials,
plotted using Eq.~(\ref{inverse local time distribution}) and $T=2$.}
\end{figure}

The inverse local time means how long one has to observe the particle until
the total time spent in the infinitesimal neighborhood of the origin is
$T$. The double Laplace transform of the pdf of the inverse local time is
obtained by simply putting $x=0$ in Eq.~(\ref{LL-I}). Corresponding $u(0)$
in Eq.~(\ref{LL-I}), which is nothing but the double Laplace transform of
the pdf of local time, has already been evaluated in Sec.~\ref{local time}
and is given by Eq.~(\ref{u(0)-local}).  Substituting $u(0)$ and replacing
$I(t|T,0)$ with the pdf of the inverse local time $\lss{I}{loc}(t|T)$, after
straightforward simplification, for $x=0$ Eq.~(\ref{LL-I}) reads
\begin{equation} 
\int_0^\infty dt\,e^{-\alpha t}\int_0^\infty dT\, e^{-pT} \lss{I}{loc}(t|T)
 = {1\over p +\lambda(\alpha)},
\end{equation}
where $\lambda(\alpha)$ is given by Eq.~(\ref{lambda-alpha}), which depends
on the force $F(x)$ through Eq.~(\ref{ordinary-y}).  Inverting the Laplace
transform with respect to $p$ gives the general formula
\begin{equation}
\int_0^\infty dt\,e^{-\alpha t} \lss{I}{loc}(t|T)=
\exp\left[-\lambda(\alpha)T\right],
\label{L-I}
\end{equation}
valid for arbitrary force $F(x)$, a result known in the mathematics
literature~\cite{bertoin, borodin:2002}.

We first consider the pure case where the force given by Eq.~(\ref{F(x)})
with $\sigma=0$.  Substituting solutions of Eq.~(\ref{ordinary-y}) for
$F(x)=\mu\sign(x)$ in Eq.~(\ref{lambda-alpha}) we obtain $\lambda(\alpha)=
\mu+\sqrt{\mu^2+2\alpha}$.  Now using this $\lambda(\alpha)$ in
Eq.~(\ref{L-I}) and making a change of the parameter $\alpha=\beta -\mu^2/2$
we get
\begin{equation} 
\int_0^\infty dt\,e^{-\beta t} \left[e^{\mu^2 t/2} \lss{I}{loc}(t|T) \right]
= e^{-\mu T} e^{-\sqrt{2\beta}T},
\end{equation}
where the right hand side is simply the Laplace transform of $e^{\mu^2 t/2}
\lss{I}{loc}(t|T) $ with respect to $t$. The Laplace transform can be
inverted to obtain the exact pdf of the inverse local time
\begin{equation}
\lss{I}{loc}(t|T)={T\over\sqrt{2\pi t^3}} \exp\left[- {(T+\mu t)^2 \over
2t}\right].
\label{inverse local time distribution}
\end{equation}
with the normalization condition
\begin{equation}
\int_0^\infty \lss{I}{loc}(t|T)\,dt=e^{-(\mu + |\mu|)T} =\left\{
\begin{array}{lll}
1 & \mbox{for} & \mu \le 0,\\
e^{-2\mu T} & \mbox{for} & \mu > 0,
\end{array}
\right.
\label{normalization inverse local time distribution}
\end{equation}
which is simply obtained by putting $\alpha=0$ in Eq.~(\ref{L-I}).  As we
infer from Eq.~(\ref{inverse local time distribution}), although in the
limit $t\rightarrow 0$ the inverse local time distribution
$\lss{I}{loc}(t|T)\sim \exp(-T^2/2t)$ is independent of $\mu$, for large $t$
it depends on the nature of the potential, as shown in Fig~\ref{I_loc plot
sigma=0}. While in the absence of any force, \ie $\mu=0$ the inverse local
time distribution has a power-law tail $\lss{I}{loc}(t|T)\sim t^{-3/2}$, for
the stable potential, \ie $\mu<0$, it decays exponentially
$\lss{I}{loc}(t|T)\sim \exp(-\mu^2 t/2)$. On the other hand when the
potential is unstable, $\mu >0$, as we see from Eq.~(\ref{normalization
inverse local time distribution}), the distribution $\lss{I}{loc}(t|T)$ is
not normalized to unity. In this case the particle escapes to $\pm\infty$
with probability $(1-e^{-2\mu T})$ and Eq.~(\ref{inverse local time
distribution}) gives the distribution only for those events where the
particle does not escape to $\pm\infty$. Therefore for $\mu>0$, it is
appropriate to represent the full normalized distribution as
\begin{multline} 
\lss{I}{loc}(t|T)={T\over\sqrt{2\pi t^3}} \exp\left[- {(T+\mu t)^2 \over
    2t}\right]\\* +\left(1-e^{-2\mu T}\right) \delta(t-\infty).
\end{multline}
Note that the second term does not show up in the Laplace transform of
$\lss{I}{loc}(t|T)$ with respect to $t$.

%%%%%%%%%%%%%%%%%%%%%%%%%%%%%%%%%%%%%%%%%%%%%%%%%%%%%%%%%%%%%%%%%%%%%%%%
\section{Inverse  Local Time with disorder ($\sigma>0$)}
\label{inverse local time disorder}
%%%%%%%%%%%%%%%%%%%%%%%%%%%%%%%%%%%%%%%%%%%%%%%%%%%%%%%%%%%%%%%%%%%%%%%%

In this section, we switch on the disorder by considering $\sigma>0$ in the
force given by Eq.~(\ref{F(x)}).  In the presence of disorder, taking the
disorder average of Eq.~(\ref{L-I}) gives
\begin{equation}
\int_0^\infty dt\,e^{-\alpha t} \overline{\lss{I}{loc}(t|T)}=
\overline{\exp\left[-\lambda(\alpha)T\right]},
\label{L-I disorder}
\end{equation}
with $\lambda(\alpha)=[z_-(0)-z_+(0)]/2$, where $-z_+(0)$ and $z_-(0)$ are
independent random variables, whose distributions are given by Eqs.~(\ref{P+
equilibrium}) and~(\ref{P- equilibrium}) respectively with $a_\pm=\alpha$.
The object $\overline{\exp\left[-\lambda(\alpha)T\right]}$ on the right hand
side of Eq.~(\ref{L-I disorder}), has already been evaluated in
Sec.~\ref{local time disorder}, which is given by Eq.~(\ref{lambda -
disorder}).  In the following subsections we will determine the behavior of
$\overline{\lss{I}{loc}(t|T)}$ in the scaling limit $t\rightarrow\infty,
T\rightarrow \infty$, while keeping their ratio $x=t/T$ fixed, for the three
qualitatively different cases: (i)~$\mu=0$, (ii)~$\mu>0$ and (iii)~$\mu<0$.

%%%%%%%%%%%%%%%%%%%%%%%%%%%%%%%%%%%%%%%%%%%%%%%%%%%%
\subsection{Flat potential ($\mu=0$) -- Sinai model}
\label{inverse local time disorder mu=0}
%%%%%%%%%%%%%%%%%%%%%%%%%%%%%%%%%%%%%%%%%%%%%%%%%%%%

\begin{figure}
\centering
\includegraphics[width=0.9\hsize]{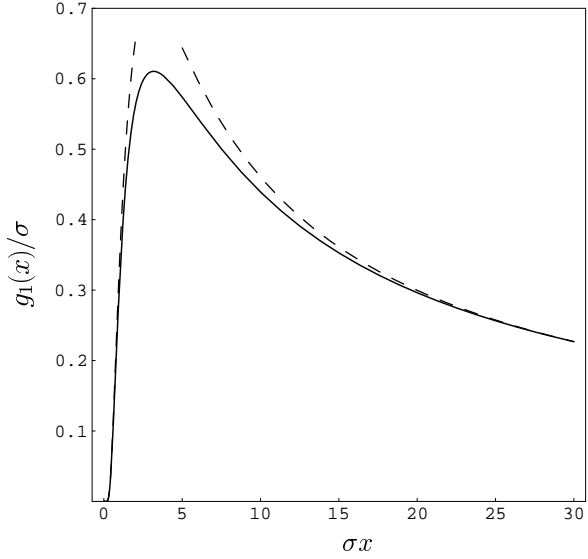}
\caption{\label{g1 plot}The scaling function $g_1(y)$ in Eq.~(\ref{scalinng
inverse local disprder mu=0}). The solid line is plotted by using
Eq.~(\ref{g1 function}), and the dash line is plotted by using the limiting
forms: $g_1(x)\approx \sqrt{2\pi\sigma} x^{-1/2} \exp(-2/\sigma x)$ for
small $x$ and $g_1(x)\sim 2\log(\sigma x)/x$ for large $x$.  }
\end{figure}

Following the analysis of Sec.~\ref{local time disorder mu=0}, in the limit
$\alpha\rightarrow 0$ with keeping $\alpha T=s$ fixed,
\begin{equation} 
\overline{\exp\left[-\lambda(\alpha)T\right]} \rightarrow
{4\over\log^2\alpha} K_0^2 \left({\sqrt{2 s}\over\sqrt{\sigma}} \right).
\label{inverse local disorder mu=0 rhs}
\end{equation} 
Therefore, substituting $t=x s/\alpha$ and $T=s/\alpha$, in the limit
$\alpha\rightarrow 0$, Eq.~(\ref{L-I disorder}) reads
\begin{equation}
\int_0^\infty dx\,e^{-s x} \left[ {s\over\alpha} \,
\overline{\lss{I}{loc}\left({sx\over\alpha}\left|
{s\over\alpha}\right.\right)} \right] = {4\over\log^2\alpha} K_0^2
\left({\sqrt{2 s}\over\sqrt{\sigma}} \right).
\label{L-I disorder alpha->0}
\end{equation}
This suggest the scaling form
\begin{equation} 
\overline{\lss{I}{loc}(t|T)} = {1\over T \log^2 T}\, g_1(t/T),
\label{scalinng inverse local disprder mu=0}
\end{equation}
in the limit $t\rightarrow \infty, T\rightarrow\infty$ but keeping $x=t/T$
fixed.  Substituting this scaling form in Eq.~(\ref{L-I disorder alpha->0}),
after straightforward simplification one obtains
\begin{equation}
\int_0^\infty dx\,e^{-sx} g_1(x) = 4K_0^2 \left({\sqrt{2
s}\over\sqrt{\sigma}} \right).
\end{equation}
Now direct comparison of the above equation with Eq.~(\ref{laplace-y (1)})
gives $g_1(x)=f_1(1/x)/x^2$, where $f_1(x)$ is given by Eq.~(\ref{f local
time disorder mu=0}). Substituting $f_1(1/x)$ one obtains the scaling
function $g_1(x)$ as
\begin{equation} 
g_1(x)={2\over x}\, e^{-1/\sigma x} K_0\left(1/\sigma x\right),
\label{g1 function}
\end{equation}
which is displayed in Fig.~\ref{g1 plot}.  The scaling function increases as
$g_1(x)\approx \sqrt{2\pi\sigma} x^{-1/2} \exp(-2/\sigma x)$ for small $x$
and decays as $g_1(x)\sim 2\log(\sigma x)/x$ at large $x$.

%%%%%%%%%%%%%%%%%%%%%%%%%%%%%%%%%%%%%%%%%
\subsection{Unstable potential ($\mu>0$)}
\label{inverse local time disorder mu>0}
%%%%%%%%%%%%%%%%%%%%%%%%%%%%%%%%%%%%%%%%%

In this case the right hand side of Eq.~(\ref{L-I disorder}) is given by
\begin{equation}
\overline{\exp\left[-\lambda(\alpha)T\right]} = (1+\sigma T)^{-\nu}
{K_{\nu}^2\left(\sqrt{2\alpha(1+\sigma T)}/\sigma\right) \over
K_{\nu}^2\left(\sqrt{2\alpha}/\sigma\right)},
\label{inverse local disorder mu>0 rhs}
\end{equation}
with $\nu=\mu/\sigma$.  Putting $\alpha=0$ in the above equation gives the
normalization condition, $\int_0^\infty \overline{\lss{I}{loc}(t|T)}\, dt =
(1+\sigma T)^{-2\nu}$, which implies that for the unstable potential, where
the force is repulsive from the origin, the particle escapes to $\pm \infty$
with probability $1-(1+\sigma T)^{-2\nu}$, and the disorder averaged pdf
$\overline{\lss{I}{loc}(t|T)}$ obtained by inverting the Laplace transform
in Eq.~(\ref{L-I disorder}) represents only those events where the particle
does not escape to $\pm\infty$.

Now in the limit of $\alpha\rightarrow 0$ with $\alpha T=s$ keeping fixed,
one gets
\begin{align}
&K_\nu \left({\sqrt{2\alpha}\over\sigma} \right) \rightarrow
{\Gamma(\nu)\over 2} T^{\nu/2}\left(
\sigma\sqrt{2}\over\sqrt{s}\right)^{\nu}, \\
&K_{\nu}\left({\sqrt{2\alpha(1+\sigma T)}\over\sigma} \right) \rightarrow
K_{\nu} \left({\sqrt{2 s}\over\sqrt{\sigma}} \right).
\end{align}
Therefore Eq.~(\ref{inverse local disorder mu>0 rhs}) becomes
\begin{equation}
\overline{\exp\left[-\lambda(\alpha)T\right]} \rightarrow {4 T^{-2\nu}\over
(2\sigma^3)^{\nu}\Gamma^2(\nu)} \left[s^{\nu/2} K_{\nu} \left({\sqrt{2
s}\over\sqrt{\sigma}} \right)\right]^2.
\end{equation}

\begin{figure}
\centering
\includegraphics[width=0.9\hsize]{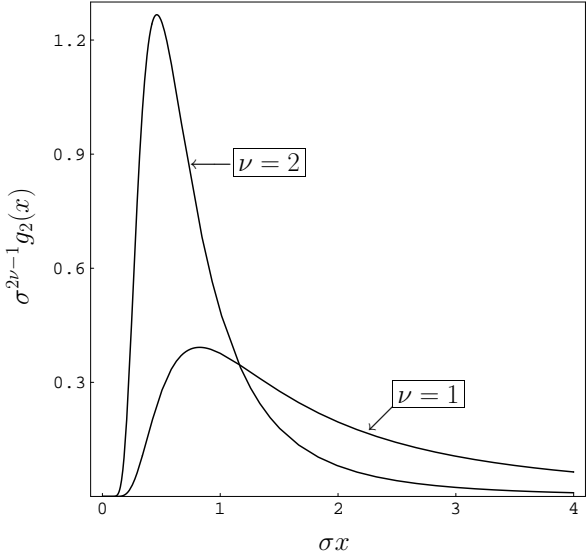}
\caption{\label{g2 plot} The scaling functions $g_2(x)$ in Eq.~(\ref{scaling
inverse local time disorder mu>0}) plotted by using Eq.~(\ref{g2 function}).
$\nu=|\mu|/\sigma$.}
\end{figure}

In the corresponding limit $T\rightarrow\infty, t\rightarrow\infty$, but
keeping their ratio $x=t/T$ fixed, using the scaling form
\begin{equation}
\overline{\lss{I}{loc}(t|T)} ={1\over T^{2\nu+1}}\, g_2(t/T)
\label{scaling inverse local time disorder mu>0}
\end{equation}
in Eq.~(\ref{L-I disorder}) one finally arrives at the Laplace transform
\begin{equation} 
\int_0^\infty e^{-sx} g_2(x)\, dx = {4 \over (2\sigma^3)^{\nu}\Gamma^2(\nu)}
\left[s^{\nu/2} K_{\nu} \left({\sqrt{2 s}\over\sqrt{\sigma}}
\right)\right]^2.
\label{L-g2}
\end{equation}
The Laplace transform can be inverted with respect to $s$ to obtain the
scaling function $g_2(x)$ and in fact the inversion has already been done in
Sec.~\ref{local time disorder mu<0}. Comparing the above equation with
Eq.~(\ref{laplace-y disorder mu<0}) readily gives $g_2(x)=\sigma^{-2\nu}
f_2(1/x)/x^2$ where $f_2(x)$ is given by Eq.~(\ref{f local time disorder
mu<0}). Substituting $f_2(1/x)$ gives
\begin{equation}
g_2(x) = \left[{2\sigma\sqrt{\pi}\over\sigma^{2\nu} \Gamma^2(\nu)}\right]
{e^{-2/\sigma x}\over (\sigma x)^{2\nu+1} }U(1/2,1+\nu,2/\sigma x),
\label{g2 function}
\end{equation}
where $U(a,b,x)$ is the Confluent Hypergeometric Function of the Second
Kind, whose its small and large $x$ behaviors are given in
Eq.~(\ref{limiting CHF}).  The scaling function $g_2(x)$ is displayed in
Fig.~\ref{g2 plot}.  The scaling function increases as $g_2(x) \sim
\exp(-2/\sigma x)$ for small $x$ and eventually decreases for large $x$ as
$g_2(x)\sim 1/x^{2\nu}$. In particular, for $\nu=1/2$ it has a very simple
form $g_2(x)=\sqrt{2/\pi\sigma^3} x^{-3/2}\exp(-2/\sigma x)$.

%%%%%%%%%%%%%%%%%%%%%%%%%%%%%%%%%%%%%%%%%%
\subsection{Stable potential ($\mu<0$)}
\label{inverse local time disorder mu<0}
%%%%%%%%%%%%%%%%%%%%%%%%%%%%%%%%%%%%%%%%%%

Following the analysis of Sec.~\ref{local time disorder mu<0}, in the limit
$\alpha\rightarrow 0$, keeping $\alpha T=s$ fixed one gets
\begin{equation} 
\overline{\exp\left[-\lambda(\alpha)T\right]} = {4\over(2\sigma)^\nu
\Gamma^2(\nu)} \left[s^{\nu/2} K_\nu \left({\sqrt{2 s}\over\sqrt{\sigma}}
\right)\right]^2,
\label{inverse local disorder mu<0 rhs}
\end{equation}
with $\nu=|\mu|/\sigma$.

On the other hand, in the corresponding limit $T\rightarrow\infty,
t\rightarrow\infty$, but keeping $t/T=x$ fixed, using the scaling form
\begin{equation} 
\overline{\lss{I}{loc}(t|T)}= {1\over T}\, g_3(t/T),
\label{scaling inverse local time disorder mu<0}
\end{equation}
one gets
\begin{equation} 
\int_0^\infty dt\,e^{-\alpha t} \overline{\lss{I}{loc}(t|T)}= \int_0^\infty
dx\,e^{-sx} g_3(x),
\label{inverse local disorder mu<0 lhs}
\end{equation}
with $s=\alpha T$.  Therefore, in this scaling limit Eq.~(\ref{L-I
disorder}) becomes
\begin{equation}
\int_0^\infty dx\,e^{-sx} g_3(x) = {4\over(2\sigma)^\nu \Gamma^2(\nu)}
\left[s^{\nu/2} K_\nu \left({\sqrt{2 s}\over\sqrt{\sigma}} \right)\right]^2.
\end{equation}
Now comparing the above equation with Eq.~(\ref{L-g2}) one gets
\begin{equation}
g_3(x)=\sigma^{2\nu} g_2(x),
\end{equation}
where the scaling function $g_2(x)$ is given by Eq.~(\ref{g2 function}) and
displayed in Fig.~\ref{g2 plot}.  While $\overline{\lss{I}{loc}(t|T)}$ has
the same scaling function (up to a multiplicative factor of $\sigma^{2\nu}$)
for both stable and unstable potential, the physical behaviors, however, are
quite different. For the stable potential, $\overline{\lss{I}{loc}(t|T)}$ is
normalized to unity. Note that the scaling function $g_3(x)$ becomes
narrower as one increases $\nu$, as expected since the particle becomes more
localized near the origin. For the unstable potential, on the other hand,
the weight of $\overline{\lss{I}{loc}(t|T)}$ decreases as $(\sigma
T)^{-2\nu}$, as one increases $\nu$, as expected since when the repulsive
force from the origin becomes stronger, the particle escapes to $\pm\infty$
with a higher probability.

%%%%%%%%%%%%%%%%%%%%%%%%%%%%%%%%%%%%%%%%%%%%%%%%%%%%%%%%
\section{Occupation time without disorder ($\sigma=0$)}
\label{occupation time}
%%%%%%%%%%%%%%%%%%%%%%%%%%%%%%%%%%%%%%%%%%%%%%%%%%%%%%%%%

In this case $V(x)=\theta(x)$, corresponds to $T$ in Eq.~(\ref{definition of
T}) being the occupation time in the region $x>0$ and $P(T|t,0)$ in
Eq.~(\ref{LL-P}) being $\lss{P}{occ}(T|t)$ --- the pdf of the occupation
time for a given observation time window of size $t$ and the initial
position of the particle $x(0)=0$. Again as before, we need to solve the
differential equation~(\ref{ordinary-u}) for $x>0$ and $x<0$ separately and
then match the solutions at $x=0$. The matching condition for the slope of
the solutions is obtained by integrating Eq.~(\ref{ordinary-u}) across
$x=0$. Thus the matching conditions are
\begin{equation} 
u_+(0)=u_-(0)=u(0), ~~\mbox{and}~~ u_+^\prime(0)=u_-^\prime(0),
\label{match-occupation}
\end{equation}
where $u_\pm(x)$ satisfy the following differential equations:
\begin{align}
&{1\over 2} u_+^{\prime\prime}(x) + F(x) u_+^\prime(x) -
(\alpha+p)u_+(x)=-1,\\* \intertext{for $x>0$, and} &{1\over 2}
u_-^{\prime\prime}(x) + F(x) u_-^\prime(x) - \alpha u_-(x)=-1,
\end{align}
for $x<0$.  The boundary conditions of $u_\pm(x)$ when
$x\rightarrow\pm\infty$ are obtained from the fact that if the starting
position goes to $\pm\infty$, the particle will never cross the origin in
finite time: $P(T|t,x\rightarrow\infty)=\delta(t-T)$ and $P(T|t,
x\rightarrow -\infty)=\delta(T)$, and hence from Eq.~(\ref{LL-P})
\begin{equation} 
u_+(\infty)={1\over \alpha+p}, ~~~\mbox{and}~~~ u_-(-\infty)={1\over\alpha}.
\end{equation}

Writing $u_+(x)=1/(\alpha+p)+B_+y_+(x)$ and $u_-(x)=1/\alpha +
B_-y_-$(x), we obtain the homogeneous differential equations for
$y_\pm(x)$ as
\begin{align}
&{1\over 2} y_+^{\prime\prime}(x) + F(x) y_+^\prime(x) - (\alpha+p)y_+(x)=0,
\label{occupation:y+}
\\*
\intertext{for $x>0$, and}
&{1\over 2} y_-^{\prime\prime}(x) + 
F(x) y_-^\prime(x) - \alpha y_-(x)=0, 
\label{occupation:y-}
\end{align}
for $x<0$, with the boundary conditions $y_+(x\rightarrow \infty)=0$ and
$y_-(x\rightarrow -\infty)=0$. The constants $B_\pm$ are determined by the
matching conditions given in Eq.~(\ref{match-occupation}), which can be
rewritten as
\begin{subequations}
\begin{align}
& {1\over\alpha+p}+B_+y_+(0)={1\over\alpha}+B_-y_-(0)=u(0), \\* &
B_+y_+^\prime(0) = B_-y_-^\prime(0).
\end{align}
\label{match-occupation.1}
\end{subequations}

Eliminating the constants from Eq.~(\ref{match-occupation.1}), we obtain the
double Laplace transform of the pdf of the occupation time
\begin{equation}
\begin{split}
 u(0)&=\int_0^\infty dt\,e^{-\alpha t}\int_0^t dT\,
e^{-pT}\lss{P}{occ}(T|t)\\* &={\ell_1(\alpha,p)\over\alpha} +
{\ell_2(\alpha,p)\over\alpha +p},
\end{split}
\label{u(0)-occupation}
\end{equation}
where
\begin{align}
\ell_1(\alpha,p) &=\left[{z_-(0)\over z_-(0)-z_+(0)}\right],
\label{ell_1}
\\ \ell_2(\alpha,p) &=\left[{-z_+(0)\over z_-(0)-z_+(0)}\right],
\label{ell_2}
\end{align}
and $z_\pm(x)=y_\pm^\prime(x)/y_\pm(x)$. Note that
\begin{equation}
\ell_1(\alpha,p) + \ell_2(\alpha,p) = 1.
\label{ell normalization}
\end{equation} 
Putting $p=0$, in Eq.~(\ref{u(0)-occupation}) gives $u(0)=1/\alpha$, and
hence inverting the Laplace transform with respect to $\alpha$ readily
verifies the normalization
\begin{equation} 
\int_0^t \lss{P}{occ}(T|t)\,dT=1.  
\end{equation}

For any symmetric deterministic potential the distribution of the
occupation time is symmetric about its mean $\langle T\rangle = t/2$, \ie
$\lss{P}{occ}(T|t)=\lss{P}{occ}(t-T|t)$. Then, it follows from this symmetry
that 
\begin{equation}
\ell_1(\alpha+p,-p) = \ell_2(\alpha,p).
\label{ell relation}
\end{equation}
In other words, the double-integral in Eq.~(\ref{u(0)-occupation}) remains
invariant under the following simultaneous replacements:
$(\alpha+p)\rightarrow\alpha$ and $\alpha\rightarrow(\alpha+p)$. Thus under
these replacements one must have $z_+(0)\rightarrow -z_-(0)$ and vice versa,
which also implies that $z_+(0)=-z_-(0)$ for $p=0$. Equivalently,
$\ell_1(\alpha,0)= \ell_2(\alpha,0) =1/2$, which also directly follows from
Eqs.~(\ref{ell normalization}) and~(\ref{ell relation}).

Therefore if one splits the distribution function into two parts:
$\lss{P}{occ}(T|t)=R_L(T|t)+R_R(T|t)$ such that
\begin{align}
\int_0^\infty dt\,e^{-\alpha t}\int_0^t dT\, e^{-pT} R_L(T|t) &=
{\ell_1(\alpha,p)\over\alpha},
\label{RL}
\\ \int_0^\infty dt\,e^{-\alpha t}\int_0^t dT\, e^{-pT}
R_R(T|t) &= {\ell_2(\alpha,p)\over\alpha +p},
\label{RR}
\end{align}
then it follows from the above discussion that $R_L(t-T|t)=R_R(T|t)$. This
symmetry of the distribution will come handy later. Moreover, putting $p=0$
and inverting the Laplace transforms with respect to $\alpha$ gives the
normalization for each part separately
\begin{equation}
\int_0^t R_L(T|t)\, dT = \int_0^t R_R(T|t)\, dT = {1\over2}.
\end{equation}

As an example, we first consider the pure case: $\sigma=0$ in the force
given by Eq.~(\ref{F(x)}). For $F(x)=\mu \sign(x)$, the solutions of
Eqs.~(\ref{occupation:y+}) and (\ref{occupation:y-}) are obtained as
\begin{align}
& y_+(x)=y_+(0)\exp\left[-\left(\mu+\sqrt{\mu^2+2(\alpha+p)}\right)x\right],
\\ \intertext{for $x>0$, and,} &
y_-(x)=y_-(0)\exp\left[\left(\mu+\sqrt{\mu^2+2\alpha}\right)x\right],
\end{align}
for $x<0$. These give the expressions for $z_\pm(0)=y'_\pm(0)/y_\pm(0)$ as
\begin{equation}
\begin{split}
z_+(0)&=-\left[\mu+\sqrt{\mu^2+2(\alpha+p)}\right], \\
z_-(0)&=\left[\mu+\sqrt{\mu^2+2\alpha}\right].
\end{split}
\label{z+-}
\end{equation}
In the following subsections we will consider the three different cases:
(i)~$\mu=0$, (ii)~$\mu>0$ and (iii)~$\mu<0$.

%%%%%%%%%%%%%%%%%%%%%%%%%%%%%%%%%%%%%
\subsection{Flat potential $(\mu=0)$}
\label{occupation time mu=0}
%%%%%%%%%%%%%%%%%%%%%%%%%%%%%%%%%%%%%

For $\mu=0$, using $z_+(0)=-\sqrt{2(\alpha+p)}$ and $z_-(0)=\sqrt{2\alpha}$
from Eq.~(\ref{u(0)-occupation}) we get
\begin{equation}
\int_0^\infty dt\,e^{-\alpha t}\int_0^t dT\, e^{-pT} \lss{P}{occ}(T|t)
={1\over\sqrt{\alpha(\alpha+p)}}.
\end{equation}
Inverting the double Laplace transform with respect to $p$ and then with
respect to $\alpha$ finally reproduces the well known L\'{e}vy's ``arcsine''
law~\cite{levy:1939} for the pdf of the occupation time of an ordinary
Brownian motion,
\begin{equation} 
\lss{P}{occ}(T|t)={1\over \pi \sqrt{T(t-T)}}, ~~~ 0<T<t.
\label{occupation-flat}
\end{equation} 
The distribution $\lss{P}{occ}(T|t)$ diverges on both ends $T=0$ and $T=t$,
which indicates that the Brownian particle ``tends'' to stay on one side of
the origin.

%%%%%%%%%%%%%%%%%%%%%%%%%%%%%%%%%%%%%%%%%% 
\subsection{Unstable potential $(\mu>0)$}
\label{occupation time mu>0}
%%%%%%%%%%%%%%%%%%%%%%%%%%%%%%%%%%%%%%%%%% 

Since for $\mu>0$, the force is repulsive from the origin $x=0$, one would
expect the occupation time distribution to be convex (concave upward), with
minimum at $T=t/2$.  Now in the limit of large window size
$t\rightarrow\infty$, the part of the distribution $\lss{P}{occ}(T|t)$ to
the left of the midpoint $T=t/2$ approaches $R_L(T|t)$, as the midpoint
itself goes to $\infty$.

By making a change of variable $z=\alpha t$, it follows from Eq.~(\ref{RL})
\begin{equation} 
\int_0^\infty dz\, e^{-z} \int_0^{z/\alpha} dT\, e^{-pT} R_L(T|z/\alpha)
=\ell_1(\alpha,p).
\end{equation}
Now the large $t$ limit of $R_L(T|t)$ can be obtained by taking
$\alpha\rightarrow 0$ in the above equation, where one realizes that
$R_L(T|t)$ approaches a steady ($t$ independent) distribution,
$R_L(T|t\rightarrow\infty)\rightarrow R_L(T)$, whose Laplace transform is
given by
\begin{equation} 
\int_0^\infty dT\, e^{-pT} R_L(T)= \ell_1(0,p),
\end{equation}
where $\ell_1(0,p)$ is obtained from Eq.~(\ref{ell_1}), by using $z_\pm(0)$
from Eq.~(\ref{z+-}), which gives
\begin{equation} 
\ell_1(0,p)= {2\mu\over3\mu+\sqrt{\mu^2+2p}}.
\end{equation}
The above Laplace transform can be inverted with respect to $p$, which gives
\begin{multline}
R_L(T)=\mu\sqrt{2}\,e^{-\mu^2T/2}\\ \times\left[ {1\over\sqrt{\pi T}}-
{3\mu\over\sqrt{2}}\exp\left({9\mu^2\over 2}T\right)
\erfc\left({3\mu\over\sqrt{2}}\sqrt{T}\right) \right].
\label{R_L unstable}
\end{multline}
with the normalization $\int_0^\infty R_L(T)\,dT=\ell_1(0,0)=1/2$.

The limiting behavior of the distribution is given by
\begin{align}
& R_L(T)\approx{\mu\sqrt{2}\over\sqrt{\pi T}},\\ \intertext{for small $T$
and decays exponentially for large $T$,} & R_L(T)\approx
{\sqrt{2}\over9\mu\sqrt{\pi}} {e^{-\mu^2T/2}\over T^{3/2}}.
\end{align}

%%%%%%%%%%%%%%%%%%%%%%%%%%%%%%%%%%%%%%%%
\subsection{Stable potential $(\mu<0)$}
\label{occupation time mu<0}
%%%%%%%%%%%%%%%%%%%%%%%%%%%%%%%%%%%%%%%%
As we discussed earlier in Sec.~\ref{local time mu<0} in the context of the
local time, for generic stable potential $U(x)$ the system eventually
becomes ergodic at large $t$ and hence the average $\langle \theta[x(t)]
\rangle$ approaches its stationary value $ \langle \theta[x(t)] \rangle
\rightarrow Z_+/Z$, where $Z=\int_{-\infty}^\infty e^{-2U(x)} \, dx$ is the
equilibrium partition function and $Z_+= \int_0^\infty e^{-2U(x)} \, dx$ is
the restricted partition function. Therefore, for large $t$ the average
occupation time $\langle T \rangle = \int_0^t \langle \theta[x(t')]\rangle
\, dt'$ scales linearly with $t$
\begin{equation}
\langle T \rangle \rightarrow \left({Z_+ \over Z}\right) t.
\end{equation}
Note that when the potential $U(x)$ is symmetric about zero, the average
occupation time $\langle T \rangle = t/2$ for all $t$.

From the definition, $T-\langle T\rangle = \int_0^t \{\theta[x(t')] -\langle
\theta[x(t')] \rangle\}\, dt'$, it follows that when $T\rightarrow \langle
T\rangle $, the random variables $ \theta[x(t')] -\langle \theta[x(t')]
\rangle$ at different times $t'$ become only weekly correlated. Then in the
limit when $t$ is much larger than the correlation time between these
variables, one expects the central limit theorem to hold, which predicts a
Gaussian form for the distribution of the occupation time $T$ near the mean
value $\langle T \rangle$,
\begin{equation} 
\lss{P}{occ}(T|t) \sim \exp \left[-{(T-\langle T \rangle)^2\over 2\sigma^2}
\right],
\end{equation}
where the variance $\sigma^2 = \langle T^2\rangle -\langle T \rangle^2$ can
be obtained from the Laplace transform of the moments
\begin{equation} 
\int_0^\infty \langle T^n \rangle\, e^{-\alpha t}\, dt = \left.(-1)^n
{\partial^n u(0) \over \partial p^n}\right|_{p=0},
\label{moments occupation stable}
\end{equation}
with $u(0)$ given by Eq.~(\ref{u(0)-occupation}).

For the particular attractive force $F(x)=-|\mu| \sign(x)$, using $z_\pm(0)$
from Eq.~(\ref{z+-}) in Eq.~(\ref{u(0)-occupation}) and taking derivatives
with respect to $p$ in we get
\begin{align}
 \left.-{\partial u(0) \over \partial p}\right|_{p=0}&={1\over 2\alpha^2},
\\ \left.{\partial^2 u(0) \over \partial p^2}\right|_{p=0} &= {1\over
2\alpha^3} + {1\over 4\mu^2\alpha^2} + \order\left({1\over\alpha}\right).
\end{align}
Therefore inverting the Laplace transform in Eq.~(\ref{moments occupation
stable}) with respect to $\alpha$ immediately gives $\langle T \rangle =
t/2$ for all $t$, and $\langle T^2 \rangle = t^2/4 + t/4\mu^2$ for large $t$
which gives $\sigma^2 = t/4\mu^2$.

%%%%%%%%%%%%%%%%%%%%%%%%%%%%%%%%%%%%%%%%%%%%%%%%%%%%%%%%%%%%%%%%%%%%%%
\section{Occupation Time with disorder ($\sigma>0$)}
\label{occupation time disorder}
%%%%%%%%%%%%%%%%%%%%%%%%%%%%%%%%%%%%%%%%%%%%%%%%%%%%%%%%%%%%%%%%%%%%%%

Now we consider the occupation time when the disorder is switched on:
$\sigma>0$ in Eq.~(\ref{F(x)}). Our aim is to calculate the disorder
averaged $\overline{\lss{P}{occ}(T|t)}$. As one realizes from
Eqs.~(\ref{u(0)-occupation}), (\ref{ell_1}) and (\ref{ell_2}), to calculate
$\overline{\lss{P}{occ}(T|t)}$ one needs the distribution of $-z_+(0)$ and
$z_-(0)$, which are given by Eqs.~(\ref{P+ equilibrium}) and~(\ref{P-
equilibrium}) with $a_+=\alpha+p $ and $a_-=\alpha$ respectively.  In the
following subsections, we will consider the three cases: (i)~$\mu=0$,
(ii)~$\mu>0$ and (iii)~$\mu<0$.

%%%%%%%%%%%%%%%%%%%%%%%%%%%%%%%%%%%%%%%%%%%%%%%%%%%%
\subsection{Flat potential ($\mu=0$) -- Sinai model}
\label{occupation time disorder mu=0}
%%%%%%%%%%%%%%%%%%%%%%%%%%%%%%%%%%%%%%%%%%%%%%%%%%%%

We first consider the diffusive motion of a particle in a continuous Sinai
potential, where the potential itself is a Brownian motion in space. In the
limit of large window size $t$ the left half of the disorder averaged pdf of
the occupation time $\overline{R_L(T|t)}$ for $0\le T\le t/2$ is obtained by
taking the disorder average in Eq.~(\ref{RL}).  The right half of the
distribution for $t/2\le T\le t$ is just the symmetric reflection of the
left part. The detailed calculations for $\overline{R_L(T|t)}$ are presented
in appendix~\ref{appendix occupation Sinai}.

We find that $\overline{R_L(T|t)}$ has a large $t$ behavior 
\begin{equation}
\overline{R_L(T|t)} \approx {1\over \log t} R(T),
\end{equation}
where the function $R(T)$ is independent of $t$. The limiting behaviors of
$R(T)$ are given by
\begin{equation}
R(T)\approx {\sqrt{2}\sigma\over\sqrt{\pi T}},
\end{equation}
as $T\rightarrow 0$ and
\begin{equation}
R(T) \sim {1\over 2T},
\end{equation}
for large $T$.

%%%%%%%%%%%%%%%%%%%%%%%%%%%%%%%%%%%%%%%
\subsection{Unstable potential ($\mu>0$)}
\label{occupation time disorder mu>0}
%%%%%%%%%%%%%%%%%%%%%%%%%%%%%%%%%%%%%%%

For $\mu>0$, we find that disorder does not change the asymptotic behavior of
the distribution for the pure case qualitatively. The calculations are
presented in appendix~\ref{appendix occupation time disorder mu>0}.  We find
that in the limit $t\rightarrow\infty$ the left half of the disorder
averaged occupation time distribution tends to a $t$ independent form
\begin{equation}
\overline{R_L(T|t)}=\overline{R_L(T)}.
\end{equation}
In fact the small $T$ limit of $\overline{R_L(T)}$ remains same as in the
pure case
\begin{equation} 
\overline{R_L(T)} \approx {\mu\sqrt{2}\over\sqrt{\pi T}}.
\end{equation}
For large $T$, the distribution $\overline{R_L(T)}$ still decays
exponentially
\begin{equation} 
\overline{R_L(T)} \sim e^{-b T},
\end{equation}
where the decay coefficient $b$ is, however, different from the pure case
(see Eq.~\ref{rhs (5) occupation disorder mu>0}).

%%%%%%%%%%%%%%%%%%%%%%%%%%%%%%%%%%%%%%%%%%
\subsection{Stable potential ($\mu<0$)}
\label{occupation time disorder mu<0}
%%%%%%%%%%%%%%%%%%%%%%%%%%%%%%%%%%%%%%%%%%

This particular situation, where one finds the interplay between two
competing processes, is a very interesting one. On one hand, as we discussed
in Sec.~\ref{occupation time mu<0}, the stable potential in the absence of
the disordered potential makes the system ergodic in the large $t$ limit,
and as a result the pdf of the occupation time is peaked at $ =t/2$ and
decays fast away from it. On the other hand, as we discussed in
Sec.~\ref{occupation time disorder mu=0}, without any underlying
deterministic potential the disorder averaged pdf of the occupation time is
convex (concave upward) with a minimum at $T=t/2$ and diverges at the both
ends $T\rightarrow 0$ and $T\rightarrow t$. Therefore, if both the stable
potential and disordered potential are included, as their relative strength
$\nu=|\mu|/\sigma$ is varied, one expects a phase transition at some
critical value $\nu_c$ where the system looses ergodicity.

\begin{figure}
\centering
\includegraphics[width=0.9\hsize]{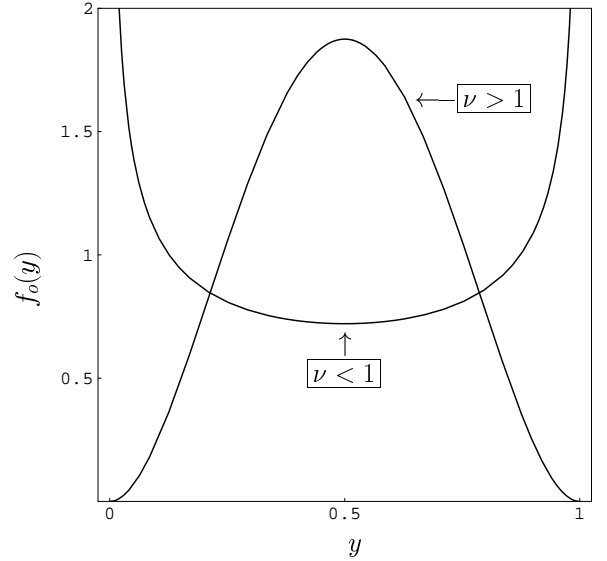}
\caption{\label{fo plot} The scaling functions $f_o(x)$ in Eq.~(\ref{scaling
occupation disorder mu=0}) plotted by using Eq.~(\ref{beta law}).}
\end{figure}

In the scaling limit where both $t\rightarrow\infty$ and
$T\rightarrow\infty$, but their ratio $y=T/t$ is kept fixed, we find that
the disorder averaged pdf of the occupation time has a scaling form
\begin{equation}
\overline{\lss{P}{occ}(T|t)}= {1\over t}f_o(T/t).
\label{scaling occupation disorder mu=0}
\end{equation}
The calculation of the scaling function $f_o(y)$ is presented in
appendix~\ref{appendix occupation time disorder mu<0}, where we find the
Beta law
\begin{equation} 
f_o(y)={1\over B(\nu,\nu)}\left[y(1-y)\right]^{\nu-1}, ~~~~ 0\le y\le 1,
\label{beta law}
\end{equation}
where $\nu=|\mu|/\sigma$ and $B(\nu,\nu)$ is the Beta
function~\cite{gradshteyn}.  Now if one tunes the parameter $\nu$ by varying
either $\mu$ or the disorder strength $\sigma$, the distribution
$\overline{\lss{P}{occ}(T|t)}$ exhibits a phase transition in the ergodicity
of the particle position at $\nu_c=1$ (Fig.~\ref{fo plot}). For $\nu<\nu_c$
the distribution $f_o(y)$ in Eq.~(\ref{scaling occupation disorder mu=0}) is
convex with a minimum at $y=1/2$ and diverges at the two ends $y=0,1$. This
means that particle tends to stay on one side of the origin such that $T$ is
close to either $0$ or $t$. In other words the paths with small number of
zero crossings carry more weight than the ones that cross many times. For
$\nu>\nu_c$ the scenario is exactly opposite, where $f_o(y)$ is maximum at
the mean value $y=1/2$ indicating that particle tends to spend equal times
on both sides of the origin $x=0$, such that paths with large number of zero
crossings, for which $T$ is closer to $t/2$ carry larger weight. Similar
phase transition in the ergodicity properties of a stochastic process as one
changes a parameter, was first noted in the context of diffusion
equation~\cite{newman:1998}, and later found for a class of Gaussian Markov
processes~\cite{dhar:6413} and in simple models of
coarsening~\cite{baldassarri:R20,0305-4470-33-42-303}.

A very interesting observation about Eq.~(\ref{beta law}) is that for
$\nu=1/2$, the result is same as L\'evy's result for the one-dimensional
Brownian motion given by Eq.~(\ref{occupation-flat}). It seems as if the
attractive force cancels the effect of disorder exactly at
$\nu=1/2$. However, this is no more true in the context of the local time.

%%%%%%%%%%%%%%%%%%%%%%%%%%%%%%%%%%%%%%%%%%%%%%%%%%%%%%%%%%%%%%%%%
\section{Inverse  occupation time without disorder ($\sigma=0$)}
\label{inverse occupation time}
%%%%%%%%%%%%%%%%%%%%%%%%%%%%%%%%%%%%%%%%%%%%%%%%%%%%%%%%%%%%%%%%%%

In this case $I(t|T,0)$ in Eq.~(\ref{LL-I}) is replaced with
$\lss{I}{occ}(t|T)$, which is the distribution of the time $t$ needed to
observe the particle with a starting position $x=0$, until the total amount
of time spent on the positive side $x>0$ is $T$.  Corresponding $u(0)$ in
Eq.~(\ref{LL-I}) with $x=0$, which is the double Laplace transform of the
pdf of the occupation time, has already been evaluated in
Sec.~\ref{occupation time}, which is given by
Eq.~(\ref{u(0)-occupation}). Substituting $u(0)$ in Eq.~(\ref{LL-I}) gives
\begin{equation}
\int_0^\infty dT\, e^{-pT} \int_T^\infty dt\,e^{-\alpha t} \lss{I}{occ}(t|T)
= {\ell_2(\alpha,p)\over\alpha+p},
\label{LL-Io}
\end{equation}
where $\ell_2(\alpha,p)$ is given by Eq.~(\ref{ell_2}).  Comparing the above
equation with Eq.~(\ref{RR}), one can infer that $\lss{I}{occ}(t|T)$ and
$R_R(T|t)$ have the same functional form, \ie $\lss{I}{occ}(t|T)=R_R(T|t)$
and especially for symmetric deterministic potential
$\lss{I}{occ}(t|T)=R_R(T|t)=R_L(t-T|t)$.

It is useful to present the above equation in the following form,
\begin{equation} 
\int_0^\infty dz\, e^{-z} \int_0^\infty d\tau\, e^{-\alpha\tau}\,
\lss{I}{occ}\left(\tau+{z\over\beta}\left|{z\over\beta}\right.\right)
=\ell_1(\beta,\alpha -\beta),
\label{LL-Io beta}
\end{equation}
which has been obtained by substituting $p=\beta-\alpha$ in
Eq.~(\ref{LL-Io}) and subsequently making the change of variables $\beta
T=z$ and $\tau=t-T$. On the right hand side, we have substituted
$\ell_2(\alpha,\beta -\alpha)=\ell_1(\beta, \alpha - \beta)$, using
Eq.~(\ref{ell relation}) and $\ell_1(\alpha,p)$ is given by
Eq.~(\ref{ell_1}).  Now by taking the limit $\beta\rightarrow 0$ in
Eq.~(\ref{LL-Io beta}), one obtains the large $T$ behavior of
$\lss{I}{occ}(t|T)$.

For the pure case, $\sigma=0$ in Eq.~(\ref{F(x)}), we have already obtained
$z_\pm(0)$ in Sec.~\ref{occupation time}, which are given by Eq.~(\ref{z+-})
and hence we can evaluate $\ell_1(\alpha,p)$ and $\ell_2(\alpha,p)$ by using
Eqs.~(\ref{ell_1}) and~(\ref{ell_2}) respectively. In the following
subsections we will analyze the behavior of $\lss{I}{occ}(t|T)$ for the
cases (i)~$\mu=0$, (ii)~$\mu>0$ and (iii)~$\mu<0$.

%%%%%%%%%%%%%%%%%%%%%%%%%%%%%%%%%%%%%%
\subsection{Flat potential $(\mu=0)$}
\label{inverse occupation time mu=0}
%%%%%%%%%%%%%%%%%%%%%%%%%%%%%%%%%%%%%%%

\begin{figure}
\centering
\includegraphics[width=0.9\hsize]{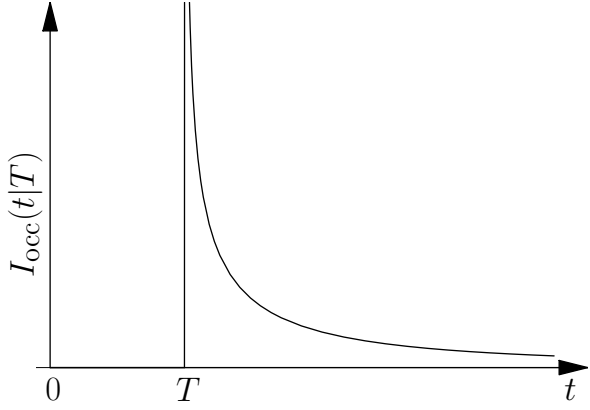}
\caption{\label{I_occ mu=0} The pdf of the inverse occupation time for
simple Brownian motion.}
\end{figure}

For $\mu=0$, which is the case of a simple Brownian motion,
$z_+(0)=-\sqrt{2(\alpha+p)}$ and $z_-(0)=\sqrt{2\alpha}$. Therefore using
Eq.~(\ref{ell_2}), from Eq.~(\ref{LL-Io}) we get
\begin{multline}
\int_0^\infty dT\, e^{-pT} \int_T^\infty dt\,e^{-\alpha t}
\lss{I}{occ}(t|T)\\ ={1\over
\sqrt{\alpha+p}(\sqrt{\alpha}+\sqrt{\alpha+p})}.
\end{multline}
Now inverting the Laplace transform with respect to $p$ gives
\begin{equation}
\int_T^\infty dt\,e^{-\alpha t} \lss{I}{occ}(t|T)=\erfc(\sqrt{\alpha T}),
\label{L-Io}
\end{equation}
and further inverting the Laplace transform with respect to $\alpha$ gives
\begin{equation} 
\lss{I}{occ}(t|T)=\frac{\sqrt{T}}{\pi t\sqrt{t-T}}\,\theta(t-T).
\label{I_occ flat}
\end{equation}
with the normalization condition $\int_T^\infty \lss{I}{occ}(t|T)\,dt=1$,
which is readily checked by putting $\alpha=0$ in Eq.~(\ref{L-Io}).  The
inverse occupation time has non-zero support only for $t>T$, as shown in
Fig.~\ref{I_occ mu=0}.

Note that, since $R_R(T|t)=\lss{I}{occ}(t|T)=\sqrt{T}/\pi t\sqrt{t-T}$ and
$R_L(T|t)=R_R(t-T|t)=\sqrt{t-T}/\pi t\sqrt{T}$, adding the two parts,
$\lss{P}{occ}(T|t)= R_L(T|t)+R_R(T|t)= 1/\pi \sqrt{T(t-T)}$, one recovers
Eq.~(\ref{occupation-flat}).

%%%%%%%%%%%%%%%%%%%%%%%%%%%%%%%%%%%%%%%%%
\subsection{Unstable potential $(\mu>0)$}
\label{inverse occupation time mu>0}
%%%%%%%%%%%%%%%%%%%%%%%%%%%%%%%%%%%%%%%%%

For $\mu>0$, although one can invert the Laplace transform in
Eq.~(\ref{LL-Io}) with respect to $p$ exactly, the other Laplace transform
with respect to $\alpha$ can be inverted only in the large $T$
limit. Therefore to keep the presentation simpler, we will consider the
large $T$ behavior of $\lss{I}{occ}(t|T)$ by analyzing Eq.~(\ref{LL-Io
beta}) in the limit $\beta \rightarrow 0$.

By using $z_\pm(0)$ from Eq.~(\ref{z+-}) in Eq.~(\ref{ell_1}), one gets
\begin{equation}
\ell_1(0,\alpha) = {2\mu\over3\mu+\sqrt{\mu^2+2\alpha}},
\end{equation}
Therefore, Eq.~(\ref{LL-Io beta}) suggest that $\lss{I}{occ}(t|T)$ should
only depend on the difference $(t-T)$ at large $T$
\begin{equation} 
\lss{I}{occ}(t|T)=I_1(t-T).
\end{equation} 
Substituting this form in Eq.~(\ref{LL-Io beta}) in the limit $\beta
\rightarrow 0$ gives
\begin{equation} 
\int_0^\infty d\tau\,e^{-\alpha \tau} \, I_1(\tau) = \ell_1(0,\alpha),
\end{equation}
where putting $\alpha=0$ gives the normalization $\int_0^\infty
I_1(\tau)\,d\tau=\ell_1(0,0)=1/2$, indicating that the particle can escape
to $-\infty$ with probability $1/2$ for the unstable potential (the force is
repulsive from the origin).  Now inverting the Laplace transform with
respect to $\alpha$ gives
\begin{multline} 
I_1(\tau)=\mu\sqrt{2}\,e^{-\mu^2 \tau /2}\\ \times\left[ {1\over\sqrt{\pi
\tau}}- {3\mu\over\sqrt{2}}\exp\left({9\mu^2\over 2} \tau \right)
\erfc\left({3\mu\over\sqrt{2}}\sqrt{\tau}\right) \right],
\end{multline}
The limiting behavior of this distribution is given by
\begin{equation}
I_1(\tau)\approx{\mu\sqrt{2}\over\sqrt{\pi \tau}},
\end{equation}
for small $\tau=(t-T)$ and decays exponentially for large $\tau=(t-T)$,
\begin{equation} 
I_1(\tau)\approx {\sqrt{2}\over9\mu\sqrt{\pi}} {e^{-\mu^2\tau/2}\over
\tau^{3/2}}.  
\end{equation}

%%%%%%%%%%%%%%%%%%%%%%%%%%%%%%%%%%%%%%%
\subsection{stable potential $(\mu<0)$}
\label{inverse occupation time mu<0}
%%%%%%%%%%%%%%%%%%%%%%%%%%%%%%%%%%%%%%%

It is reasonable to consider the difference variable $t-T$ instead of $t$,
as $t\ge T$. Therefore, we write
\begin{equation} 
\lss{I}{occ}(t|T)=I_2(t-T,T).
\end{equation}
Substituting this form and $p=\beta-\alpha$ in Eq.~(\ref{LL-Io}) one gets
\begin{equation} 
\int_0^\infty dT\,e^{-\beta T}\int_0^\infty d\tau\, e^{-\alpha \tau}
I_2(\tau,T) = {\ell_1(\beta,\alpha-\beta)\over \beta},
\label{LL-Io mu<0}
\end{equation}
where we have substituted $\ell_2(\alpha,\beta -\alpha)=\ell_1(\beta, \alpha
- \beta)$ on the right hand side, using Eq.~(\ref{ell relation}).  Using
$z_\pm(0)$ from Eq.~(\ref{z+-}) for $\mu<0$, in Eq.~(\ref{ell_1}) gives
\begin{equation} 
\ell_1(\beta,\alpha-\beta) = \left[{ \sqrt{\mu^2+2\beta} - |\mu| \over
\sqrt{\mu^2+2\beta} + \sqrt{\mu^2+2\alpha} - 2|\mu| }\right],
\end{equation}
Therefore, taking the small $\beta$ limit in Eq.~(\ref{LL-Io mu<0}) gives
\begin{multline}
 \int_0^\infty dT\,e^{-\beta T}\int_0^\infty d\tau\, e^{-\alpha \tau}
I_2(\tau,T)\\ = {1\over \beta +|\mu|\sqrt{\mu^2+2\alpha} - \mu^2},
\end{multline}
and inverting the Laplace transform with respect to $\beta$ gives
\begin{equation} 
\int_0^\infty d\tau\, e^{-\alpha \tau} I_2(\tau,T) = \exp\left(\mu^2 T
-|\mu|T\sqrt{\mu^2+2\alpha}\right),
\end{equation}
where putting $\alpha=0$ confirms the normalization condition $\int_0^\infty
I_2(\tau,T)\,d\tau=1$. Now by inverting the other Laplace transform with
respect to $\alpha$ one gets the distribution
\begin{equation} 
I_2(\tau,T)={|\mu| T\over \sqrt{2\pi \tau^3}} e^{-\mu^2(\tau-T)^2/2\tau},
\end{equation}
where $\tau=t-T$.

%%%%%%%%%%%%%%%%%%%%%%%%%%%%%%%%%%%%%%%%%%%%%%%%%%%%%%%%%%%%%%%%%%%%%%%%%%%%%
\section{Inverse occupation time with disorder ($\sigma>0$)}
\label{inverse occupation time disorder}
%%%%%%%%%%%%%%%%%%%%%%%%%%%%%%%%%%%%%%%%%%%%%%%%%%%%%%%%%%%%%%%%%%%%%%%%%%%%%

In the presence of disorder, \ie $\sigma>0$ in Eq.~(\ref{F(x)}), taking the
disorder average in Eq.~(\ref{LL-Io}) gives
\begin{equation} 
\int_0^\infty dT\,e^{-pT}\int_T^\infty dt\,e^{-\alpha
t}\,\overline{\lss{I}{occ}(t|T)} =
{\overline{\ell_2(\alpha,p)}\over\alpha+p},
\label{LL-Io disorder}
\end{equation}
where $\overline{\ell_2(\alpha,p)}$ is obtained by taking the disorder
average of Eq.~(\ref{ell_2}), using the distributions of $-z_+(0)$ and
$z_-(0)$ given by Eqs.~(\ref{P+ equilibrium}) and~(\ref{P- equilibrium})
respectively with $a_+=\alpha+p$ and $a_-=\alpha$.

It is useful to consider a different form of the above equation, which is
obtained by taking the disorder average of Eq.~(\ref{LL-Io beta})
\begin{equation} 
\int_0^\infty dz\, e^{-z} \int_0^\infty d\tau\, e^{-\alpha\tau}\,
\overline{\lss{I}{occ}\left(\tau+{z\over\beta}
\left|{z\over\beta}\right.\right)} =\overline{\ell_1(\beta,\alpha -\beta)},
\label{LL-Io disorder beta}
\end{equation}
where by taking the limit $\beta\rightarrow 0$, one obtains the large $T$
behavior of $\overline{\lss{I}{occ}(t|T)}$.

%%%%%%%%%%%%%%%%%%%%%%%%%%%%%%%%%%%%%%%%%%%%%%%%%%%%
\subsection{Flat potential ($\mu=0$) -- Sinai model}
\label{inverse occupation time disorder mu=0}
%%%%%%%%%%%%%%%%%%%%%%%%%%%%%%%%%%%%%%%%%%%%%%%%%%%%

We will now study the large $T$ behavior of $\overline{\lss{I}{occ}(t|T)}$,
for the Sinai potential ($\mu=0$), by analyzing Eq.~(\ref{LL-Io disorder
beta}) in the limit $\beta\rightarrow 0$.

It follows from Eq.~(\ref{rhs (0) occupation disorder mu=0}) that
\begin{equation} 
\overline{\ell_1(\beta,\alpha -\beta)} = { m_1(\beta,\alpha-\beta)\over
\Omega_+\Omega_-},
\label{rhs (0) inverse occupation disorder mu=0}
\end{equation}
where $m_1(\alpha,p)$ is given by Eq.~(\ref{rhs (1) occupation disorder
mu=0}), and
\begin{equation} 
\Omega_+ = 2K_0\left({\sqrt{2\beta}\over\sigma} \right) ~~~\mbox{and}~~~
\Omega_- = 2K_0\left({\sqrt{2\alpha}\over\sigma} \right).  
\end{equation}

In the limit $\beta\rightarrow 0$, since $\Omega_+ \sim -\log\beta$, hence
\begin{equation} 
\overline{\ell_1(\beta,\alpha -\beta)} \sim {m_1(0,\alpha) \over
\left[-2K_0\left({\sqrt{2\alpha}/\sigma} \right)\,\log\beta\right]},
\label{rhs (2) inverse occupation disorder mu=0}
\end{equation}
which suggest the following scaling form at large $T$,
\begin{equation}
\overline{\lss{I}{occ}(t|T)}={1\over\log T}\,I_3(t-T).
\end{equation}
Therefore in the limit $\beta\rightarrow 0$, substituting the above scaling
form in Eq.~(\ref{LL-Io disorder beta}) and using Eq.~(\ref{rhs (2) inverse
occupation disorder mu=0}) one gets
\begin{equation} 
\int_0^\infty d\tau\, e^{-\alpha \tau}\, I_3(\tau) ={m_1(0,\alpha)\over 2
K_0\left({\sqrt{2\alpha}/\sigma} \right)}.
\end{equation}
However, the above Laplace transform is the same one given by Eq.~(\ref{(1)
scaling limit disorder mu=0}) in appendix~\ref{appendix occupation Sinai},
where $\alpha$ is replaced by $p$.  Therefore we can directly borrow the
results obtained there. Using the results from Eq.~(\ref{scaling function
occupation disorder mu=0 small T}) gives
\begin{equation} 
I_3(\tau)\approx{\sqrt{2}\sigma\over\sqrt{\pi\tau}}
~~~\mbox{as}~~\tau\rightarrow 0, 
\end{equation}
and results from Eq.~(\ref{scaling function occupation disorder mu=0 large
T}) gives
\begin{equation}
I_3(\tau)\sim{1\over 2\tau}~~~\mbox{as}~~\tau\rightarrow\infty.
\end{equation}

%%%%%%%%%%%%%%%%%%%%%%%%%%%%%%%%%%%%%%%%%%%%%%
\subsection{Unstable potential ($\mu>0$)}
\label{inverse occupation time disorder mu>0}
%%%%%%%%%%%%%%%%%%%%%%%%%%%%%%%%%%%%%%%%%%%%%%%

For $\mu>0$, Eq.~(\ref{LL-Io disorder beta}) suggests that in the large $T$
limit, $\overline{\lss{I}{occ}(t|T)}$ will only depend on the difference
$(t-T)$,
\begin{equation}
\overline{\lss{I}{occ}(t|T)}=I_4(t-T).
\end{equation}
Therefore in the limit $\beta\rightarrow 0$, using the above form in
Eq.~(\ref{LL-Io disorder beta}) one gets
\begin{equation} 
\int_0^\infty d\tau\, e^{-\alpha\tau} I_4(\tau) =
\overline{\ell_1(0,\alpha)},
\label{inverse occupation disorder mu>0}
\end{equation}
However, the above Laplace transform is the same one given by Eq.~(\ref{L
occupation disorder mu>0}) where $\alpha$ is replaced by $p$. Therefore
borrowing the results from appendix~\ref{appendix occupation time disorder
mu>0} readily gives
\begin{equation} 
I_4(\tau)\approx {\mu\sqrt{2}\over\sqrt{\pi \tau}},
\end{equation}
for small $\tau$ and
\begin{equation} 
I_4(\tau) \sim e^{-b\tau}, 
\end{equation}
for large $\tau$, with the same constant $b$ as in Eq.~(\ref{R_L large T}).

The normalization condition $\int_0^\infty I_4(\tau)\, d\tau =
\overline{\ell_1(0,0)}$=1/2, indicates that the particle escapes to
$-\infty$ with probability $1/2$.

%%%%%%%%%%%%%%%%%%%%%%%%%%%%%%%%%%%%%%%%%%%%%
\subsection{Stable potential ($\mu<0$)}
\label{inverse occupation time disorder mu<0}
%%%%%%%%%%%%%%%%%%%%%%%%%%%%%%%%%%%%%%%%%%%%%

\begin{figure}
\centering
\includegraphics[width=0.9\hsize]{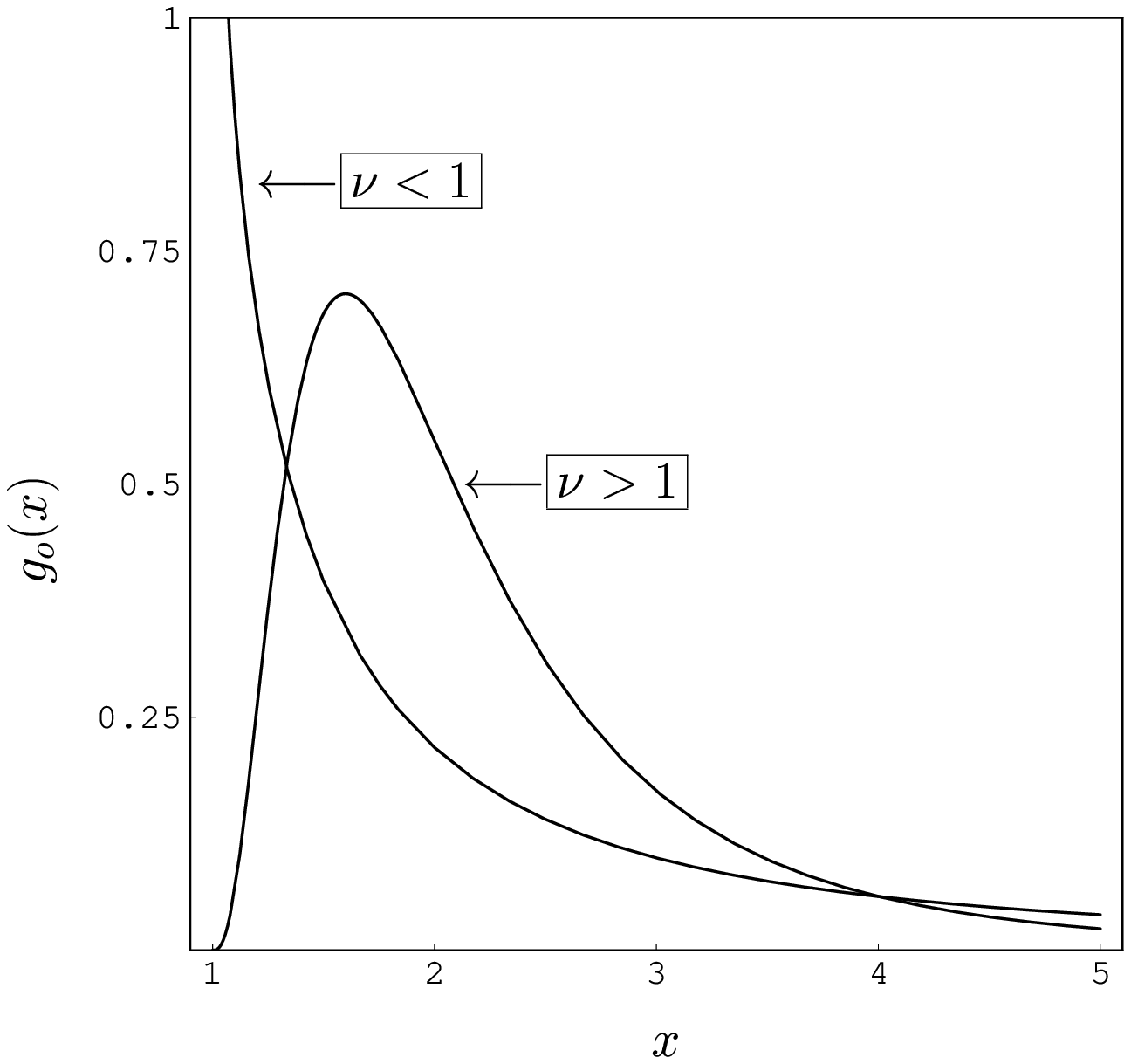}
\caption{\label{go plot} The scaling functions $g_o(x)$ in Eq.~(\ref{scaling
inverse occupation disorder mu=0}) plotted by using Eq.~(\ref{go
function}).}
\end{figure}

We are interested in the behavior of $\overline{\lss{I}{occ}(t|T)}$ in the
scaling limit where $t\rightarrow \infty$ and $T\rightarrow \infty$, but the
ratio $x=t/T$ is kept fixed.  Substituting $T=z/\alpha$, $t=x z/\alpha$ and
$p=\alpha s$ in Eq.~(\ref{LL-Io disorder}), we get
\begin{equation} 
\int_1^\infty dx\, \int_0^\infty dz\, e^{-(s+x) z}
\left[{z\over\alpha}\,\overline{\lss{I}{occ} \left({xz\over\alpha}\left|
{z\over\alpha}\right. \right)} \right] = {1-m_3(\alpha,s) \over 1+s},
\label{xz Io disorder mu<0 alpha}
\end{equation}
where $m_3(\alpha,s)=\overline{\ell_1(\alpha,\alpha s)}=1 -
\overline{\ell_2(\alpha.p)}$, given by Eq.~(\ref{rhs (1) occupation disorder
mu<0}). The above equation suggest the scaling form
\begin{equation} \overline{\lss{I}{occ}(t|T)}={1\over T}\, g_o(t/T),
\label{scaling inverse occupation disorder mu=0}
\end{equation}
with the normalization $\int_1^\infty g_o(x)\,dx =1$, which follows directly
from the normalization $\int_T^\infty \overline{\lss{I}{occ}(t|T)} \, dt
=1$.  By substituting the above scaling form in Eq.~(\ref{xz Io disorder
mu<0 alpha}) in the limit $\alpha\rightarrow 0$, after simplification one
gets
\begin{equation} 
\int_1^\infty \left[{x-1\over x+s}\right]\, g_o(x)\,dx = m_3(0,s),
\label{inverse occupation disorder mu<0 (1)}
\end{equation}
where $m_3(0,s)$ is given by Eq.~(\ref{rhs (6) occupation disorder
mu<0}). By making a change of variable $y=1/x$, Eq.~(\ref{rhs (6) occupation
disorder mu<0}) reads
\begin{equation} 
m_3(0,s) ={1\over B(\nu,\nu)}\int_1^\infty \left[{x-1\over x+s}\right]
{(x-1)^{\nu-1}\over x^{2\nu}}\,dx.
\label{inverse occupation disorder mu<0 (2)}
\end{equation}
Therefore comparing Eq.~(\ref{inverse occupation disorder mu<0 (1)}) and
Eq,~(\ref{inverse occupation disorder mu<0 (2)}) readily gives the inverted Beta law 
\begin{equation} 
g_o(x) = {1\over B(\nu,\nu)} {(x-1)^{\nu-1}\over x^{2\nu}}, \quad x>1,
\label{go function}
\end{equation}
which is displayed in Fig.~\ref{go plot}.  The scaling function $g_o(x)$ has
a maximum at $x=2\nu/(\nu+1)$ for $\nu>1$.  However, $g_o(x)$ diverges near
$x=1$ for $\nu<1$. Note that for $\nu=1/2$, Eq.~(\ref{go function}) gives
identical results to that of a pure Brownian motion ($\mu=0$ and
$\sigma=0$), given by Eq.~(\ref{I_occ flat}).

%%%%%%%%%%%%%%%%%%%%%%%%%%%
\section{Concluding Remarks}
\label{summary}
%%%%%%%%%%%%%%%%%%%%%%%%%%%

In this paper we have considered the motion of a particle in a one
dimensional random potential. We have presented a general formalism for
computing statistical properties of functionals and the inverse functionals
of this process.  We have used a backward Fokker-Planck equation approach to
calculate the pdf of these functionals for each realization of the quenched
random potential. The most difficult part of the problem is to carry out the
disorder average on these pdfs. Thus to demonstrate the formalism
explicitly, we have chosen the external potential to be the combination of a
deterministic part and a random part, $U(x) = -\mu |x| + \sqrt{\sigma}
B(x)$, where $B(x)$ is the trajectory of a Brownian motion in space. The
case $\mu=0$ in the potential corresponds to the Sinai model. The
deterministic part of the external potential is stable for $\mu<0$ and
unstable for $\mu>0$.  The pdfs of the functional and the inverse functional
vary from one realization of $B(x)$ to another, and in this paper we have
shown how to carry out the disorder average on them, for two particular
functionals, namely, the local time and the occupation time, and their
inverse. Despite the simplicity of the model, we get very rich and
interesting behaviors by tuning the parameter $\mu/\sigma$, which we have
summarized in Tables~\ref{table-flat}, \ref{table-unstable} and
\ref{table-stable}, for $\mu=0$, $\mu>0$ and $\mu<0$ respectively.  In many
cases the disorder changes the behavior of the pdf drastically from the pure
case ($\sigma=0$).

A very interesting phase transition in the ergodicity of the particle
position occurs at a critical value of the parameter $|\mu|/\sigma=1$, when
the deterministic part of the potential is stable ($\mu<0$). For
$|\mu|/\sigma<1$, when the particle gets trapped in the wells of the random
potential, the deterministic force $-|\mu|\sign(x)$ is not strong enough to
lift it from the well and push it towards the origin and hence there are
small number of zero crossings. On the other hand, for $|\mu|/\sigma>1$, the
strong deterministic force sends the particle frequently towards the origin,
and hence the system becomes ergodic. This change in the ergodic properties
shows up in the qualitative change in the curvatures of the disorder
averaged pdfs when the parameter $\nu=|\mu|/\sigma$ passes through
unity. While for $\nu<1$, the disorder averaged pdf of the occupation time
$\overline{\lss{P}{occ} (T|t)}$ is concave upward with a minimum at $T=t/2$
and diverges at both ends $T=0$ and $T=t$; for $\nu>1$ it is concave
downward, which goes to zero at the two ends $T=0$ and $T=t$, and has a
maximum at $T=t/2$ (see Fig.~\ref{fo plot}).  In the context of inverse
occupation time, while for $\nu<1$, the disorder averaged pdf
$\overline{\lss{I}{occ}(t|T)}$ diverges near its lower end $t=T$ and
decreases monotonically as $t$ increases, for $\nu>1$ it has a maximum at
$t=[2\nu/(\nu+1)]T$ and goes to zero at both ends $t=T$ and
$t\rightarrow\infty$ (see Fig.~\ref{go plot}).  Similarly, the disorder
averaged pdf of the local time $\overline{\lss{P}{loc}(T|t)}$ diverges near
the lower end $T=0$ and decreases monotonically as $T$ increases for
$\nu<1$. On the other hand for $\nu>1$, it has a maximum and goes to zero at
both ends $T=0$ and $T\rightarrow\infty$ (see Fig.~\ref{f2 plot}).

For the stable potential, another very interesting observation is that at
$|\mu|/\sigma=1/2$, in the limit $T\rightarrow\infty$ and
$t\rightarrow\infty$ while keeping the ratio $T/t$ fixed, the exact
asymptotic disorder averaged pdfs of the occupation time
$\overline{\lss{P}{occ} (T|t)}$ and inverse occupation time
$\overline{\lss{I}{occ}(t|T)}$ become exactly identical to the respective
pdfs $\lss{P}{occ} (T|t)$ and $\lss{I}{occ}(t|T)$ for the simple Brownian
motion ($\mu=0$ and $\sigma=0$). It looks as if at the particular value
$|\mu|/\sigma=1/2$, the effect of disorder is exactly canceled by the
deterministic stable potential. However, similar conclusion is not true in
the context of the local time and inverse local time. Therefore, a physical
understanding of what exactly happens at this particular value of the
parameter will be extremely useful.

There are several directions open for pursuing research farther in this
area. In this paper we have considered only the average of the pdfs over
disorder. However, in many cases, as we have seen in this paper, the
disorder broadens the distributions considerably. For example, for the
unstable potential ($\mu>0$), even though for each realization of random
potential the local time has a narrow exponential distribution, by taking
the disorder average one gets a broad power law distribution, which is the
indication of large sample to sample fluctuations and lack of
self-averaging. Therefore, in this situations the knowledge about the
disorder averaged pdf (first moment) is not enough, and one requires to
compute the other higher moments (over disorder). Thus extending our
formalism to compute the full distribution (over disorder) of pdf will be
very useful.

The random part of the potential we have considered in this paper is very
particular, where the barrier heights grow as $\sqrt{x}$. However, in
realistic systems the random potential remains of order one throughout the
sample. Therefore, it will be very interesting to extend this formalism for
more realistic random potentials.

Recently several asymptotically exact long time results for other quantities
in Sinai model were obtained by using a real space renormalization group
method~\cite{fisher:1998}. Using that method, reproducing the exact results
obtained in this paper remains as challenging open problem.  Another
interesting direction is to study the properties of functionals of a more
general non-Markovian stochastic process in random media, and to extend our
results to higher dimensions.

%%%%%%%%%%%%%%%%%%%%%%%
\section{Appendixes}
\appendix
%%%%%%%%%%%%%%%%%%%%%%

%%%%%% TABLE %%%%%%
\input{table.tbl}

%%%%%%%%%%%%%%%%%%%

%%%%%%%%%%%%%%%%%%%%%%%%%%%%%%%%%%%%%%%%%%%%%%%%%%%%%%%%%%%%%%%%%%%%%%%%
\section{pdf of the local time in the case of  the stable 
potential, $\mu<0$ and $\sigma=0$ in Eq.~(\ref{F(x)})}
\label{appendix local time stable}
%%%%%%%%%%%%%%%%%%%%%%%%%%%%%%%%%%%%%%%%%%%%%%%%%%%%%%%%%%%%%%%%%%%%%%%%

In this appendix we will derive the pdf of the local time
$\lss{P}{loc}(T|t)$, for the stable potential ($\mu<0$) in the absence of
disorder ($\sigma=0$).  In this case by solving Eq.~(\ref{ordinary-y}) with
the boundary conditions $y_+(x\rightarrow\infty)\rightarrow 0$ and
$y_-(x\rightarrow -\infty)\rightarrow 0$ we get
\begin{equation} 
y_\pm(x)=y_\pm(0) \exp \left[ \mp \left( -\nu+ \sqrt{\nu^2+2\alpha}
\right)x \right],  
\end{equation}
where $\nu=|\mu|$.  Substituting these results in Eq.~(\ref{lambda-alpha})
we get $\lambda(\alpha)=-\nu+\sqrt{\nu^2+2\alpha}$. Therefore the Laplace
transform $G(\alpha)$ in Eq.~(\ref{laplace-a}) becomes
\begin{equation}
G(\alpha)={-\nu+\sqrt{\nu^2+2\alpha} \over \alpha}
\exp\left[-\left(-\nu+\sqrt{\nu^2+2\alpha}\right)T \right].
\end{equation}
Now making a shift $\alpha=\beta-\nu^2/2$, in Eq.~(\ref{laplace-a}) yields
\begin{equation}
\int_0^\infty dt\,e^{-\beta t} [e^{\nu^2 t/2} \lss{P}{loc}(T|t)] =
\sqrt{2}e^{\nu T}{e^{-\sqrt{2\beta}T}\over\sqrt{\beta} + \nu /\sqrt{2}},
\end{equation}
where the right hand side is the Laplace transform of $e^{\nu^2 t/2}
\lss{P}{loc}(T|t)$.  Inverting the Laplace transform with respect to $\beta$
and after simplification gives the exact distribution of the local time for
all $T$ and $t$,
\begin{equation}
\lss{P}{loc}(T|t)={\sqrt{2}\over\sqrt{\pi t}} e^{-(T-\nu t)^2/2t} -\nu
e^{2\nu T}\erfc\left({\nu \over 2}\sqrt{t}+{T\over\sqrt{2t}} \right),
\label{local-2}
\end{equation}
where $\erfc(x)$ is the complementary error function.  Note that
Eq.~(\ref{local-2}) reduces to Eq.~(\ref{local-1}) for $\nu=0$.

For large $t$, since
\begin{multline}
\erfc\left( {\nu \over\sqrt{2}}\sqrt{t}+{T\over\sqrt{2t}}\right) \sim
{1\over\sqrt{\pi}} \left[ {\nu
\over\sqrt{2}}\sqrt{t}+{T\over\sqrt{2t}}\right]^{-1} \\ \times
\exp\left(-\left[{\nu
\over\sqrt{2}}\sqrt{t}+{T\over\sqrt{2t}}\right]^2\right),
\end{multline}
Eq.~(\ref{local-2}) simplifies to
\begin{equation} 
\lss{P}{loc}(T|t)=\left[{T\over T+\nu t}\right] {\sqrt{2}\over\sqrt{\pi t}}
e^{-(T-\nu t)^2/2t}.
\end{equation}
Putting $\nu=0$ in the above equation one still recovers the result given by
Eq.~(\ref{local-1}). For non-zero $\nu$, near the mean $\langle T \rangle =
\nu t$, the pdf of the local time reduces to a Gaussian one
\begin{equation} 
\lss{P}{loc}(T|t)\approx{1\over\sqrt{2\pi t}} e^{-(T-\nu t)^2/2t}.
\end{equation}

%%%%%%%%%%%%%%%%%%%%%%%%%%%%%%%%%%%%%%%%%%%%%%%%%%%%%%%%%%%%%%%%%%%%%%%
\section{pdf of the slope variables $z_\pm(0)$, that appear in the 
disorder average computations}
\label{z+- distributions}
%%%%%%%%%%%%%%%%%%%%%%%%%%%%%%%%%%%%%%%%%%%%%%%%%%%%%%%%%%%%%%%%%%%%%%%%

Both in the contexts of local and occupation time we have a homogeneous
differential equation of the type
\begin{equation} 
{1\over 2} y_\pm^{\prime\prime}(x) + F(x) y_\pm^\prime(x) -a_\pm
y_\pm(x)=0,
\label{homogeneous-y}
\end{equation}
with the boundary conditions $y_+(x\rightarrow\infty) \rightarrow 0$ and
$y_-(x\rightarrow-\infty) \rightarrow 0$, and the force
\begin{equation}
F(x)=\mu\sign(x)+ \sqrt{\sigma}\xi(x),
\end{equation}
with $\langle \xi(x)\rangle =0$ and $\langle \xi(x) \xi(x') \rangle =
\delta(x-x')$. For each realization of $\{\xi(x)\}$ in the force $F(x)$, the
solution of Eq.~(\ref{homogeneous-y}) is different, and for the disorder
averaged computations performed in this paper we finally require the
distributions of the stochastic variables $y'_\pm(0)/y_\pm(0)$ and in this
appendix our goal is to find these distributions.

By defining the variables
\begin{equation}
z_\pm(x) = {y'_\pm(x) \over y_\pm(x)}.
\end{equation}
we find from Eq.~(\ref{homogeneous-y}) that $z_\pm(x)$, satisfy the
stochastic Riccati equation
\begin{equation} 
z_\pm^\prime(x)=-z_\pm^2(x)-2 F(x) z_\pm +2 a_\pm.
\label{differential-z}
\end{equation}

However, now the boundary conditions for $z_\pm$ in Eq.~(\ref{ordinary-y})
are not specified. Therefore for each realizations of $\{\xi(x)\}$, the
solutions of $z_\pm(x)$ involve one unknown each that can not be eliminated
due to the lack of the boundary conditions. In other words, to find the
distributions of $z_\pm(x)$, we need the respective distributions at some
initial points, unfortunately which are not specified.

It turns out, however, that this difficulty can be bypassed by a
method~\cite{majumdar:061105,comtet:jpa:1998, 1990PhyA..164...52A} which
lets us to compute the distributions of $z_+(0)$ and $z_-(0)$ without having
the knowledge of the boundary conditions on $z_+(\infty)$ and
$z_-(-\infty)$. We will present the method below for the present context.

First we consider Eq.~(\ref{differential-z}) for $x>0$, \ie
\begin{equation}
z_+^\prime(x)=-z_+^2(x) -2[\mu +\sqrt{\sigma}\xi(x)]z_+(x) +2 a_+.
\label{differential-z+}
\end{equation}
Note that $z_+(x)=y_+^\prime(x)/y_+(x)$ is negative. We make a change of
variable $x=-\tau$ and substitute $z_+(-\tau)=-\exp[\phi(\tau)]$ in
Eq.~(\ref{differential-z+}) to find that the new variable $\phi(\tau)$
satisfies a much simpler stochastic differential equation
\begin{equation}
{d\phi\over d\tau}=b(\phi)+2\sqrt{\sigma}\tilde{\xi}(\tau),
\label{Langevin-phi}
\end{equation}
where $\tilde{\xi}(\tau)=\xi(-\tau)$ and thus $\langle\tilde{\xi}(\tau)
\rangle=0$ and $\langle \tilde{\xi}(\tau)\tilde{\xi}(\tau^\prime)
\rangle=\delta(\tau-\tau^\prime)$. The source term $b(\phi)$ is given by
\begin{equation}
b(\phi)=-e^{\phi}+2 a_+ e^{-\phi} +2\mu.
\end{equation}
Now we can interpret Eq.~(\ref{Langevin-phi}) as a simple Langevin equation
describing the evolution of a Brownian particle starting at time
$\tau=-\infty$, in a classical stable potential $U_{\mbox{\footnotesize
cl}}(\phi) = -\int_0^\phi b(\varphi)\,d\varphi= e^\phi+2 a_+
e^{-\phi}-2\mu\phi-(2 a_+ +1)$. Even though we do not know the starting
position of the particle $\phi(-\infty)$, it is completely irrelevant. No
matter what the initial position is, eventually after a long time, \ie when
$\tau$ is far away from $-\infty$, the system will reach equilibrium and
hence the stationary probability distribution of $\phi$ is simply given by
the Gibbs measure
\begin{equation} 
P_{\mbox{\footnotesize st}} (\phi)=A\,\exp\left[-{1\over
    2\sigma}U_{\mbox{\footnotesize cl}} (\phi) \right] =A\,\exp\left[{1\over
    2\sigma}\int_0^\phi b(\varphi)\,d\varphi \right],
\label{equilibrium phi distribution}
\end{equation}
where $A$ is a normalization constant such that $\int_{-\infty}^\infty
P_{\mbox{\footnotesize st}} (\phi) \,d\phi=1$. Now changing back to the
original variable $z_+(x)$ we obtain the distribution of $z_+(0)$ as
\begin{equation} 
P^+\left(-z_+(0)=w\right) = {1\over \Omega_+} w^{\mu/\sigma -1}
\exp\left[-{1\over 2\sigma}\left\{w + {2 a_+ \over w} \right\} \right],
\label{P+ equilibrium}
\end{equation}
where
\begin{equation}
\begin{split}
\Omega_+ &= \int_0^\infty w^{\mu/\sigma -1} \exp\left[-{1\over
2\sigma}\left\{w + {2 a_+ \over w} \right\} \right] \,dw \\ &= 2(2
a_+)^{\mu/2\sigma} K_{\mu/\sigma} \left({\sqrt{2 a_+}\over\sigma} \right).
\end{split}
\label{omega+}
\end{equation}

Similarly for $x<0$, by putting $F(x)=-\mu +\sqrt{\sigma}\xi(x)$ in
Eq.~(\ref{differential-z}) and substituting $z_-(x)=\exp[\phi(x)]$ one finds
that $\phi(x)$ satisfies the same differential equation in $x$ as
Eq.~(\ref{Langevin-phi}) with $\tilde{\xi}(x)=-\xi(x)$ and $a_+$ is replaced
with $a_-$. Therefore $\phi(x)$ has the same stationary distribution as
Eq.~(\ref{equilibrium phi distribution}) and consequently the distribution
of $z_-(0)$ is same as that of $-z_+(0)$, namely
\begin{equation} 
P^-\left(z_-(0)=w\right) = {1\over \Omega_-} w^{\mu/\sigma -1}
\exp\left[-{1\over 2\sigma}\left\{w + {2 a_- \over w} \right\} \right],
\label{P- equilibrium}
\end{equation}
with
\begin{equation}
\begin{split}
\Omega_- &= \int_0^\infty w^{\mu/\sigma -1} \exp\left[-{1\over
2\sigma}\left\{w + {2 a_- \over w} \right\} \right] \,dw \\ &= 2(2
a_-)^{\mu/2\sigma} K_{\mu/\sigma} \left({\sqrt{2 a_-}\over\sigma} \right).
\end{split}
\label{omega-}
\end{equation}

Note that the distributions of $-z_+(0)$ and $z_-(0)$ given by Eqs.~(\ref{P+
equilibrium}) and ~(\ref{P- equilibrium}) have maxima at $(\mu-\sigma) +
\sqrt{(\mu-\sigma)^2+ 2 a_\pm}$ respectively and in the limit
$\sigma\rightarrow 0$ the distributions tend to delta functions around their
maxima. Therefore in the limit $\sigma\rightarrow 0$ one recovers the pure
case results by using the distributions $P^+\left(z_+(0)\right)
=\delta\left(z_+(0)+[\mu + \sqrt{\mu^2+ 2 a_+}] \right)$ and
$P^-\left(z_-(0)\right)=\delta\left(z_-(0)-[\mu + \sqrt{\mu^2+ 2
a_-}]\right)$.

%%%%%%%%%%%%%%%%%%%%%%%%%%%%%%%%%%%%%%%%%%%%%%%%%%%%%%%%%%%%%%%%%
\section{Left half of the disorder averaged pdf of the occupation 
time for Sinai potential ($\mu=0$ and $\sigma>0$)}
\label{appendix occupation Sinai}
%%%%%%%%%%%%%%%%%%%%%%%%%%%%%%%%%%%%%%%%%%%%%%%%%%%%%%%%%%%%%%%%%%%%%%%%

By taking the disorder average of Eq.~(\ref{RL}) one gets
\begin{equation} 
\int_0^\infty dt\,e^{-\alpha t}\int_0^t dT\, e^{-pT}\,\overline{R_L(T|t)}
={1\over \alpha}\, \overline{\ell_1(\alpha,p)}.
\label{LL occupation disorder mu=0}
\end{equation}
Using the distributions of $-z_+(0)$ and $z_-(0)$ from Eqs.~(\ref{P+
equilibrium}) and~(\ref{P- equilibrium}) respectively with $a_+=\alpha + p$
and $a_-=\alpha$, from Eq.~(\ref{ell_1}) one gets
\begin{equation}
\overline{\ell_1(\alpha,p)} = { m_1(\alpha,p)\over \Omega_+\Omega_-},
\label{rhs (0) occupation disorder mu=0}
\end{equation}
where,
\begin{multline}
m_1(\alpha,p)= \int_0^\infty
  {dw_1\over w_1} \exp\left[-{1\over 2\sigma}\left(w_1+{2(\alpha+p)\over
  w_1}\right) \right]\\\times \int_0^\infty {dw_2\over
  w_1+w_2} \exp\left[-{1\over 2\sigma}\left(w_2+{2\alpha\over
  w_2}\right)\right],
\label{rhs (1) occupation disorder mu=0}
\end{multline}
and
\begin{equation}
\Omega_+ = 2K_0\left({\sqrt{2(\alpha+p)}\over\sigma} \right), ~~~ \Omega_- =
2K_0\left({\sqrt{2\alpha}\over\sigma} \right).
\end{equation}

Before we proceed further, let us take a detour to check the normalization
condition of $\overline{R_L(T|t)}$.  By putting $p=0$ in the above equations
we get
\begin{equation}
\Omega_+ = \Omega_- = \Omega = 2K_0\left({\sqrt{2\alpha}\over\sigma} \right)
\end{equation}
and
\begin{align} 
m_1(\alpha,0)= \int_0^\infty\int_0^\infty &{dw_1\over w_1} {dw_2\over w_2}
\left[{w_2\over w_1+w_2}\right] \nonumber\\ \times&\exp\left\{{-{1\over
2\sigma}\left[ w_1 +{2\alpha\over w_1}\right]}\right\} \nonumber\\ \times&
\exp\left\{{-{1\over 2\sigma}\left[ w_2 +{2\alpha\over w_2}\right]}\right\}.
\end{align}
Note that the above integral must remain invariant under the transformation
$w_1 \leftrightarrow w_2$ of the dummy variables. Therefore we get
\begin{equation}
2 m_1(\alpha,0)= \left[\int_0^\infty {dw\over w} \exp\left\{{-{1\over
 2\sigma}\left[ w +{2\alpha\over w}\right]}\right\}\right]^2 = \Omega^2.
\end{equation}
Therefore we have $\overline{\ell_1(\alpha,0)} = 1/2$, and inverting the
Laplace transform in Eq.~(\ref{LL occupation disorder mu=0}) with respect to
$\alpha$ for $p=0$ gives the normalization condition
$\int_0^t\overline{R_L(T|t)}\, dT = 1/2$.

Now we analyze the large $t$ behavior of $\overline{R_0(T|t)}$.  By making a
change of variable $z=\alpha t$, it follows from Eq.~(\ref{LL occupation
disorder mu=0})
\begin{equation} 
\int_0^\infty dz\, e^{-z} \int_0^{z/\alpha} dT\, e^{-pT}
\overline{R_0(T,z/\alpha)} =\overline{\ell_1(\alpha,p)}.
\label{LL occupation disorder mu=0 alpha}
\end{equation}
In the $\alpha\rightarrow 0$ limit, $ \Omega_+ =
2K_0\left({\sqrt{2p}/\sigma} \right) $ and $\Omega_- \sim
-\log\alpha$. Therefore from Eq.~(\ref{rhs (0) occupation disorder mu=0})
\begin{equation} 
\overline{\ell_1(\alpha\rightarrow 0,p)} ={ m_1(0,p)\over
\left[-2K_0\left({\sqrt{2p}/\sigma} \right)\,\log\alpha\right]},
\label{rhs (0) occupation disorder mu=0 alpha=0}
\end{equation}
which suggest the following form for $\overline{R_L(T|t)}$ at large $t$
\begin{equation} 
\overline{R_L(T|t)} = {1\over \log t} R(T),
\label{RL scaling disorder mu=0}
\end{equation}
where $R(T)$ is independent of $t$.

Now using Eqs.~(\ref{rhs (0) occupation disorder mu=0 alpha=0}) and~(\ref{RL
scaling disorder mu=0}), in the limit $\alpha\rightarrow 0$, Eq.~(\ref{LL
occupation disorder mu=0 alpha}) gives
\begin{equation} 
\int_0^\infty dT\,e^{-pT} R(T) = {m_1(0,p)\over 2
K_0\left({\sqrt{2p}/\sigma} \right)},
\label{(1) scaling limit disorder mu=0}
\end{equation}
where $m_1(0,p)$ is obtained from Eq.~(\ref{rhs (1) occupation disorder
mu=0}),
\begin{multline}
 m_1(0,p) = \int_0^\infty {dw_1\over w_1} \exp\left[-{1\over
    2\sigma}\left(w_1+{2p\over w_1}\right) \right]\\ \times \int_0^\infty
    dw_2\, {e^{-w_2/2\sigma}\over w_1+w_2}.
\label{rhs (2) occupation disorder mu=0}
\end{multline}
By making change of variables $w_1=2\sigma x$ and $w_2=2\sigma y$ in the
integrals in the above equation one gets
\begin{multline}
 m_1(0,p)= \int_0^\infty {dx\over x} \exp\left[-\left(x+{p\over 2\sigma^2
x}\right) \right]\\ \times \int_0^\infty dy\, {e^{-y}\over y+x},
\label{rhs (3) occupation disorder mu=0}
\end{multline}
where now the integral over $y$ can be expressed in terms of the incomplete
Gamma function~\cite{gradshteyn}
\begin{equation}
\int_0^\infty dy\, {e^{-y}\over y+x} = e^x\,\Gamma(0,x).
\end{equation}
Therefore, after straightforward simplification, Eq.~(\ref{rhs (3)
occupation disorder mu=0}) becomes
\begin{equation} 
m_1(0,p) = \int_0^\infty {dx\over x}\, e^{-x}\, \Gamma(0,p/2\sigma^2 x).
\label{rhs (4) occupation disorder mu=0}
\end{equation}

Now we will analyze the limiting behavior of $R(T)$ for small and large $T$
by taking the limit of large and small $p$ respectively.

Since for large $p$
\begin{equation}
\Gamma(0,p/2\sigma^2 x)\approx {2\sigma^2 x \over p}\,e^{-p/2\sigma^2 x},
\end{equation}
from Eq.~(\ref{rhs (4) occupation disorder mu=0}), one gets
\begin{align} 
m_1(0,p) &\approx {2\sigma^2\over p}\int_0^\infty
dx\,\exp\left[-\left(x+{p\over 2\sigma^2 x}\right) \right] \nonumber\\ &=
{2\sqrt{2}\sigma\over\sqrt{p}} K_1\left({\sqrt{2 p}\over\sigma} \right).
\end{align}
Since the asymptotic behavior of $K_\nu(x)$ is independent of $\nu$,
substituting $m_1(0,p)$ from above in Eq.~(\ref{(1) scaling limit disorder
mu=0}) gives
\begin{equation} 
\int_0^\infty dT\,e^{-pT} R(T) \approx {\sqrt{2}\sigma\over\sqrt{p}},
\end{equation}
for small $p$, and by inverting the Laplace transform with respect to $p$
one obtains
\begin{equation}
R(T)\approx {\sqrt{2}\sigma\over\sqrt{\pi T}}, ~~~~\mbox{as}~T\rightarrow 0.
\label{scaling function occupation disorder mu=0 small T}
\end{equation}

To obtain the large $T$ behavior, we first consider the following integral
\begin{equation} 
\mathcal{D}(z)=\int_0^\infty {dx\over x}\, e^{-xz}\,
\Gamma(0,p/2\sigma^2 x), 
\end{equation}
where $\mathcal{D}(1)= m_1(0,p)$, follows from Eq.~(\ref{rhs (4) occupation
disorder mu=0}). Now by differentiating $\mathcal{D}(z)$ with respect to
$z$, one can express it in terms of the modified Bessel function
as~\cite{gradshteyn}
\begin{equation} 
\mathcal{D}'(z)=-\int_0^\infty dx\, e^{-x z}\,\Gamma(0,p/2\sigma^2 x)
=-{2\over z}K_0\left({\sqrt{2 p z}\over\sigma} \right).
\end{equation}
Now by integrating back again with respect to $z$, we obtain $m_1(0,p)$ as
\begin{align} 
m_1(0,p)=\mathcal{D}(1)&=2\int_1^\infty {dz\over z} K_0\left({\sqrt{2 p
z}\over\sigma} \right)\nonumber\\ &= 4\int_{\sqrt{2p}/\sigma}^\infty
{dx\over x} K_0(x),
\end{align}
where we have made the change of variable $2pz/\sigma^2=x^2$. The
$p\rightarrow 0$ limit can be obtained from the limiting behavior of the
integral~\cite{abramowitz}
\begin{equation} 
\int_y^\infty {K_0(x)\,dx\over x} \sim {1\over 2}\left(\log y \right)^2
~~~~\mbox{as}~y\rightarrow 0,
\end{equation}
which gives
\begin{equation} 
m_1(0,p)\sim {1\over 2} \left(\log p \right)^2, ~~~~\mbox{as}~p\rightarrow
0.
\end{equation}
Since $ K_0\left({\sqrt{2p}/\sigma} \right) \sim -{1\over 2}\log p$, as
$p\rightarrow 0$, Eq.~(\ref{(1) scaling limit disorder mu=0}) gives
\begin{equation} 
\int_0^\infty dT\,e^{-pT} R(T) \sim -{1\over 2} \log p,
\end{equation}
as $p\rightarrow 0$.  Thus, inverting the Laplace transform with respect to
$p$ one obtains
\begin{equation} 
R(T) \sim {1\over 2T}, ~~~~\mbox{as}~T\rightarrow\infty.
\label{scaling function occupation disorder mu=0 large T}
\end{equation}

%%%%%%%%%%%%%%%%%%%%%%%%%%%%%%%%%%%%%%%%%%%%%%%%%%%%%%%%%%%%%%%%%%%%%%%%
\section{Left half of the disorder averaged pdf of the occupation 
time for $\mu>0$}
\label{appendix occupation time disorder mu>0}
%%%%%%%%%%%%%%%%%%%%%%%%%%%%%%%%%%%%%%%%%%%%%%%%%%%%%%%%%%%%%%%%%%%%%%%%%

By taking the disorder average of Eq.~(\ref{RL}) one gets
\begin{equation} 
\int_0^\infty dt\,e^{-\alpha t}\int_0^t dT\, e^{-pT}\,\overline{R_L(T|t)} =
{1\over \alpha}\, \overline{\ell_1(\alpha,p)}.
\label{LL occupation disorder mu>0}
\end{equation}
As in the pure case ($\sigma=0$), one also expects the $t\rightarrow\infty$
behavior of the disorder averaged distribution to tend to a $t$ independent
form
\begin{equation}
\lim_{t\rightarrow\infty} \overline{R_L(T|t)}=\overline{R_L(T)}.
\end{equation}
Therefore Eq.~(\ref{LL occupation disorder mu>0}) becomes
\begin{equation}
\int_0^\infty dT\, e^{-pT}\, \overline{R_L(T)} =\overline{\ell_1(0,p)}.
\label{L occupation disorder mu>0} 
\end{equation}
Using the distributions of $-z_+(0)$ and $z_-(0)$ from Eqs.~(\ref{P+
equilibrium}) and~(\ref{P- equilibrium}) respectively with $a_+=p$ and
$a_-=0$, from Eq.~(\ref{ell_1}) one gets
\begin{align} 
\overline{\ell_1(0,p)} ={1\over \Omega_+\Omega_-} &\int_0^\infty
dw_1\,w_1^{\mu/\sigma-1} \exp\left[-{1\over 2\sigma}\left(w_1+{2p\over
w_1}\right) \right]\nonumber\\ &\times \int_0^\infty dw_2\,
{w_2^{\mu/\sigma}e^{-w_2/2\sigma}\over w_1+w_2},
\label{rhs (2) occupation disorder mu>0}
\end{align}
with
\begin{equation} 
\begin{split}
&\Omega_+ = 2(2p)^{\mu/2\sigma}K_{\mu/\sigma}\left({\sqrt{2p}\over\sigma}
\right),\\ &\Omega_- = (2\sigma)^{\mu/\sigma}\Gamma(\mu/\sigma).
\end{split}
\label{omega+- occupation disorder mu>0}
\end{equation}
Now the integral over $w_2$ in Eq.~(\ref{rhs (2) occupation disorder
  mu>0}) can be expressed as~\cite{gradshteyn}
\begin{multline} 
\int_0^\infty dw_2\, {w_2^{\mu/\sigma}e^{-w_2/2\sigma}\over w_1+w_2}\\=
  w_1^{\mu/\sigma}e^{w_1/2\sigma} \Gamma\left({\mu\over\sigma}+1\right)
  \Gamma\left(-{\mu\over\sigma},{w_1\over2\sigma}\right),
\end{multline}
where $\Gamma(\alpha,x)$ is the incomplete Gamma function.  Therefore
Eq.~(\ref{rhs (2) occupation disorder mu>0}) becomes
\begin{equation} 
\overline{\ell_1(0,p)} = {\mu m_2(p)\over
  (2\sigma)^{\mu/\sigma+1}(2p)^{\mu/2\sigma}
  K_{\mu/\sigma}(\sqrt{2p}/\sigma)}
\label{rhs (3) occupation disorder mu>0}
\end{equation}
where
\begin{equation} 
m_2(p)=\int_0^\infty dw_1\, w_1^{2\mu/\sigma-1}e^{-p/\sigma
w_1}\Gamma\left(-{\mu\over\sigma},{w_1\over2\sigma}\right).
\label{m_2(p)}
\end{equation}
The small and large $T$ behavior of $\overline{R_L(T)}$ can be found by
analyzing Eq.~(\ref{rhs (3) occupation disorder mu>0}) in the limiting cases
$p\rightarrow\infty$ and $p\rightarrow 0$ respectively.

Making a change of variable $w_1=p/\sigma x$ in Eq.~(\ref{m_2(p)}), and then
taking the $p\rightarrow\infty$ in the incomplete Gamma function,
\begin{equation} 
\Gamma\left(-{\mu\over\sigma},{p\over2\sigma^2 x}\right)\sim
\left({p\over2\sigma^2 x}\right)^{-\mu/\sigma-1} \exp\left(-{p\over2\sigma^2
x}\right),
\end{equation}
gives
\begin{equation} 
m_2(p)\approx (2\sigma)^2(2p)^{\mu/\sigma-1} \int_0^\infty dx\,
x^{-\mu/\sigma}\exp\left[-\left(x+{p\over 2\sigma^2 x} \right) \right],
\end{equation}
where the integral above on the right hand side can further be expressed in
terms of the modified Bessel function as~\cite{gradshteyn}
\begin{multline} 
\int_0^\infty dx\, x^{-\mu/\sigma} \exp\left[-\left(x+{p\over 2\sigma^2 x}
      \right) \right]\\ = 2 (2\sigma)^{\mu/\sigma -1} (2p)^{-\mu/2\sigma
      +1/2} K_{\mu/\sigma-1}\left({\sqrt{2p}\over\sigma} \right).
\end{multline}
Since the large $x$ behavior or $K_\nu(x)$ is independent of $\nu$,
Eq.~(\ref{rhs (3) occupation disorder mu>0}) simplifies to
\begin{equation}
\overline{\ell_1(0,p)} \approx {\mu\sqrt{2}\over\sqrt{p}}
~~~\mbox{as}~~p\rightarrow\infty.
\end{equation}
Therefore by inverting the Laplace transform in Eq.~(\ref{L occupation
    disorder mu>0}) with respect to $p$, one gets
\begin{equation} 
\overline{R_L(T)} \approx {\mu\sqrt{2}\over\sqrt{\pi T}},
~~~\mbox{for small}~T. 
\end{equation}

Now we will analyze the the large $T$ behavior, by taking the limit
$p\rightarrow 0$ in Eq.~(\ref{rhs (3) occupation disorder mu>0}). It is
straightforward to obtain from Eq.~(\ref{m_2(p)}) that
\begin{equation} 
m_2(0) ={(2\sigma)^{2\mu/\sigma+1}\Gamma(\mu/\sigma)\over 4\mu},
\end{equation}
which gives
\begin{equation} 
\overline{\ell_1(0,p)}\approx {1\over4}\Gamma\left({\mu\over\sigma}\right)
\left({\sigma\sqrt{2}\over\sqrt{p}}\right)^{\mu/\sigma}
K_{\mu/\sigma}^{-1}\left({\sqrt{2p}\over\sigma}\right),
\label{rhs (4) occupation disorder mu>0}
\end{equation}
for small $p$.  However if one takes the limit $p\rightarrow 0$ now in
$K_{\mu/\sigma}(\sqrt{2p}/\sigma)$ in the above expression, it only gives
the normalization condition $\int_0^\infty \overline{R_L(T)}\, dT =1/2$ and
does not provide any information about the large $T$ behavior of
$\overline{R_L(T)}$.

We make the following ansatz
\begin{equation} 
\overline{R_L(T)} \sim e^{-b T},
\label{R_L large T}
\end{equation}
for large $T$. Then the Laplace transform
\begin{equation} 
\int_0^\infty dT\, e^{-pT}\, \overline{R_L(T)} \approx {1\over p+b}.
\label{rhs (5) occupation disorder mu>0}
\end{equation}
for small $p$. Therefore substituting Eqs.~(\ref{rhs (4) occupation disorder
mu>0}) and~(\ref{rhs (5) occupation disorder mu>0}) in Eq.~(\ref{L
occupation disorder mu>0}), one can conclude that $b$ is given by the zero
of $K_{\mu/\sigma}(\sqrt{2p}/\sigma)$ closest to origin in the left part of
the complex--$p$ plane.

%%%%%%%%%%%%%%%%%%%%%%%%%%%%%%%%%%%%%%%%%%%%%%%%%%%%%%%%%%%%%%%%%%
\section{Disorder averaged pdf of the occupation time for $\mu<0$}
\label{appendix occupation time disorder mu<0}
%%%%%%%%%%%%%%%%%%%%%%%%%%%%%%%%%%%%%%%%%%%%%%%%%%%%%%%%%%%%%%%%%%

Taking the disorder average of Eq.~(\ref{u(0)-occupation}) gives
\begin{multline} 
\int_0^\infty dt\,e^{-\alpha t}\int_0^t dT\,
e^{-pT}\,\overline{\lss{P}{occ}(T|t)}\\ = {1\over\alpha +p} +{p \over \alpha
(\alpha+p)} \,\overline{\ell_1(\alpha,p)},
\label{LL occupation disorder mu<0}
\end{multline}
where we have substituted $\overline{\ell_2(\alpha,p)} = 1 -
\overline{\ell_1(\alpha,p)}$.

We are interested in finding the behavior of $\overline{\lss{P}{occ}(T|t)}$,
in the scaling limit $t\rightarrow\infty, T\rightarrow\infty$, but keeping
$T/t=y$ fixed, which corresponds to the limit of conjugate variables:
$\alpha\rightarrow 0, p\rightarrow 0$, keeping $p/\alpha=s$ fixed.
Substituting $z=\alpha t$, $T=y z/\alpha$ and $p=\alpha s$ in Eq,~(\ref{LL
occupation disorder mu<0}), we get
\begin{equation} 
\int_0^1 dy\, \int_0^\infty dz\, e^{-(1+sy)z} \left[{z\over\alpha}\,
\overline{\lss{P}{occ}\left({yz\over\alpha}\left|
{z\over\alpha}\right. \right)} \right] = {1+s m_3(\alpha,s)\over (1+s)},
\label{yz occupation disorder mu<0}
\end{equation}
where $m_3(\alpha,s)=\overline{\ell_1(\alpha,\alpha s)}$. Equation~(\ref{yz
occupation disorder mu<0}) suggest the form,
\begin{equation}
\overline{\lss{P}{occ}(T|t)}= {1\over t}\, f_o(T/t),
\end{equation}
in the scaling limit $t\rightarrow\infty, T\rightarrow\infty$, while their
ratio $T/t$ is kept fixed. In the limit $\alpha\rightarrow 0$, by
substituting the above scaling form in Eq.~(\ref{yz occupation disorder
mu<0}), it is straightforward to obtain
\begin{equation} 
\int_0^1 dy\,{f_o(y)\over 1+sy} = {1+s m_3(0,s)\over (1+s)}, 
\end{equation}
where putting $s=0$, gives the normalization condition
$\int_0^1f_o(y)\,dy=1$.  Using this normalization, the above equation can be
simplified to the following elegant form
\begin{equation} 
\int_0^1{1-y\over 1+sy}\,f_o(y)\,dy = m_3(0,s).
\label{L scaling occupation disorder mu<0}
\end{equation}

Now by using the distributions of $z_\pm(0)$ from Eqs.~(\ref{P+
equilibrium}) and~(\ref{P- equilibrium}) with $a_-=\alpha$ and
$a_+=\alpha(1+s)$, from Eq.~(\ref{ell_1}) we get
\begin{align}
 m_3(\alpha,s)&=\overline{\ell_1(\alpha, \alpha s)}\nonumber\\ &= {1\over
\Omega_+\Omega_-}\int_0^\infty dw_1\int_0^\infty dw_2\, {w_2\over w_1+w_2}
\nonumber\\ &\qquad\times w_1^{-\nu-1}\exp\left[-{1\over 2\sigma}\left\{w_1
+ {2\alpha(1+s) \over w_1} \right\}\right] \nonumber\\ &\qquad\times
w_2^{-\nu-1}\exp\left[-{1\over 2\sigma}\left\{w_2 + {2\alpha \over w_2}
\right\} \right],
\label{rhs (1) occupation disorder mu<0}
\end{align}
where 
\begin{align}
&\Omega_+ = 2(2\alpha)^{-\nu/2}(1+s)^{-\nu/2} K_\nu
\left({\sqrt{2\alpha(1+s)}\over\sigma} \right),\\
&\Omega_- = 2(2\alpha)^{-\nu/2} K_\nu
\left({\sqrt{2\alpha}\over\sigma} \right),
\end{align}
with $\nu=|\mu|/\sigma$.  Note that we simply can not take the limit
$\alpha\rightarrow 0$ in the integrals in Eq.~(\ref{rhs (1) occupation
disorder mu<0}), as it diverges in that limit. However, it is possible to
extract the divergent contribution outside the integrals which finally
cancels exactly with the divergence of $\Omega_\pm$. This is done by making
the change of variables
\begin{equation}
{\alpha(1+s)\over\sigma w_1} = x,\quad\mbox{and}\quad
{\alpha\over\sigma w_2} = y,
\end{equation}
in the integral to get
\begin{multline} 
m_3(\alpha,s) = {\sigma^{2\nu}\alpha^{-2\nu}(1+s)^{-\nu}\over
\Omega_+\Omega_-} \int_0^\infty dx\int_0^\infty dy\, \\ \times
\frac{x^{\nu}y^{\nu-1}}{x+(1+s)y} \\\shoveright{\times \exp\left[-\left\{x +
{\alpha(1+s)\over 2\sigma^2 x} \right\}\right]} \\ \times
\exp\left[-\left\{y + {\alpha \over 2\sigma^2 y} \right\} \right].
\label{rhs (2) occupation disorder mu<0}
\end{multline}
Now the limit $\alpha\rightarrow 0$ can be taken in the above equation, as,
in this limit $\Omega_+
\rightarrow\sigma^\nu\alpha^{-\nu}(1+s)^{-\nu}\Gamma(\nu)$ and $\Omega_-
\rightarrow\sigma^\nu\alpha^{-\nu}\Gamma(\nu)$.  Therefore from
Eq.~(\ref{rhs (2) occupation disorder mu<0}) we get
\begin{equation} 
m_3(0,s) = {1\over\Gamma^2(\nu)}\int_0^\infty
dy\,y^{\nu-1}e^{-y}\int_0^\infty { x^\nu e^{-x} \over x+(1+s)y}\, dx.
\label{rhs (3) occupation disorder mu<0}
\end{equation}
Now the integration over $x$ can be expressed in terms of the incomplete
Gamma function~\cite{gradshteyn}
\begin{equation}
\Gamma(\rho,\lambda)={e^{-t}t^\rho\over\Gamma(1-\rho)} \int_0^\infty
{x^{-\rho}e^{-x}\over x+\lambda}\,dx, \quad [\mbox{Re}\,\rho <1,\,
\lambda>0],
\end{equation}
which gives
\begin{equation} 
m_3(0,s) = {\nu(1+s)^\nu\over\Gamma(\nu)}\int_0^\infty y^{2\nu-1}
e^{sy}\Gamma\left(-\nu,(1+s)y\right)\,dy.
\label{rhs (4) occupation disorder mu<0}
\end{equation}
The right hand side, however, is one of the integral representation of the
Gauss's hypergeometric function
$F(\alpha,\beta;\gamma;z)$~\cite{gradshteyn}, which gives
\begin{equation} 
m_3(0,s) = {1\over 2}\, F(1,\nu;2\nu+1;-s).
\label{rhs (5) occupation disorder mu<0}
\end{equation}
Now by using another integral representation~\cite{gradshteyn}
\begin{equation}
F(\alpha,\beta;\gamma;-s)={1\over B(\beta,\gamma-\beta)} \int_0^1
{y^{\beta-1} (1-y)^{\gamma-\beta-1}\over(1+sy)^\alpha}\,dy,
\end{equation}
we get
\begin{equation} 
m_3(0,s) = {1\over B(\nu,\nu)}\int_0^1 {1-y\over 1+sy}
  \left[y(1-y)\right]^{\nu-1}\,dy,
\label{rhs (6) occupation disorder mu<0}
\end{equation}
where $B(\alpha,\beta)=\Gamma(\alpha)\Gamma(\beta)/\Gamma(\alpha+\beta)$ is
the Beta function~\cite{gradshteyn}.

Now by comparing Eq.~(\ref{rhs (6) occupation disorder mu<0}) with
Eq.~(\ref{L scaling occupation disorder mu<0}), one immediately gets
\begin{equation} 
f_o(y)={1\over B(\nu,\nu)}\left[y(1-y)\right]^{\nu-1}, ~~~~ 0\le y\le 1.
\end{equation}

%%%%%%%%%%%%%%%%%%%%%%
%%%%% REFERENCES %%%%%
%\bibliography{../bibliography}

\input{main.bbl}
%%%%%%%%%%%%%%%%%%%%%%

\end{document}

%% file: table.tbl
%auto-ignore

% \documentclass[pre,twocolumn]{revtex4}
% \usepackage{amsmath}
% \newcommand{\lss}[2]{\ensuremath{#1_{\mbox{\scriptsize #2}}}}
% \newcommand{\erfc}{\text{erfc}}
% \begin{document}

\newcommand{\dm}[1]{
\begin{math}
\smallskip
\displaystyle #1
\smallskip
\end{math}
} 
\newcommand{\entry}[1]{
\begin{minipage}[c]{\hsize}
\raggedright
\smallskip
#1
\smallskip
\end{minipage}
}

%****************** mu=0 ********************
\begin{table*}
\caption{\label{table-flat}{\sl Flat potential.} --- Disorder averaged pdf's
of the local, inverse local, occupation and inverse occupation times of a
particle starting at the origin, diffusing in the Sinai potential
$U(x)=\sqrt{\sigma} B(x)$, where $B(x)$ represents the trajectory of a
Brownian motion in space with the initial condition $B(0)=0$.}
\begin{ruledtabular}
\begin{tabular}{p{.49\hsize}|p{.49\hsize}}
PURE CASE ($\sigma=0$) & DISORDERED CASE ($\sigma>0$)
\\ \hline
%
%%%%%%%%%%%%% LOCAL TIME %%%%%%%%%%%%%%%%%%%%%%%
\entry{\dm{\lss{P}{loc}(T|t)={\sqrt{2}\over\sqrt{\pi t}}
\exp\left[-{T^2\over 2t}\right]}}
&\entry{\dm{\overline{\lss{P}{loc}(T|t)} \xrightarrow[T/t\;
\text{fixed}]{t\rightarrow\infty, T\rightarrow\infty} {1\over t \log^2t}
f_1(T/t)}

\qquad\dm{
f_1(y)={2\over y}\, e^{-y/\sigma} K_0\left(y/\sigma\right)
}
}\\ \hline
%
%%%%%%%%%%%% INVERSE LOCAL TIME %%%%%%%%%%%%%%%%
\entry{\dm{\lss{I}{loc}(t|T)={T\over\sqrt{2\pi t^3}} \exp\left[- {T^2 \over
    2t}\right]}}
&\entry{\dm{\overline{\lss{I}{loc}(t|T)} 
\xrightarrow[t/T\; \text{fixed}]{t\rightarrow\infty, T\rightarrow\infty}
{1\over T \log^2 T}\, g_1(t/T)}

\qquad
\dm{g_1(x)={2\over x}\, e^{-1/\sigma x} K_0\left(1/\sigma x\right)}
}\\ \hline
%
%%%%%%%%%%% OCCUPATION TIME %%%%%%%%%%%%%%%%%%%%
\entry{\dm{\lss{P}{occ}(T|t)={1\over \pi \sqrt{T(t-T)}}, \quad 0<T<t}}
&\entry{ \dm{\overline{\lss{P}{occ}(T|t)} = \overline{R_L(T|t)} +
\overline{R_L(t-T|t)}}

\qquad \dm{\overline{R_L(T|t)} \xrightarrow{t\rightarrow\infty} {1\over \log
t} R(T)}

\qquad\qquad
\begin{math}
\begin{alignedat}{2}
& R(T)\approx {\sqrt{2}\sigma\over\sqrt{\pi
T}}, &\quad& \text{for small}\;   T  \\
& R(T) \sim {1\over 2T},  && \text{for large}\;  T 
\end{alignedat}
\end{math}
}\\ \hline
%
%%%%%%%%%%% INVERSE OCCUPATION TIME %%%%%%%%%%%%
\entry{\dm{\lss{I}{occ}(t|T)={\sqrt{T}\over\pi t\sqrt{t-T}}\,\theta(t-T)}}
&\entry{
\dm{\overline{\lss{I}{occ}(t|T)} 
\xrightarrow[t>T]{T\rightarrow \infty}
{1\over\log T}\,I_3(t-T)}

\qquad\qquad
\begin{math}
\begin{alignedat}{2}
& I_3(\tau)\approx{\sqrt{2}\sigma\over\sqrt{\pi\tau}}, &\quad& \text{for
small}\; \tau \\ & I_3(\tau)\sim{1\over 2\tau}, && \text {for large} \; \tau
\end{alignedat}
\end{math}
}
\end{tabular}
\end{ruledtabular}
\end{table*}

%************ mu>0 ***********************
\begin{table*}
\caption{\label{table-unstable}{\sl Unstable potential.} --- Disorder
averaged pdf's of the local, inverse local, occupation and inverse
occupation times of a particle starting at the origin, diffusing in the
unstable random potential $U(x)=-\mu |x| + \sqrt{\sigma} B(x)$, where
$\mu>0$ and $B(x)$ represents the trajectory of a Brownian motion in space
with the initial condition $B(0)=0$.  We denote $\nu=\mu/\sigma$.}
\begin{ruledtabular}
\begin{tabular}{p{.49\hsize}|p{.49\hsize}}
PURE CASE ($\sigma=0$) & DISORDERED CASE ($\sigma>0$) \\ \hline
%
%%%%%%%%%%%%% LOCAL TIME %%%%%%%%%%%%%%%%%%%%%%%
\entry{ \dm{\lss{P}{loc}(T|t) \xrightarrow{t\rightarrow\infty}
\lss{P}{loc}(T)},
\quad \dm{\lss{P}{loc}(T)= 2\mu e^{-2\mu T}}} 
&\entry{ \dm{ \overline{\lss{P}{loc}(T|t)} \xrightarrow{t\rightarrow\infty}
\overline{\lss{P}{loc}(T)}, \quad 
\overline{\lss{P}{loc}(T)} =
2\mu(1+\sigma T)^{-(2\nu +1)}}
}
\\ \hline
%
%%%%%%%%%%%% INVERSE LOCAL TIME %%%%%%%%%%%%%%%%
\entry{
\begin{math}
\begin{aligned}
\lss{I}{loc}(t|T)={T\over\sqrt{2\pi t^3}} &\exp\left[- {(T+\mu
    t)^2 \over 2t}\right] \\
&+\left(1-\exp[-2\mu T]\right)
\delta(t-\infty)
\end{aligned}
\end{math}
}
&\entry{
\begin{math}
\begin{aligned}
\overline{\lss{I}{loc}(t|T)} \xrightarrow[t/T
\;\text{fixed}]{t\rightarrow\infty, T\rightarrow\infty} {1\over T^{2\nu+1}}\,
& g_2(t/T) \\
&+\left(1-[1+\sigma T]^{-2\nu}\right) \delta(t-\infty)
\end{aligned}
\end{math}

\medskip
\quad
\begin{math}
\displaystyle
g_2(x)= \left[ {2\sqrt{\pi}\over\sigma^{2\nu-1} \Gamma^2(\nu)} \right]
{e^{-2/\sigma x}\over (\sigma x)^{2\nu+1}}\, U(1/2,1+\nu,2/\sigma x)
\end{math}
}
\\ \hline
%
%%%%%%%%%%% OCCUPATION TIME %%%%%%%%%%%%%%%%%%%%
\entry{
\dm{\lss{P}{occ}(T|t) = R_L(T|t) + R_L(t-T|t)}

\quad\dm{R_L(T|t) \xrightarrow{t\rightarrow\infty} R_L(T)}

\qquad
\begin{math}
\begin{aligned}
R_L(T)=&\mu\sqrt{2}\,\exp\left[-\frac{\mu^2 T}{2}\right]\\ 
&\times\left[ {1\over\sqrt{\pi T}}-
{3\mu\over\sqrt{2}}\exp\left({9\mu^2\over 2}T\right)
\erfc\left({3\mu\over\sqrt{2}}\sqrt{T}\right) \right]
\end{aligned}
\end{math}

\medskip
\qquad\quad
\begin{math}
\begin{alignedat}{2}
& R_L(T)\approx{\mu\sqrt{2}\over\sqrt{\pi T}},
&\quad&\text{for small}\; T\\
& R_L(T)\approx {\sqrt{2}\over9\mu\sqrt{\pi}}
{e^{-\mu^2T/2}\over T^{3/2}}, && \text{for large}\; T
\end{alignedat}
\end{math}
}
&\entry{
\dm{\overline{\lss{P}{occ}(T|t)} = \overline{R_L(T|t)} +
\overline{R_L(t-T|t)}}

\qquad\dm{\overline{R_L(T|t)} \xrightarrow{t\rightarrow\infty} 
\overline{R_L(T)}}

\medskip
\qquad\qquad
\begin{math}
\begin{alignedat}{2}
& \overline{R_L(T)} \approx {\mu\sqrt{2}\over\sqrt{\pi T}},
&\quad& \text{for small}\; T\\
&\overline{R_L(T)} \sim e^{-b T}, &&\text{for large}\; T 
\end{alignedat}
\end{math}

\bigskip
$b$ is given by the zero of $K_{\nu}(\sqrt{2p}/\sigma)$ closest to
origin in the left part of the complex--$p$ plane.  
} 
\\ \hline
%
%%%%%%%%%%% INVERSE OCCUPATION TIME %%%%%%%%%%%%
\entry{
\dm{\lss{I}{occ}(t|T)\xrightarrow[t>T]{T\rightarrow\infty}I_1(t-T)
+\frac{1}{2} \delta(t-\infty)
}

\qquad
\begin{math} 
\begin{aligned}
I_1(\tau)=&\mu\sqrt{2}\,\exp\left[-\frac{\mu^2 \tau}{2}\right]\\ 
&\times\left[ {1\over\sqrt{\pi
\tau}}- {3\mu\over\sqrt{2}}\exp\left({9\mu^2\over 2} \tau \right)
\erfc\left({3\mu\over\sqrt{2}}\sqrt{\tau}\right) \right]
\end{aligned}
\end{math}

\medskip
\qquad\quad
\begin{math}
\begin{alignedat}{2}
&I_1(\tau)\approx{\mu\sqrt{2}\over\sqrt{\pi \tau}},
&\quad& \text{for small}\; \tau \\
&I_1(\tau)\approx {\sqrt{2}\over9\mu\sqrt{\pi}} {e^{-\mu^2\tau/2}\over
\tau^{3/2}}, && \text{for large}\; \tau
\end{alignedat}
\end{math}
}
&\entry{
\dm{\overline{\lss{I}{occ}(t|T)}\xrightarrow[t>T]{T\rightarrow\infty}I_4(t-T)
+\frac{1}{2} \delta(t-\infty)
}

\qquad
\begin{math}
\begin{alignedat}{2} 
& I_4(\tau)\approx {\mu\sqrt{2}\over\sqrt{\pi \tau}}, &\quad& \text{for
small}\; \tau \\ & I_4(\tau) \sim e^{-b\tau}, &&\text{for large}\; \tau
\end{alignedat}
\end{math}

\bigskip
$b$ is the same constant as above.
}
\end{tabular}
\end{ruledtabular}
\end{table*}

%************ mu<0 ***********************
\begin{table*}
\caption{\label{table-stable}{\sl Stable potential} --- Disorder averaged
pdf's of the local, inverse local, occupation and inverse occupation times
of a particle starting at the origin, diffusing in the stable random
potential $U(x)=-\mu |x| + \sqrt{\sigma} B(x)$, where $\mu<0$ and $B(x)$
represents the trajectory of a Brownian motion in space with the initial
condition $B(0)=0$.  We denote $\nu=|\mu|/\sigma$.}
\begin{ruledtabular}
\begin{tabular}{p{.49\hsize}|p{.49\hsize}}
PURE CASE ($\sigma=0$) & DISORDERED CASE ($\sigma>0$) 
\\ \hline
%%%%%%%%%%%%% LOCAL TIME %%%%%%%%%%%%%%%%%%%%%%%
\entry{ \dm{\lss{P}{loc}(T|t) \sim \exp\left[-t
\Phi\left({T\over t}\right) \right] },
\quad$\Bigglb\{$\parbox[c]{3cm}{\raggedright
$t\rightarrow\infty, T\rightarrow\infty$\\ $T/t\;$ fixed}

\qquad \dm{\Phi(r)={1\over 2} \left(r-|\mu|\right)^2},
\quad near $r=|\mu|$
}
&\entry{ \dm{\overline{\lss{P}{loc}(T|t)} \xrightarrow[T/t\;
\text{fixed}]{t\rightarrow\infty, T\rightarrow\infty} {1\over t} f_2(T/t)}

\begin{math}
\displaystyle
f_2(y)=
\left[{2\sqrt{\pi}\over\sigma \Gamma^2(\nu)}\right]
\left(\frac{y}{\sigma}\right)^{2\nu-1} e^{-2y/\sigma} U(1/2,1+\nu,2y/\sigma)
\end{math}
}
\\ \hline
%
%%%%%%%%%%%% INVERSE LOCAL TIME %%%%%%%%%%%%%%%%
\entry{
\dm{\lss{I}{loc}(t|T)={T\over\sqrt{2\pi t^3}} \exp\left[-
    {(T-|\mu| t)^2 \over 2t}\right]}}
&\entry{ \dm{\overline{\lss{I}{loc}(t|T)} \xrightarrow[t/T\;
\text{fixed}]{t\rightarrow\infty, T\rightarrow\infty} {1\over T}\, g_3(t/T)}

\qquad
\begin{math}
\displaystyle g_3(x)= \left[{2\sigma\sqrt{\pi}\over \Gamma^2(\nu)} \right]
{e^{-2/\sigma x}\over (\sigma x)^{2\nu+1} }U(1/2,1+\nu,2/\sigma x)
\end{math}
} \\ \hline
%
%%%%%%%%%%% OCCUPATION TIME %%%%%%%%%%%%%%%%%%%%
\entry{
\dm{\lss{P}{occ}(T|t) \sim \exp\left[-t
\Phi\left({T\over t}\right) \right] },
\quad$\Bigglb\{$\parbox[c]{3cm}{\raggedright
$t\rightarrow\infty, T\rightarrow\infty$\\ $T/t\;$ fixed}

\qquad \dm{\Phi(r)=2\mu^2\left(r-\frac{1}{2}\right)^2},
\quad near \dm{r=\frac{1}{2}}
}
&\entry{
\dm{\overline{\lss{P}{occ}(T|t)} \xrightarrow[T/t\;
\text{fixed}]{t\rightarrow\infty, T\rightarrow\infty} {1\over t} f_o(T/t)} 

\qquad\dm{f_o(y)={1\over B(\nu,\nu)}\left[y(1-y)\right]^{\nu-1}, \quad 0\le
y\le 1} } \\ \hline
%
%%%%%%%%%%% INVERSE OCCUPATION TIME %%%%%%%%%%%%
\entry{\dm{\lss{I}{occ}(t|T)=I_2(t-T,T)\,\theta(t-T)}

\qquad
\dm{I_2(\tau,T) \xrightarrow{\text{large}\; T}{|\mu| T\over \sqrt{2\pi \tau^3}}
\exp\left[-\frac{\mu^2(\tau-T)^2}{2\tau}\right]}
}
&\entry{
\dm{\overline{\lss{I}{occ}(t|T)} 
\xrightarrow[t/T\;\text{fixed}]{t\rightarrow\infty, T\rightarrow\infty}
{1\over T}\, g_o(t/T)}

\qquad
\dm{g_o(x) = {1\over B(\nu,\nu)} {(x-1)^{\nu-1}\over x^{2\nu}}, \quad x>1}
}
%
%%%%%%%%%%%%%%%%%%%%%%%%%%%%%%%%%%%%%%%%%%%%%%%%%%
\end{tabular}
\end{ruledtabular}
\end{table*}

% \end{document}